\documentclass{book}
\usepackage{amsfonts}
\usepackage{amsmath}

\input{tcilatex}
\usepackage{graphicx}

\begin{document}

\frontmatter
\title{\textbf{Cycles in Nonlinear Macroeconomics}}
\author{\textbf{Anatoly V. Voronin} \\
\textit{Kharkiv National Economic University}\\
\textit{9a Lenin Ave., 61001 Kharkiv, Ukraine}\\
\textit{E-mail: voronin@hneu.edu.ua} \and \textbf{Sergey I. Chernyshov} \\
\textit{Lemma Joint Stock Insurance Company}\\
\textit{3 Kolomenskaya St., 61166 Kharkiv, Ukraine}\\
\textit{Tel.: +38 (057) 759 12 \ 99}\\
\textit{Fax: +38 (057) 759 12 \ 98}\\
\textit{E-mail: press@traffic.net.ua}\\
\textit{URL: www.lemma.ua }}
\date{}
\maketitle
\tableofcontents

\chapter*{\newpage}

\bigskip

\bigskip

\bigskip

\begin{quotation}
'\textit{All linear systems are equally happy, whereas each nonlinear system
is unhappy in its own way.'}

\qquad Yu. Bolotin, A. Tur, and V. Yanovskii

\qquad \qquad "Constructive Chaos"
\end{quotation}

\chapter{Preface}

Nonlinear dynamics is one of the most important and prospective trends of
the development of economic science. Powerful modern techniques of
qualitative theory of differential equations and related subjects of
mathematical topology provide broad possibilities of obtaining substantial
results of qualitative character, in the first place, in solving the
problems of economic forecasting. Mathematical economics is characterized by
two principally different ways of modeling, i.e., static and dynamic ones.
Following N. D. Kondratev \cite{[25]}, we shall dwell on a more detailed and
concrete characterization of these two purely theoretical approaches to the
study of economic reality.

The static theory considers economic processes in terms of their instant
manifestation, without any regard to inertial changes in time. The static
approach to the modeling of economic reality is based on the concept of the
equilibrium of interrelated elements of an economic system. The concept of
equilibrium itself had been sufficiently well familiar to scientists
concerned with mechanics before the appearance in 1776 of Adam Smith's
prominent work "An Inquiry into the Nature and Causes of the Wealth of
Nations" wherein, as it seems, the author managed to find an analogy between
the economic balance and the resultant force in mechanics. A. Smith put
forward the most substantial postulate of the general theory of equilibrium:
namely, an ability of the system of competition to achieve such a
distribution of resources that, in a sense, proves to be efficient.

On positions analogous to those of A. Smith stand also the constructions of
D. Ricardo who put forward the thesis of the freedom of competition and the
freedom of movement of labor and capital from one sphere of economic
relations to another. D. Ricardo was aware of the fact that the actual
price, the actual level of wages and incomes are variables with respect to
time, and, at that, he suggested that there exist a tendency towards
attaining a certain natural level of equilibrium for the above mentioned
characteristics.

An exhaustive formulation of the general concept of equilibrium rightfully
belongs to L. Walras, a representative of the school of marginalism (or
marginal utility). Works by L. Walras, S. Jevons, and V. Pareto unified the
theory of equilibrium with regard to an application to the spheres of
exchange, production, capital and money. They were subsequently elaborated
by J. Hicks and P. Samuelson. On the whole, these works are of rather broad,
comprehensive character: economy is considered as a set of individual
consumers and producers, and the number of involved variables is absolutely
unlimited. The system of general equilibrium is closed in the sense that the
whole set of variables is determined by given conditions. In order to verify
the compatibility of the system with the state of equilibrium, one only has
to compare the number of equations with the number of unknown variables.
However, there exists the problem of the existence of the point of
equilibrium and of its uniqueness.

The dynamic analysis in economics formed in parallel with economic theory
itself. A confrontation between the dynamic and static approaches can be
easily traced throughout the whole history of the economic thought.
Apparently, the reason lies in principal differences between the
understanding of the balance of the equilibrium of forces and casual
dynamics. There is a vast choice of literature references concerned with the
theory of economic dynamics. Among a large number of different problems of
evolutionary economics, the problems of economic growth and of business
cycles are the most important ones. In the treatment of N. D. Kondratev,
these distinctions between dynamic processes are interpreted as evolutionary
(nonrecurrent, irreversible) and wave-like (recurrent, reversible). By
evolutionary (or irreversible) processes one means those changes that in the
absence of external perturbative interactions flow in a certain, one and the
same, direction. Examples of such processes are given by tendencies of
population growth, increase in the total volume of production, etc. N. D.
Kondratev terms as wave-like (or reversible) those processes of changes that
at each point in time have their specific direction and, consequently,
change it permanently. In these processes, the phenomenon, being at a given
moment in a given state and changing it afterwards, sooner or later may
return to the initial state. The processes of changes in the prices of
different consumer goods, interest rates, the level of unemployment, etc.
may serve as examples.

In our view, more attention should be paid to the issue of slow
(low-frequency) oscillations in economics, i.e., to the so-called long
cycles (waves). N. D. Kondratev himself, while singling out long waves of
economic activities, related them to the industrial revolution at the end of
the XVIIIth - beginning of the XIXth centuries, the construction of railway
networks, the dissemination of new communication means (telephone,
telegraph) and of electric power, as well as rapid development of automotive
industry \cite{[2]}. Nowadays, the most wide-spread concept is that of five
long cycles whose length is approximately equal to 50 years:

\begin{itemize}
\item the end of the XVIIIth century - the first third of the XIXth century;

\item the second third of the XIXth century - the early 90s of the XIXth
century;

\item the end of the XIXth century - the 30s of the XXth century;

\item the 40s of the XXth century - the 70s of the XXth century;

\item from the 80s of the XXth century up to the present time.
\end{itemize}

Let us try to understand the very nature of the mechanism of the long cycle.
To initiate the expansion phase of the cycle, it is necessary to accumulate
not only inventions but also capital as well as a desire of entrepreneurs to
increase investments. For the industrialist, the dynamics of profit is a
factor of primary importance. As a matter of fact, the expansion wave of the
long cycle develops as a system of mutually related and mutually stimulating
phenomena: an innovation provides a possibility to improve the production
conditions, to reduce production costs and to increase profit, which
stimulates entrepreneurs to introduce innovations under the condition of
availability of necessary resources. Innovations give rise to an increase in
profit, which generates additional investments, an increase in the volume of
demand and a general positive movement of the growth rate of business
factors.

However, at a certain moment the dynamics of the process exhibits a return
point. A technological basis for this is provided by substantial weakening
of the factors that initiated the expansion phase of the cycle. A cessation
of their action slows down the growth of profitability and then decreases
it, which reduces the interest of business structures in further innovations
and investments. The industry slows down its growth, and negative effects of
the economic life, typical of long-wave decline, appear. In the process of
this decline, an increase in the number of new inventions takes place, which
creates a prerequisite for the completion of the decline and the beginning
of a new expansion wave.

N. D. Kondratev's undoubted merit consists in the fact that he based his
conclusions on an analysis of long temporal series of prices of the
commodity output, of interest rates, wages, etc. The above-described
dynamics is merely a simplified scheme, because the actual changes in the
economy are much more complicated and diverse.

The modern status of macroeconomics cannot be understood without an
evaluation of the contribution of J. M. Keynes. As a matter of fact , he
created macroeconomics as a science in the 1930's in order to explain the
causes of the Great Depression that proved to be the most large-scale
recession of the XXth century and, as such, the most important event in the
modern history of business cycles \cite{[18]}. The theory, existing at that
time, could not explain why the GDP of the USA had fallen by one third from
1929 to 1933 and the unemployment level had risen to one fourth of the total
work force. The classical theories were bases on the assumption that the
economy was at competitive equilibrium, with the market regulating
everything. In particular, high unemployment had to give rise to the
reduction in wage rates down to the level where the employers would agree to
employ all those who were willing to work for these wages and unemployment
would disappear by itself. However, in practice, this was not the case.
Therefore, Keynes put forward radically new ideas whose essence could be
reduced to two major postulates.

\textit{Firstly}, the economy is not at competitive equilibrium at each
separate moment; that is, the "invisible hand" of the market does not fulfil
its duties. The basic reason for this equilibrium is the conservation of
fixed prices and wage rates for a long time and the absence of adaptation to
the current market conditions. Secondly, the level of the development of the
economy is determined by the aggregate demand, and the latter, in its turn,
depends on some unexplained factors that were vaguely termed by Keynes as
"the brute of investors". On the basis of these two assumptions, the edifice
of the theory of macroeconomics was erected. According to its postulate, the
volume of the GDP of the country is influenced by the scale of expenditure
of consumers, investors and the government on commodities and services.
Therefore, business cycles are stipulated exactly by oscillations of the
demand rather than by resources of the country. The first and main formal
Keynesian model IS-LM was formulated on the basis of the Keynesian theory by
J. Hicks in 1937.

In the course of the next several decades, up to the mid-70s, the discussion
went on mostly in the mainstream of the Keynesian theory. The issue was
whether the government should at all try to revive the economy in the
periods of decline and, if it were so, by what means. According to the
above-mentioned position, the government had to react to the decline by
increasing government spending. From a point of view of other scientists and
specialists, stabilization should be achieved by means of control over money
supply: this point of view gave rise to the development of the doctrine of
monetarism according to which the main objective of the state is to avoid
strong oscillations of the money supply. Nevertheless, the positions of both
the antagonistic scientific trends agreed on the point that the basis
driving force of the cycle were oscillations of the volume of demand, and,
therefore, the main differences between "Keynesians" and "monetarists" were
almost completely obliterated.

However, in the mid-1970s the word economy faced a new phenomenon, i.e.,
stagflation, that could not be satisfactorily explained within the framework
of the Keynesian concept. At that time, there appeared critical works of R.
Lucas, subsequently the 1995 Nobel Prize winner, who criticized not only the
economic policy of the authorities but also the whole Keynesian theory of
business cycles for the disregard of optimum behavior of business agents
including the formation of rational expectations. He suggested that, in
contrast to investors and consumers in Keynes's models that followed certain
formal rules of behavior, business agents made, on the average, correct
forecasts of the future state of the economy and adhered to the strategy of
maximizing their own profits. All this created a demand for some alternative
theory of business cycles.

This niche was occupied by the American economists\ F. Kydland and E.
Prescott, who won the 2004 Nobel Prize for Economics. In their seminal paper 
\cite{[45]}, they proposed a new description of a real business cycle based
on the fact that firms maximized the profit and made decisions to invest
taking into account the expectation of future demand for their product and
of the development of technologies.

Kydland and Prescott presented a series of dynamic models and showed what
kind of behavior of basic economic variables (GDP, investment, and savings)
was to be expected depending on the effect of technological shocks upon
labour productivity and changes in external market conditions. The authors
demonstrated that the results of modelling were in satisfactory agreement
with the observed regularities. Besides, they drew an important conclusion
that a considerable part of oscillations of GDP in many countries
corresponded to the predictions of equilibrium models. In other words, there
is no need to introduce into these models deviations from market equilibrium
in the Keynesian spirit and to realize governmental stabilization policy.

For fairness, it should be noted that the theory of cycles of F. Kydland and
E. Prescott by no means explains all the phenomena of actual economic
reality: it is permanently subjected to constructive criticism by "new
Keynesianists". In particular, the most striking example of disagreement
between representatives of these two schools is an attempt to explain the
technological boom of the 1990s. One is just left with expectations that in
not too remote future a consensus will be achieved concerning the actual
sources of business cycles.

In our point of view, the achievement of this goal is impossible without
accepting the fact that economics, in essence, is a developing system and
should be constructed within the framework of the theory of developing
systems whose constructiveness is convincingly proved by the example of
chemical kinetics, biology and other natural sciences. In this theory, it is
shown that in the process of proceeding to the goal in the presence of
substantially nonlinear feedback couplings, there emerges a whole hierarchy
of instabilities that leads to the appearance of limit cycles, homoclinic
structures and \ to spontaneous formation of chaos. As a result of such
transformations (bifurcations), several different states of business
equilibrium may appear (the so-called \textit{effect of bistability}). The
methods of nonlinear mechanics allow us to predict the moment of the
occurrence of a chaotic regime in the system under investigation, the number
of possible states of equilibrium and to determine the character of their
stability. All this, in its turn, generates a principal general problem of
the construction of alternative scenarios of complex, irreversibly
developing system. It would be in order to mention here the statement of G.
Malinetskii \cite{[28]}: "Indeed, social-technological objects are complex
hierarchy systems, with various processes in them developing at different
characteristic time scales. The rate of their instability, the limits of
their predictability are different as well. In the economic system, the
horizon of the forecast has fallen sharply: whereas just 15 years ago 5-year
directive or indicative planning was a norm in the world, nowadays this is
out of question. In the world, there is more and more supply of 'quick
money' and less and less supply of 'slow money'. However, on the other hand,
stable development of the society requires slowly changing strategic goals,
scales of social values and norms, culture and ideology. One needs
technique, theory and formalism that would allow one to analyze possible
dynamics of such 'different-time-scale' systems and to direct their
development on this basis."

One can hardly question the fact that exactly the mathematical technique of
nonlinear dynamics provides the very tools that allow us to approach closely
the solution of the problem of "designing the future", of finding stable and
safe ways of social and economic evolution. The experience of the
application of methods and models of nonlinear dynamics has shown that many
complex developing systems can be satisfactorily described with the help of
a small number of variables, \textit{the order parameters}. The
determination of the order parameters is realized by reduction of the
multidimensional system to a subspace of a small dimension owing to methods
of the theory of bifurcations and of the theory of central manifolds.
However, exactly this fact predetermines the locality of the carried out
analysis of dynamic behavior of the studied system. Its applicability is
admissible only in small neighborhood of the bifurcation point, to solutions
of small amplitude. In what follows, we shall present other periodic
solutions of small amplitude generated as a result of the Andronov-Hopf
bifurcation of limit cycles and shall determine the character of their
stability.

In this book, the choice of the discussed models is made more or less
arbitrarily. The authors consider the models that are rooted in basic
principles of traditional economics, neoclassical synthesis and Keynesianism.

\mainmatter

\chapter{Instability and cycles in the Walras-Marshall model}

Economics operates such notions as the quantity of goods (productive
factors) and their price. In every market, there exist groups of sellers and
buyers. In this chapter, a model of the market for one kind of goods will be
considered. In the model of a single market the variables, i.e., functions
of time, are the volumes of bought and sold goods as well as their prices.
The basic principle of modeling of the market interaction is the formation
of balance relations between the volumes of the demand and the supply of
goods and, accordingly, the prices of the demand and the supply.

The problem of joint action of demand and supply as indicators determining
quantitative relations between the volume of a commodity and its price in a
given market is very precisely characterized by A. Marshall \cite{[29]}: "We
could ask on equal grounds whether the price is regulated by utility or
production costs, or whether a sheet of paper is cut by the upper or the
lower blade of the scissors. Indeed, if one blade is kept motionless and
cutting is carried out by the motion of the other blade, we can, without a
good deal of thinking, argue that cutting is done by the second blade.
However, such an argument is not completely exact, and it may be justified
only by a pretension to mere popularity rather than to an exact scientific
description of the realized process."

For more concrete understanding of the modern phenomenological basis of
demand and supply, we should \ present a definition of these notions, using
the formulations given, e.g., in \cite{[16]}.

By \textbf{the commodity demand} one means \textit{the quantity of this
commodity that an individual, a group of individuals or the population on
the whole are ready to buy per unit time under certain conditions}. A list
of these conditions includes the tastes and the preferences of the buyers,
the price of this commodity, the income rate, etc. By \textbf{the demand
price} one means the \textit{maximum price the buyers agree to pay for a
fixed quantity of a given commodity}. At the same time, \textit{the
dependence of the volume of the demand on its determining factors} is called 
\textbf{the demand function}.

Analogously, \textbf{the supply} \textit{serves as a characteristic of the
readiness of the seller to sell a certain quantity of the commodity in a
fixed period of time}.

By \textbf{the volume of the supply} one means \textit{the quantity of a
certain commodity that one seller or a group of sellers are willing to sell
in the market per unit time under certain conditions}.

These conditions, as a rule, include the properties of the applied
manufacturing technology, the price of the given commodity, the price rates
of the employed resources, tax rates, subventions, etc. \textbf{The supply
price} is \textit{the minimum price at which the seller agrees to sell a
certain quantity of a given commodity. The dependence of the volume of the
supply on the structure of its determining factors} is called \textbf{the
supply function}. \ Let us point out that the supply function as well as the
demand function can be represented in three ways: in the form of numerical
tables, graphically, and analytically. In what follows, we shall use only
analytical representations for the functions of the demand and the supply.

\section{Nonlinearity in the Walras model}

In classical economic theory, one employs two equally admissible but
principally different versions of the description of the mechanism of an
interaction between the demand and the supply. The first approach, worked
out by L. Walras, postulates that the driving force of changes in the price
is the volume of excess demand under a given instant value of the price. In
a dynamic aspect, the process of finding the equilibrium in L. Walras's
spirit can be represented in the form of the differential equation%
\begin{equation}
\frac{dP}{dt}=m\left( Y^{D}\left( P\right) -Y^{S}\left( P\right) \right) ,
\label{1.1}
\end{equation}%
where $P=P\left( t\right) $ is the price of the commodity;

$Y^{D}=Y^{D}\left( P\right) $ is the volume of the demand;

$Y^{S}=Y^{S}\left( P\right) $ is the volume of the supply;

$m>0$ is a constant of the time of the limit process;

$t$ is time.

The sign of the quantity $\Delta Y=Y^{D}-Y^{S}$, called the volume of excess
demand, determines the direction of the changes in the price. It is obvious
that for $\Delta Y>0$ the market price rises, whereas for $\Delta Y<0$ it
falls. The condition of the existence of an equilibrium price $P_{E}$ is the
existence of the solution to the equation%
\begin{equation}
Y^{D}\left( P_{E}\right) -Y^{S}\left( P_{E}\right) =0.  \label{1.2}
\end{equation}

There exists also a different approach to the problem under consideration,
attributed to A. Marshall. Its essence is that a change of the volume of the
mass of commodities in a given market is determined by the influence of the
difference between the demand price and the supply price, to which the
sellers (or the manufacturers) respond by an increase or a decrease in the
volume of the supply of the commodity. In a mathematical form, this
statement is expressed by means of the following differential equation:%
\begin{equation}
\frac{dP}{dt}=m\left( P^{D}\left( Y\right) -P^{S}\left( Y\right) \right) ,
\label{1.3}
\end{equation}%
where $Y=Y\left( t\right) $ is the volume of the commodity;

$P^{D}\left( Y\right) $ is the demand price;

$P^{S}\left( Y\right) $ is the supply price;

$m>0$ is a time constant.

In Eq. (\ref{1.3}), a surplus of the demand price over the supply price
stimulates an increase in $Y$; and if the supply price is higher than the
demand price, the value of $Y$ decreases. An equilibrium value of the volume
of the commodity $Y_{E}$ is determined from the equation%
\begin{equation}
P^{D}\left( Y\right) -P^{S}\left( Y\right) =0.  \label{1.4}
\end{equation}

The algebraic equations (\ref{1.2}) and (\ref{1.3}) may have only one or
several solutions. It means that both a unique state of equilibrium as well
as a set of equilibrium states is possible. It is obvious that the
nonuniqueness of equilibrium values of the volume and of the price of the
commodity is explained by the presence of nonlinear relations in the basic
equations.

An important problem is an analysis of the stability of the available states
of equilibrium. It is necessary to ascertain the reasons why an equilibrium
volume of the market remains constant under certain, remaining within
certain limit values, fluctuations of the price, or on the other hand, why,
under a given equilibrium price rate, changes in the volume of the commodity
also take place. In what follows, by the stability of equilibrium we
understand an ability of the overbalanced market to return again to the
initial state owing to the action of endogenous factors. Besides, the
problem of the stability of market equilibrium is directly related with the
problem of the necessity to employ additional measures to regulate market
relations.

First of all, let us consider the problem of stability of the economic model
(\ref{1.1}) described L. Walras's theory. Let us set the coefficient $m=1$
in Eq. (\ref{1.1}). In the neighborhood of the equilibrium point $P=P_{E}$
determined by the solution of Eq. (\ref{1.2}), we can approximately
represent the functions of the demand $Y^{D}\left( P\right) $ and of the
supply $Y^{S}\left( P\right) $ in the form of polynomials obtained by the
truncation of the corresponding Taylor series%
\begin{equation}
Y^{D}\left( P\right) \approx \sum_{i=0}^{k}\frac{d_{i}}{i!}\left(
P-P_{E}\right) ^{i},\quad Y^{S}\left( P\right) \approx \sum_{i=0}^{k}\frac{%
S_{i}}{i!}\left( P-P_{E}\right) ^{i},  \label{1.5}
\end{equation}%
\begin{equation*}
d_{i}=\frac{d^{i}Y^{D}\left( P_{E}\right) }{dP^{i}},\quad S_{i}=\frac{%
d^{i}Y^{S}\left( P_{E}\right) }{dP^{i}},\quad i=\overline{0,k}.
\end{equation*}

If we introduce a new variable $x=P-P_{E}$, which is a deviation of the
price from its equilibrium value, equation (\ref{1.1}) takes the form%
\begin{equation}
\dot{x}=F_{k}\left( x\right) ,  \label{1.6}
\end{equation}%
where $\dot{x}=\frac{dx}{dt}$, $F_{k}\left( x\right) =\sum_{i=0}^{k}\frac{%
a_{i}}{i!}x^{i}$, $a_{i}=d_{i}-S_{i}$. Note that from (\ref{1.2}) it follows 
$a_{0}=0$, because $d_{0}=S_{0}=Y_{E}$, which is an equilibrium volume of
the market. At the same time, $x=0$ is a stationary point (the state of
equilibrium) of Eq. (\ref{1.6}).

The differential equation (\ref{1.6}) is called a dynamic system of the
first order. The phase space of the considered system is one-dimensional,
therefore, the studied process of change in the price can be represented by
the motion of an image point on the phase straight \cite{[8]}.

Indeed, in general, the main elements that determine the partition of the
phase straight into trajectories are the states of equilibrium of the
system. The values $x=x_{j}$ that make the function $F_{k}\left( x\right) $
vanish are themselves independent phase trajectories. The rest of the
trajectories consist of line segments between the roots of the equation $%
F_{k}\left( x\right) =0$, or of rays forming half-intervals between one of
the roots and infinity. The direction of the motion of the image point along
these trajectories is determined by the sign of the function $F_{k}\left(
x\right) $: for $F_{k}\left( x\right) >0$ the image point moves to the
right, whereas for $F_{k}\left( x\right) <0$ it moves to the left. If the
form of the curve $z=F_{k}\left( x\right) $ is known, it is not difficult to
establish concrete partition of the phase straight into trajectories.

An example of such partition is given in Fig. 1.1, where the arrows show the
direction of the motion of the image point. From the structure of the
partition of the phase straight into trajectories, it follows directly that
the states of equilibrium of the system at the points $x_{1}$, $x_{4}$ are
stable, whereas they are unstable at the points $x_{2}$, $x_{3}$, $x_{5}$.
It is directly seen in Fig. 1.1 that in the stable states of equilibrium the
derivative $F_{k}^{\prime }\left( x\right) <0$, whereas in the unstable
states $F_{k}^{\prime }\left( x\right) >0$. The value $F_{k}^{\prime }\left(
x\right) =0$ may occur at points of both the stable and unstable state of
equilibrium. (This situation itself deserves independent consideration,
because it requires some additional conditions for the determination of the
type of stability of the stationary point.)

\FRAME{ftbpFU}{4.0421in}{2.3428in}{0pt}{\Qcb{The dependence of the
excess-demand function on price deviations.}}{}{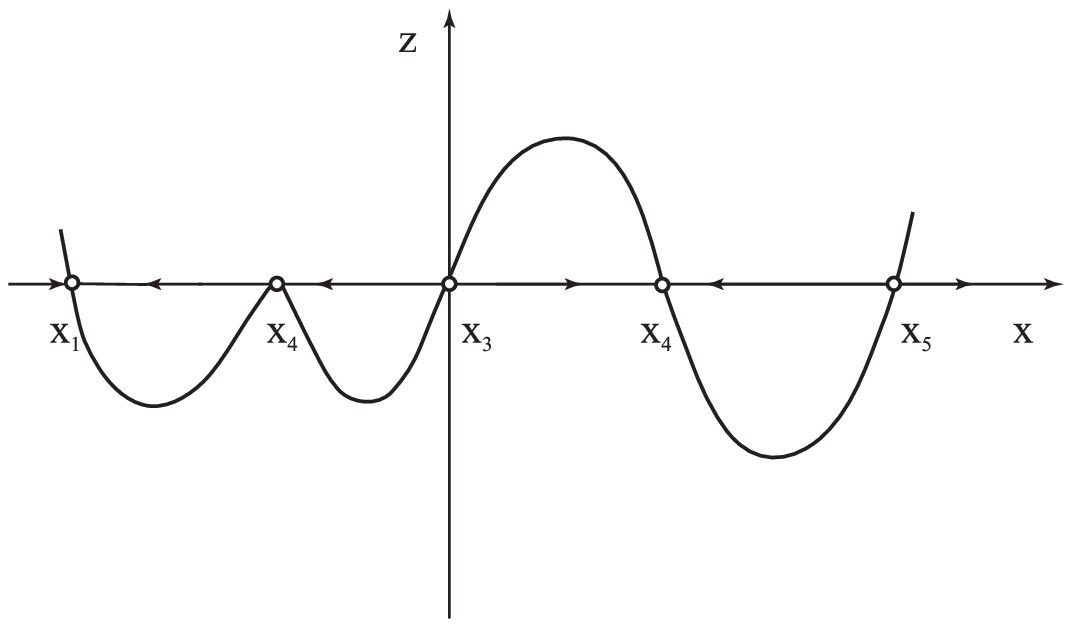}
{\raisebox{-2.3428in}{\includegraphics[height=2.3428in]{fig_1-1.ps}}}

As the character of the change of the variable in the first-order system (%
\ref{1.6}) is completely determined by the explicit form of the function $%
F_{k}\left( x\right) $, it is of interest to consider cases of different
values of the order of the polynomial $k$.

Let $k=1$. Then $F_{1}\left( x\right) =a_{1}x$ is a linear function, and
there exists the single state of equilibrium $x_{E}=0$. The stability
condition in this case is $F_{1}^{\prime }\left( 0\right) <0$. This
inequality reduces to the relation $d_{1}<S_{1}$ which is the classical
condition of stability of L. Walras.

Let us try to interpret the linear stability of L. Walras using the notion
of the elasticity of the demand and supply functions to price.

According to the definition of elasticity, in our notation, we have:%
\begin{equation*}
\eta _{D}=\frac{d_{1}P_{E}}{d_{0}},\quad \eta _{S}=\frac{S_{1}P_{E}}{S_{0}},
\end{equation*}%
or, taking into account that $d_{0}=S_{0}=Y_{E}$,%
\begin{equation}
\eta _{D}=\frac{d_{1}P_{E}}{Y_{E}},\quad \eta _{S}=\frac{S_{1}P_{E}}{Y_{E}},
\label{1.7}
\end{equation}%
where $\eta _{D}$, $\eta _{S}$ are coefficients of the demand and supply
elasticities to price. They are dimensionless, i.e., relative, quantities.

Therefore, the inequality $a_{1}=d_{1}-S_{1}<0$ can be easily reduced to the
form%
\begin{equation}
a_{1}=\frac{Y_{E}}{P_{E}}\left( \eta _{D}-\eta _{S}\right) <0.  \label{1.8}
\end{equation}

Given that $Y_{E}$, $P_{E}$ are always positive, the condition of L. Walras
is formulated as follows: for the stability of the linear system (\ref{1.6})
with $k=1$, it is necessary that the elasticity of the volume of the supply
to price should exceed the corresponding demand elasticity, i.e., $\eta
_{S}>\eta _{D}$. In other words, if we introduce the quantity $\eta =\eta
_{D}-\eta _{S}$ conditionally termed an excess-demand elasticity to price,
the stability of (\ref{1.6}) is determined by the sign of $\eta $: for $\eta
<0$, we have stability, and, on the contrary, for $\eta >0$, we have
instability.

Let us consider the peculiarities of the behavior of the system (\ref{1.6})
in the case $k=2$. Here, $F_{2}\left( x\right) =a_{1}x+\frac{a_{2}}{2}x^{2}$
is a quadratic function of the initial variable.

The equation $F_{2}\left( x\right) =0$, or $a_{1}x+\frac{a_{2}}{2}x^{2}=0$,
has two roots: $x_{E}^{1}=0$ and $x_{E}^{2}=-\frac{2a_{1}}{a_{2}}$. To
determine the character of stability of each singular point, it is necessary
to evaluate $F_{2}^{\prime }\left( x_{E}\right) $.

As a result of differentiation, we have:%
\begin{equation}
F_{2}^{\prime }\left( x_{E}\right) =a_{1}+a_{2}x_{E}\text{.}  \label{1.9}
\end{equation}

The substitution of the values of $x_{E}^{1}$ and $x_{E}^{2}$ in expression (%
\ref{1.9}) yields%
\begin{equation*}
F_{2}^{\prime }\left( 0\right) =a_{1}\text{,\quad }F_{2}^{\prime }\left( -%
\frac{2a_{1}}{a_{2}}\right) =-a_{1}.
\end{equation*}

It is thus obvious that the stability of both the states of equilibrium is
completely characterized by the sign of the quantity $a_{1}$ (or $\eta $).

In this case, if $a_{1}>0$, i.e., if the demand is more elastic than the
supply, the state of equilibrium $x_{E}^{1}=0$ is unstable, whereas $%
x_{E}^{2}=-\frac{2a_{1}}{a_{2}}$ is stable.

\FRAME{ftbpFU}{3.039in}{1.7417in}{0pt}{\Qcb{The diagram of the transcritical
bifurcation.}}{}{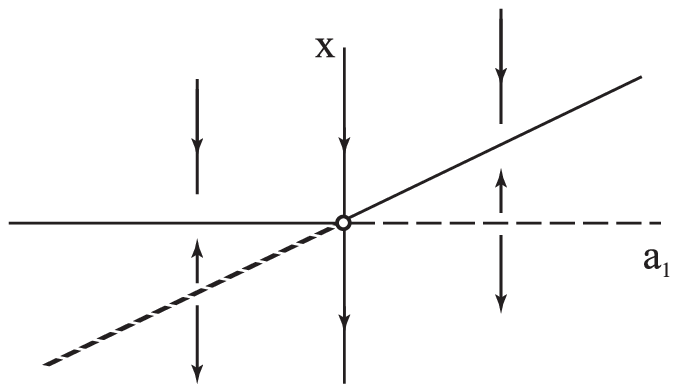}{\raisebox{-1.7417in}{\includegraphics[height=1.7417in]{fig_1-2.ps}}}

On the contrary, for $a_{1}<0$ (the demand is less elastic than the supply), 
$x_{E}^{1}$ is a stable state of equilibrium, and, accordingly, $x_{E}^{2}$
is an unstable one.

In a noncoarse situation, when $a_{1}$ is a small quantity changing its sign
in the neighborhood of zero, the so-called transcritical bifurcation appears
illustrating a change of stability of the states of equilibrium: see Fig.
1.2, where $a_{2}<0$.

As a result of this bifurcation, the singular points $x_{E}^{1}$ and $%
x_{E}^{2}$ merge for $a_{1}=0$, i.e., when the elasticity of the demand is
equal to that of the supply, to form the single two-fold state of
equilibrium $x_{E}=0$. At the same time, the condition $a_{2}\neq 0$ is
important.

Here, we can observe a considerable difference between the behavior of the
nonlinear system from that of the linear model, which manifests itself in
the present of two equilibrium values that are transformed into one and the
same point of equilibrium as a result of a transcritical bifurcation.

It seems to be reasonable to attribute pithy economic meaning to the
coefficient of the quadratic term, $a_{2}$, in terms of elasticities of the
demand and supply functions. To this end, it is necessary to find the
derivatives of the corresponding types of elasticity with respect to the
price at the point $P_{E}$. Skipping over intermediate transformations, we
present the following expressions for the quantities $d_{2}$ and $S_{2}$ as
functions of $\eta _{D}^{\prime }$, $\eta _{S}^{\prime }$, $\eta _{D}$, $%
\eta _{S}$:%
\begin{equation*}
d_{2}=\frac{Y_{E}}{P_{E}^{2}}\left( \eta _{D}^{\prime }P_{E}-\eta _{D}+\eta
_{D}^{2}\right) ,
\end{equation*}%
\begin{equation}
S_{2}=\frac{Y_{E}}{P_{E}^{2}}\left( \eta _{S}^{\prime }P_{E}-\eta _{S}+\eta
_{S}^{2}\right) .  \label{1.10}
\end{equation}

Subtracting the second equation of (\ref{1.10}) from the first one, we get%
\begin{equation*}
a_{2}=\frac{Y_{E}}{P_{E}^{2}}\left( \left( \eta _{D}^{\prime }-\eta
_{S}^{\prime }\right) P_{E}+\left( \eta _{D}+\eta _{D}-1\right) \left( \eta
_{D}-\eta _{S}\right) \right) ,
\end{equation*}%
or%
\begin{equation}
a_{2}=\frac{Y_{E}}{P_{E}^{2}}\left( \eta ^{\prime }P_{E}+\left( \eta
_{D}+\eta _{D}-1\right) \eta \right) .  \label{1.11}
\end{equation}

\textit{It is worth noting that the dependence of the coefficient }$a_{2}$%
\textit{\ on the excess-demand elasticity }$\eta $\textit{\ and on its
derivative with respect to the price }$\eta ^{\prime }$\textit{\ is a linear
function.}

Let us consider the case when the function of an excess demand in the system
(\ref{1.6}) is cubic. This takes place for $k=3$, and, accordingly,%
\begin{equation*}
F_{3}\left( x\right) =a_{1}x+\frac{a_{2}}{2}x^{2}+\frac{a_{3}}{6}x^{3}\quad
\left( a_{3}\neq 0\right) .
\end{equation*}

The cubic equation $F_{3}\left( x\right) =0$ may have, depending on the
coefficients, one or three real roots, and, accordingly, the system (\ref%
{1.6}) may have one or three states of equilibrium.

By analogy with the previous case, the stability of each state of
equilibrium is determined by the sign of $F_{3}\left( x_{E}\right) $.

Let the system (\ref{1.6}) have the representation%
\begin{equation}
\dot{x}=a_{1}x+\frac{a_{2}}{2}x^{2}+\frac{a_{3}}{6}x^{3}.  \label{1.12}
\end{equation}

Assuming the parameters $a_{1}$, $a_{2}$ to be small, sign-alternating
quantities, we consider the deformation of the saddle-node bifurcation with
an additional degeneracy in the quadratic term \cite{[4]}. In the
saddle-node case, the truncated system with $a_{1}=0$, $a_{2}=0$ takes the
form%
\begin{equation}
\dot{x}=\frac{a_{3}}{6}x^{3}.  \label{1.13}
\end{equation}

After some thinking, we can arrive at the conclusion that there exist small
perturbations of the function $\frac{a_{3}}{6}x^{3}$ when the system
possesses one or three hyperbolic fixed points (the states of equilibrium)
in the neighborhood of $x_{E}=0$, as well as certain "unusual" perturbations
when the system possesses two fixed points, with one of them being
nonhyperbolic. All the above-mentioned possibilities can be accounted for by
adding low-order terms. It is customary to represent this deformation in the
form of the equation%
\begin{equation}
\dot{u}=\theta _{1}+\theta _{2}u+\frac{a_{3}}{6}u^{3},  \label{1.14}
\end{equation}%
where $u=x+\frac{a_{2}}{a_{3}}$, $\theta _{1}=\frac{a_{2}}{a_{3}}\left( 
\frac{a_{2}^{2}}{3a_{3}}-a_{1}\right) $, $\theta _{2}=a_{1}-\frac{a_{2}^{2}}{%
2a_{3}}$.

Here, $\theta _{1}$, $\theta _{2}$ are also small quantities. The dynamics
of this vector field is formed, with the accuracy of topological
equivalence, by its fixed points and the types of their stability. Generally
speaking, mathematical theory of singularities provides the means for a
systematic study of zeros of families of mappings, with one of the examples
being given by the right-hand side of Eq. (\ref{1.14}).

Let us find the bifurcation set of parameters in the parameter plane $\theta
_{1}$, $\theta _{2}$ by demanding that the right-hand side of (\ref{1.14})
and its derivative with respect to the variable $u$ vanish:%
\begin{equation*}
\psi \left( u\right) =\theta _{1}+\theta _{2}u+\frac{a_{3}}{6}u^{3}=0,
\end{equation*}%
\begin{equation}
\psi ^{\prime }\left( u\right) =\theta _{2}+\frac{a_{3}}{2}u^{2}=0.
\label{1.15}
\end{equation}

By eliminating the variable $u$ from both the equations of the system (\ref%
{1.15}), we arrive at the following bifurcation set:%
\begin{equation}
8\theta _{2}^{3}+9a_{3}\theta _{1}^{2}=0.  \label{1.16}
\end{equation}

The bifurcation diagram of the system (\ref{1.14}) is presented in Fig. 1.3
for the case $a_{3}<0$. The system (\ref{1.14}) may possess one or three
coarse states of equilibrium. These states of equilibrium merge in pairs on
bifurcation lines $G_{1}$ and $G_{2}$ [formula (\ref{1.16})] that form
Neile's semicubical parabola with the origin at the point $\mathbf{A}=\left(
0,0\right) $. The point $\mathbf{A}$ corresponds to merging of all the three
states of equilibrium into one state. For the values of the parameters in
the plane $\theta _{1}$, $\theta _{2}$ that lie inside the parabola (region
2), the system (\ref{1.14}) possesses three states of equilibrium (two
stable states and one unstable state in between), whereas for the "outside"
values of the parameters (region 1) there exists one (stable) state of
equilibrium.

\FRAME{ftbpFU}{4.0421in}{4.3647in}{0pt}{\Qcb{Bifurcation diagrams of the
"triple-equilibrium" bifurcation.}}{}{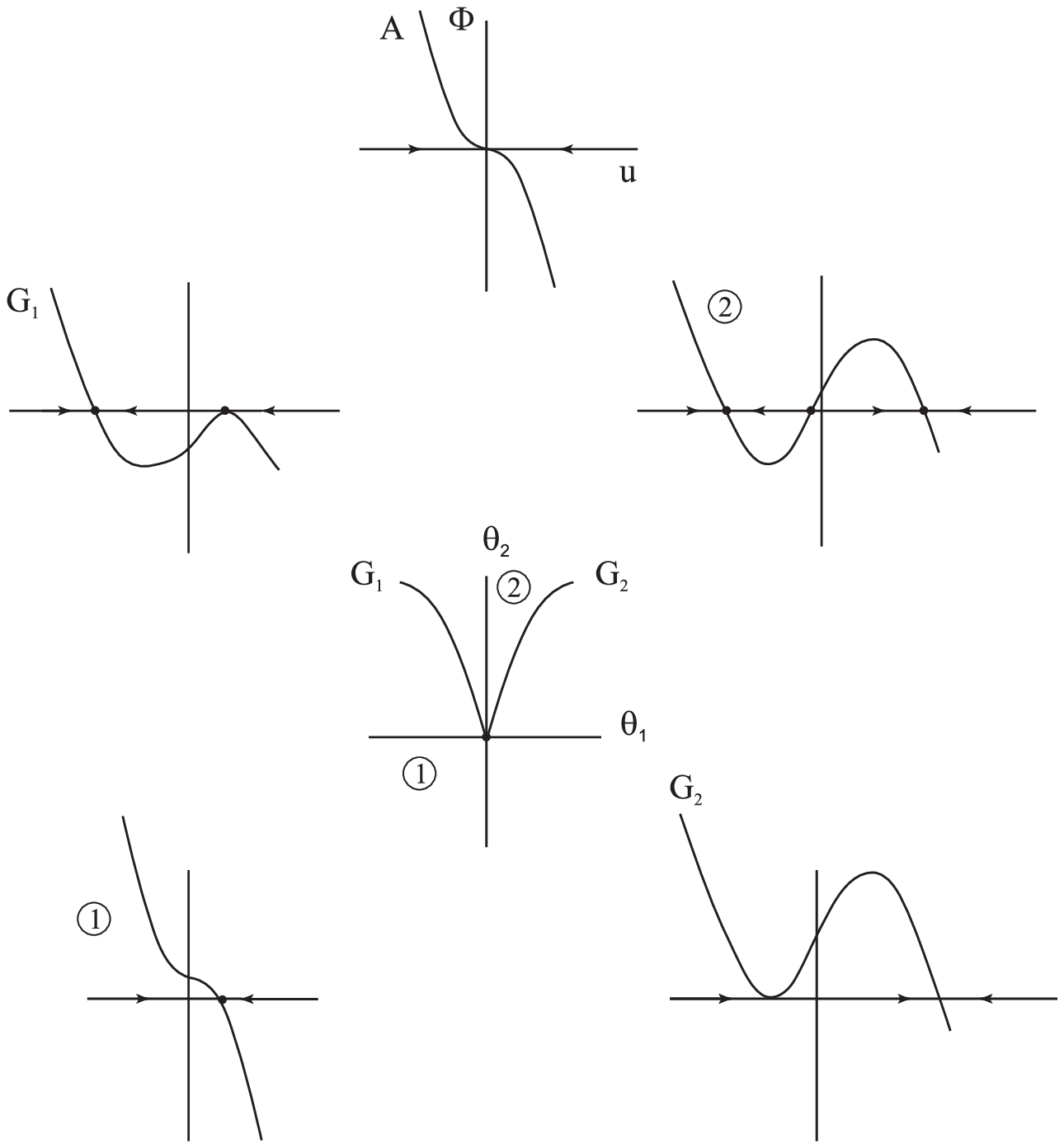}{\raisebox{-4.3647in}{\includegraphics[height=4.3647in]{fig_1-3.ps}}}

It should be noted that, for bifurcation values of the parameters, a
projective mapping of the manifold $\psi \left( u,\theta _{1},\theta
_{2}\right) =0$ onto the parameter space has fold-type singularities. The
dynamic system in the neighborhood of the bifurcation exhibits hysteresis.
Let us change the parameters in order to cross the semicubical parabola,
while keeping watch on the stable equilibrium regime (Fig. 1.3). In the
process of motion from left to right, the "breakdown of equilibrium" of
equilibrium occurs on the right branch $G_{2}$, whereas in the case of the
reverse motion it occurs on the left branch $G_{1}$. This phenomenon is
called "a hysteresis loop".

From formula (\ref{1.16}), it is not difficult to obtain an explicit form of
the bifurcation set in terms of the initial parameters $a_{1}$, $a_{2}$.
Using (\ref{1.14}) and (\ref{1.16}), we derive the relation%
\begin{equation}
8a_{1}a_{3}=3a_{2}^{2}.  \label{1.17}
\end{equation}

Furthermore, if we take into account the expressions for $a_{1}$ and $a_{2}$
in terms of the demand and supply elasticities to price, using (\ref{1.8})
and (\ref{1.11}), we obtain the following expression for the bifurcation
line:%
\begin{equation}
8P_{E}^{3}a_{3}\eta =3Y_{E}\left( \eta ^{\prime }P_{E}+\left( \eta _{D}+\eta
_{S}-1\right) \eta \right) ^{2}.  \label{1.18}
\end{equation}

In the equality (\ref{1.18}), the value of the excess-demand elasticity $%
\eta $ and its derivative with respect to the price $\eta ^{\prime }$ are
small quantities. Therefore, we can argue that the above-mentioned
bifurcation can be observed in the initial dynamic system for very close
values of the demand and supply elasticities, and of their derivatives, i.e.,%
\begin{equation}
\eta _{D}\approx \eta _{S},\quad \eta _{D}^{\prime }\approx \eta
_{S}^{\prime }.  \label{1.19}
\end{equation}

Such a type of behavior cannot be explained by means of the methods of
comparative statistics traditionally applied in economic analysis. Exactly
the analysis of the dynamics of the system in the neighborhood of the state
of equilibrium has shown that the behavior of the system is no longer
characterized by a unique and smooth reaction to small shifts of the
parameters. At the same time, there emerge a variety of states of
equilibrium, including multiple and sudden jumps stipulated by the
irreversibility of the flowing processes.

Analogously, one can carry out an investigation into the dynamics of the
process of establishing the equilibrium value of production according to the
concept of A. Marshall, described by the differential equation (\ref{1.3}.)
However, in contrast to the Walras model, in this case an analysis of the
stability of the studied dynamic system involves such substantial economic
characteristics as the values of elasticities of the prices of the demand
and of the supply to the production volume.

It should be noted that the above results ask for more profound
understanding of P. Samuelson's principle of correspondence whose validity
relies on the suggestion of predetermined stability of the economic system,
with changeability bearing a smooth character.

\section{A modified Walras-Marshall model}

Up to now, we have considered the processes of changes of the market price
and of the volume of commodity production as independent, and the
mathematical models (\ref{1.8}) and (\ref{1.11}) have been studied
separately. Therefore, in what follows, in order to study the dynamics of a
model of a certain industrial object, we shall make an attempt to unify the
equations of Walras and of Marshall into a single economic system, where the
processes of production and of formation of the price are mutually related 
\cite{[13]}.

The issue of price formation in productive economic systems has always been
and still remains relevant both for theoretical economic analysis and for
the solution of concrete practical problems of the enterprise as well. In
our view, it is important to synthesize two major factors of price
formation. Thus, on the one hand, classical theory of market price formation
argues that the market price corresponds to the equality of the demand and
of the supply in the commodity market. On the other hand, according to the
theory of the firm, it is known that, in the case of production balance, the
price of the products manufactured by the enterprise corresponds to the
marginal production costs. Thus, the first approach treats price formation
from the point of view of the consumer, whereas the second one does it from
the point of view of the producer. However, realities of the economic life
witness that the processes of changes in the price and in the volume of
production flow simultaneously and are interrelated. Therefore, it seems to
be reasonable to consider the market mechanism of balancing the demand with
the supply and the production process of accounting for the profit and costs
simultaneously, within the framework of a unified dynamic system, according
to the methodology presented in the work \cite{[5]}.

As the starting point, we consider a mathematical model describing the
dynamics of the interaction between the prices and the volume of production
(manufacturing):%
\begin{equation*}
\alpha \dot{P}=D\left( P\right) -Y,
\end{equation*}%
\begin{equation}
\beta \dot{Y}=P-P_{S}\left( Y\right) ,  \label{1.20}
\end{equation}%
where $P$ is the price of the produced and sold product;

$Y$ is the volume of the product in natural terms (the market supply of the
commodity);

$D\left( P\right) $ is the market demand for the product in natural terms;

$P_{S}\left( Y\right) $ is the supply price, equal to the marginal
production costs, i.e., $P_{S}\left( Y\right) =C^{\prime }\left( Y\right) $;

$C^{\prime }\left( Y\right) $ are the production costs (expenses);

$\alpha $, $\beta $ are constant positive parameters describing
characteristic times of transient processes.

The first equation of the system of the two ordinary differential equations (%
\ref{1.20}) is the classical model of market price formation in the form of
L. Walras (or P. Samuelson). They are based on the scheme of price formation
searching for the balance between the demand and the supply: for $D\left(
P\right) >Y$ the price rises, whereas for the opposite sign of the
inequality it falls. The second equation of (\ref{1.20}) describes the
process of establishing the balance between the price and the marginal
production costs with respect to the production (the value of production).
Here, it is assumed that the balance is disturbed and it is necessary to
regulate the volume of production: if $P>P_{S}\left( Y\right) $, the profit
of the producer $P\cdot Y-C\left( Y\right) $ rises with an increasing the
production volume, whereas in the opposite case one should decrease
production activities.

The model is based on substantial simplifications. \textit{Firstly}, the
production is assumed to be single-product. \textit{Secondly}, a local
outlet without competition is considered, when the whole supply is formed by
one producer. However, in spite of the above-mentioned assumptions, the
model (\ref{1.20}) admits complicated types of behavior, and their analysis
will be the subject of further consideration.

A formal analysis of qualitative properties of system (\ref{1.20}) should
start with the consideration of singular solutions that characterize the
states of equilibrium of the economic model.

By making the left-hand sides of (\ref{1.20}) vanish, we obtain two
constraint relations between equilibrium values of the price $P^{\ast }$ and
of the volume $Y^{\ast }$:%
\begin{equation*}
D\left( P^{\ast }\right) =Y^{\ast },
\end{equation*}%
\begin{equation}
P^{\ast }=P_{S}\left( Y^{\ast }\right) .  \label{1.21}
\end{equation}

Let us assume that the system of algebraic equations (\ref{1.20}) has, at
least, one positive solution $\left( P^{\ast },Y^{\ast }\right) $.
concerning the volume of the demand $D\left( P\right) $, we point out that
the dependence on the price is a substantially nonlinear function and there
exists a Taylor expansion up to the third order in the neighborhood of the
point $P^{\ast }$:%
\begin{equation*}
D\left( P\right) =d_{0}+d_{1}\left( P-P^{\ast }\right) +d_{2}\frac{\left(
P-P^{\ast }\right) ^{2}}{2}+d_{3}\frac{\left( P-P^{\ast }\right) ^{3}}{6}%
+O\left( \left\vert P-P^{\ast }\right\vert ^{4}\right) ,
\end{equation*}%
where $d_{i}=\frac{dD^{i}\left( P^{\ast }\right) }{dP^{i}}$, $i=\overline{0,3%
}$.

The cost (expenses) function is represented by a quadratic function of the
volume of production:%
\begin{equation*}
C\left( Y\right) =S_{1}\frac{Y^{2}}{2}+S_{0}Y+C_{0},
\end{equation*}%
where $S_{1}$, $S_{0}$, $C_{0}$ are constant parameters.

Accordingly, the marginal costs (the supply price) are described by the
formula%
\begin{equation*}
C^{\prime }\left( Y\right) =P_{S}\left( Y\right) =S_{1}Y+S_{0}.
\end{equation*}

Then, the system of equations (\ref{1.21}) can be represented as follows:%
\begin{equation*}
S_{1}D\left( P^{\ast }\right) +S_{0}-P^{\ast }=0,
\end{equation*}%
\begin{equation}
Y^{\ast }=\frac{P^{\ast }-S_{0}}{S_{1}}.  \label{1.22}
\end{equation}

Having preliminarily changed the time scale, it is convenient to study the
system (\ref{1.20}) in terms of new variables that represent deviations from
the equilibrium values $\tilde{P}=P-P^{\ast }$, $\tilde{Y}=Y-Y^{\ast }$. In
this case, the system (\ref{1.20}) reduces to the form%
\begin{equation*}
\overset{\cdot }{\tilde{P}}=d_{1}\tilde{P}+d_{2}\frac{\tilde{P}^{2}}{2}+d_{3}%
\frac{\tilde{P}^{3}}{6}-\tilde{Y},
\end{equation*}%
\begin{equation}
\overset{\cdot }{\tilde{Y}}=\gamma \left( \tilde{P}-S_{1}\tilde{Y}\right) ,
\label{1.23}
\end{equation}%
where $\gamma =\frac{\alpha }{\beta }$.

As is obvious, equations (\ref{1.23}) possess the trivial state of
equilibrium $\tilde{P}=0$, $\tilde{Y}=0$.

To study the stability of the trivial state of equilibrium, we write down an
explicit expression for the characteristic equation of the linear part of
the system (\ref{1.23}):%
\begin{equation}
\lambda ^{2}+\left( \gamma S_{1}-d_{1}\right) \lambda +\gamma \left(
1-d_{1}S_{1}\right) =0.  \label{1.24}
\end{equation}

The quadratic equation (\ref{1.24}) has negative real parts under the
conditions%
\begin{equation*}
\gamma S_{1}<d_{1},
\end{equation*}%
\begin{equation}
S_{1}d_{1}<1.  \label{1.25}
\end{equation}

The inequality (\ref{1.25}) determines restrictions on the parameters of the
initial system for stability in the linear approximation.

Let us consider in greater detail the situation in the vicinity of the
boundary of the region of stability, taking into account the equality%
\begin{equation}
\gamma C_{1}=d_{1}-\mu ,  \label{1.26}
\end{equation}%
where $\mu $ is small, sign-alternating quantity. It is obvious that, in
this case, the divergence of the vector field of the system (\ref{1.23}) is
equal to the small parameter $\mu $. Therefore, for $\mu <0$ the type of the
singular point (the state of equilibrium) is a stable focus, whereas for $%
\mu >0$ it is an unstable focus. In other words, for $\mu =0$, in the
neighborhood of the equilibrium state, there occurs the formation
(annihilation) of a limit cycle as a result of the Hopf bifurcation.

Let us verify the validity of the conditions of Hopf's theorem as applied to
the system (\ref{1.23}). The eigenvalues are determined (for $\mu =0$) by
the equality%
\begin{equation}
\lambda _{1,2}=\pm i\omega ,  \label{1.27}
\end{equation}%
where $i^{2}=-1$, $\omega ^{2}=\gamma -d_{1}^{2}$, i.e., they are purely
imaginary. Upon the substitution of (\ref{1.26}) in the quadratic equation (%
\ref{1.24}) and subsequent differentiation with respect to the parameter $%
\mu $, we obtain for $\mu =0$:%
\begin{equation}
\frac{d\lambda }{d\mu }=\lambda ^{\prime }\left( 0\right) =\frac{1}{2}-i%
\frac{d_{1}}{2\omega }.  \label{1.28}
\end{equation}

From (\ref{1.28}), it follows that the real part of the eigenvalue with
respect to the parameter does not vanish, i.e., the eigenvalues in the
complex plane cross the imaginary axis with a nonzero velocity. As a result,
all the conditions of the Hopf bifurcation theorem are fulfilled.

Let us turn once again to (\ref{1.26}) in order to give meaningful
interpretation of this equality.

The above-mentioned condition can be fulfilled is the parameters $d_{1}$ and 
$S_{1}$ have the same sign. As in the following the value $S_{1}=C^{\prime
\prime }\left( Y\right) $ will figure as the bifurcation parameter, its
positivity characterizes concavity of the function of expenses $C\left(
Y\right) $, whereas its\ negativity, accordingly, characterizes convexity.
From an economic point of view, $S_{1}<0$ determines a positive effect of
the volume of production ($C^{\prime \prime }\left( Y\right) <0$), whereas $%
S_{1}>0$ means that a rise in the costs outruns the output of the products ($%
C^{\prime \prime }\left( Y\right) >0$), i.e., the production is
resource-consuming.

In order to determine essential parameters of the limit cycle that
characterize its stability and the structure of periodic solutions, we
reduce the system of differential equations (\ref{1.23}) to the Poincar\'{e}
normal form by means of a corresponding change of variables $\tilde{P}=x_{1}$%
, $\tilde{Y}=d_{1}x_{1}+\omega x_{2}$. As a result of the reduction, we
obtain for $\mu =0$:%
\begin{equation*}
\dot{x}_{1}=-\omega x_{2}+\frac{d_{2}x_{1}^{2}}{2}+\frac{d_{3}x_{1}^{3}}{6},
\end{equation*}%
\begin{equation}
\dot{x}_{2}=\omega x_{1}-\frac{d_{1}d_{2}}{\omega }\frac{x_{1}^{2}}{2}-\frac{%
d_{1}d_{3}}{\omega }\frac{x_{1}^{3}}{6}.  \label{1.29}
\end{equation}

Using the explicit form of the coefficients of the nonlinear terms of the
system (\ref{1.29}), we derive an expression for the first Lyapunov quantity:%
\begin{equation}
l_{1}\left( 0\right) =\frac{d_{3}\left( \gamma -d_{1}^{2}\right)
+d_{2}^{2}d_{1}}{16\left( \gamma -d_{1}^{2}\right) }.  \label{1.30}
\end{equation}

For $l_{1}\left( 0\right) <0$, a stable limit cycle takes place, and a
corresponding regime of self-oscillations is called "soft". On the contrary,
if $l_{1}\left( 0\right) >0$, the limit cycle is unstable, the
self-oscillations break down "rigidly", with a manifestation of
irreversibility (hysteresis). The case $l_{1}\left( 0\right) =0$ is the most
complicated one in the sense of a variety of phase-plane structures of the
system (\ref{1.29}), because there appears a possibility of simultaneous
coexistence of two limit cycles (with one being stable and the other one
unstable) that subsequently merge into a single multiple limit cycle. This
bifurcation has codimension two and will not be studied in detail in this
Chapter.

The periodic solution of small amplitude $\varepsilon $ (up to a choice of
the initial phase) is itself written down in the form \cite{[37]}%
\begin{equation*}
P\left( t\right) =P^{\ast }+x_{1}\left( t\right) ,\quad Y\left( t\right)
=Y^{\ast }+d_{1}x_{1}\left( t\right) +\omega x_{2}\left( t\right) ,
\end{equation*}%
\begin{equation*}
x_{1}\left( t\right) =\varepsilon \cos \left( \frac{2\pi t}{T}\right)
\end{equation*}%
\begin{equation*}
+\frac{\varepsilon ^{2}d_{2}}{12\omega ^{2}}\left[ 3d_{1}-d_{1}\cos \left( 
\frac{4\pi t}{T}\right) +2\omega \sin \left( \frac{4\pi t}{T}\right) \right]
+O\left( \varepsilon ^{3}\right) ,
\end{equation*}%
\begin{equation*}
x_{2}\left( t\right) =\varepsilon \sin \left( \frac{2\pi t}{T}\right)
\end{equation*}%
\begin{equation}
+\frac{\varepsilon ^{2}d_{2}}{12\omega ^{2}}\left[ 3d_{1}-\omega \cos \left( 
\frac{4\pi t}{T}\right) -2d_{1}\omega \sin \left( \frac{4\pi t}{T}\right) %
\right] +O\left( \varepsilon ^{3}\right) .  \label{1.31}
\end{equation}

Here, $\varepsilon ^{2}=\frac{2\gamma \left( S_{1}-d_{1}\right) }{%
l_{1}\left( 0\right) }$ is the amplitude; $T\left( \varepsilon \right) =%
\frac{2\pi }{\omega }\left( 1+\tau _{2}\varepsilon ^{2}+O\left( \varepsilon
^{3}\right) \right) $, $\tau _{2}=\frac{d_{2}^{2}}{48\omega ^{2}}\left( 8+53%
\frac{d_{1}^{2}}{\omega ^{2}}\right) $ is the period of the cycle depending,
generally speaking, on the amplitude.

Thus, by the example of the system of two nonlinear differential equations (%
\ref{1.29}), it is easy to establish that, in contrast to a linear system,
periodic solutions are no longer harmonic, and the period and the amplitude
of the oscillations are interrelated.

As an illustration of the obtained results, consider examples of
economic-capacity cycles for different groups of commodities with regard to
the dependence of the demand functions on the income, following the
classification of the Swedish economist L. Tornquest \cite{[36]}.

\textbf{Example 1}. The demand function for essential commodities has a
representation%
\begin{equation*}
E=\frac{q_{1}D}{D+C_{1}},
\end{equation*}%
which reflects the fact that an increase in demand for these essential
commodities gradually slows down with an increase in the income and has a
has a limit $q_{1}>0$. The parameter $C_{1}$ is called \textit{the constant
of half-saturation of the income}.

Assuming that an equilibrium income is a function of the price,%
\begin{equation*}
E=PD\left( P\right) ,
\end{equation*}%
we express the demand in the form $D=D\left( P\right) $. After corresponding
transformations, we get%
\begin{equation}
D\left( P\right) =\frac{q_{1}-C_{1}P}{P}.  \label{1.32}
\end{equation}

With the help of (\ref{1.22}) and (\ref{1.32}), we write down the equation
for an equilibrium price:%
\begin{equation}
\left( P^{\ast }\right) ^{2}-\left( S_{0}-S_{1}C_{1}\right) P^{\ast
}-S_{1}q_{1}=0.  \label{1.33}
\end{equation}

Differentiating (\ref{1.32}) with respect to the price, we obtain the
coefficients of the demand function:%
\begin{equation}
d_{1}=-\frac{q_{1}}{\left( P^{\ast }\right) ^{2}},\quad d_{1}=\frac{2q_{1}}{%
\left( P^{\ast }\right) ^{3}},\quad d_{1}=-\frac{6q_{1}}{\left( P^{\ast
}\right) ^{4}}.  \label{1.34}
\end{equation}

From (\ref{1.34}), it follows that the quantities $d_{1}$, $d_{2}$ are
negative numbers, and, therefore, by (\ref{1.30}), the emerging limit cycle
is stable. At the same time, the reason for the emergence of the cycle is
the fact of positive influence of the effect of the scale of production of
the given group of commodities, i.e., $S_{1}<0$.

\textbf{Example 2}. The demand-for-luxury-goods function is represented in
the form%
\begin{equation*}
E=\frac{q_{1}D\left( D-b_{2}\right) }{D+C_{2}}.
\end{equation*}

By analogy with Example 1, we express the demand as a function of the price:%
\begin{equation}
D\left( P\right) =\frac{C_{2}P+q_{2}b_{2}}{q_{2}-P},  \label{1.35}
\end{equation}%
where $q_{2}$, $b_{2}$, $C_{2}$ are positive parameters.

The equation for an equilibrium price has the form%
\begin{equation}
\left( P^{\ast }\right) ^{2}-\left( q_{2}+S_{0}-S_{1}C_{2}\right) P^{\ast
}-q_{2}\left( S_{1}d+S_{0}\right) =0.  \label{1.36}
\end{equation}

Assuming that (\ref{1.30}) has at least one positive root $P^{\ast }$, we
evaluate the coefficients of the powers of $P$:%
\begin{equation}
d_{1}=\frac{q_{2}\left( b_{2}+C_{2}\right) }{\left( q_{2}-P^{\ast }\right)
^{2}},\quad d_{2}=\frac{2q_{2}\left( b_{2}+C_{2}\right) }{\left(
q_{2}-P^{\ast }\right) ^{3}},\quad d_{3}=\frac{6q_{2}\left(
b_{2}+C_{2}\right) }{\left( q_{2}-P^{\ast }\right) ^{4}}.  \label{1.37}
\end{equation}

In this case, the coefficients $d_{1}$ and $d_{2}$ are positive numbers, and
the substitution of their values in (\ref{1.30}) ensures the condition $%
l_{1}\left( 0\right) >0$,$\ $which is an indication of a catastrophic loss
of stability of the limit cycle. As $d_{1}>0$, the emergence of a limit
cycle requires the fulfillment of the condition $S_{1}>0$, which is possible
only for resource-consuming production with an outrunning increase in the
costs.

Let us consider one more version of the model (\ref{1.20}), assuming, as a
preliminary, that the supply price $P_{S}$ is a nonlinear function of the
volume of production. For simplicity, we consider the quantities $P$ and $Y$
to be deviations from certain equilibrium values $P^{\ast }$ and $Y^{\ast }$.

Concerning the demand function $D\left( P\right) $ and the price, we put
forward an assumption that they quadratically depend on their arguments,
i.e.,%
\begin{equation*}
D\left( P\right) =d_{2}\frac{P^{2}}{2}-d_{1}P,
\end{equation*}%
\begin{equation*}
P_{S}\left( Y\right) =S_{2}\frac{Y^{2}}{2}-S_{1}Y.
\end{equation*}

Then, the system (\ref{1.20}) can be represented in the form%
\begin{equation*}
\dot{P}=d_{2}\frac{P^{2}}{2}-d_{1}P-Y,
\end{equation*}%
\begin{equation}
\dot{Y}=b^{2}\left( P+S_{1}Y-S_{2}\frac{Y^{2}}{2}\right) ,  \label{1.38}
\end{equation}%
where $b^{2}=\gamma $.

One can argue that a linear analysis of the stability of (\ref{1.38})
completely corresponds to the previously obtained results for the system (%
\ref{1.23}), including a verification of the validity of Hopf's theorem, up
to the substitution of $b^{2}$ for the parameter $\gamma $. Therefore,
assuming the closeness of the bifurcation parameter $S_{1}$ to the quantity $%
\frac{d_{1}}{b^{2}}$, we transform (\ref{1.38}) to the normal form of the
given bifurcation with the help of the change of variables $x_{1}=P$, $x_{2}=%
\frac{Y-d_{1}P}{\omega }$ ($\omega ^{2}=b^{2}-d_{1}^{2}$). After some
transformations and the introduction of a new time scale $\tau =\omega t$,
we obtain:%
\begin{equation*}
\dot{x}_{1}=-x_{2}+\frac{d_{2}x_{1}^{2}}{2},
\end{equation*}%
\begin{equation}
\dot{x}_{2}=x_{1}+\frac{d_{1}\left( d_{2}-b^{2}d_{1}S_{2}\right) }{\omega
^{2}}\frac{x_{1}^{2}}{2}+\frac{b^{2}d_{1}S_{2}}{\omega ^{2}}%
x_{1}x_{2}-b^{2}S_{2}\frac{x_{2}^{2}}{2}.  \label{1.39}
\end{equation}

Let us represent the system of two ordinary differential equations (\ref%
{1.39}) in the form of a differential equation for the complex variable $%
Z=x_{1}+ix_{2}$, $\bar{Z}=x_{1}-ix_{2}$:%
\begin{equation}
\dot{Z}=iZ+g_{20}\frac{Z^{2}}{2}+g_{11}Z\bar{Z}+g_{02}\frac{\bar{Z}^{2}}{2},
\label{1.40}
\end{equation}%
where $g_{11}=\frac{1}{4\omega }\left( d_{2}+\frac{i}{\omega }\left(
d_{1}d_{2}-b^{4}S_{2}\right) \right) $;

$g_{20}=\frac{1}{4\omega }\left( d_{2}+2b^{2}d_{1}S_{2}+\frac{i}{\omega }%
\left( d_{1}d_{2}+b^{2}S_{2}\left( b^{2}-2d_{1}^{2}\right) \right) \right) $;

$g_{02}=\frac{1}{4\omega }\left( d_{2}-2b^{2}d_{1}S_{2}+\frac{i}{\omega }%
\left( d_{1}d_{2}+b^{2}S_{2}\left( b^{2}-2d_{1}^{2}\right) \right) \right) $.

Now, we possess all the necessary information for the evaluation of the
first Lyapunov quantity%
\begin{equation*}
l_{1}=-\frac{1}{2}\func{Im}g_{20}g_{11}.
\end{equation*}

Taking into account the explicit form of the coefficients of Eq. (\ref{1.40}%
), we get:%
\begin{equation}
l_{1}=\frac{d_{1}}{16\omega ^{3}}\left( b^{6}S_{2}^{2}-d_{2}^{2}\right) .
\label{1.41}
\end{equation}

From (\ref{1.41}), it is obvious that for $d_{2}^{2}>b^{6}S_{2}^{2}$ the
limit cycle is stable, whereas for $d_{2}^{2}>b^{6}S_{2}^{2}$ the
instability of the limit cycle takes place.

In Fig. 1.4, we present cyclic changes of the price and of the volume of the
commodity of the considered state of equilibrium for the following values of
the parameters of the system (\ref{1.38}):%
\begin{equation*}
b=1;\quad d_{1}=0.5;\quad d_{2}=2;\quad S_{1}=0.5;\quad S_{2}=1.5.
\end{equation*}

The condition $d_{2}^{2}=\pm b^{3}S_{2}$ makes the expression for the first
Lyapunov quantity vanish. This may mean that the initial system (\ref{1.38})
possesses two limit cycles that can merge into one two-fold cycle by means
of trajectory compaction. This situation is possible, if the second Lyapunov
quantity, $l_{2}$, does not vanish. Setting for definiteness $S_{2}=\frac{%
d_{2}}{b^{3}}$, as a result of evaluation, we obtain $l_{2}=0$. Moreover, $%
l_{3}$, the next (third) Lyapunov quantity also vanishes under the given
conditions. The fact that all the first three Lyapunov quantities vanish
means that the state of equilibrium is a center, not a focus. In other
words, the initial system (\ref{1.38}), for%
\begin{equation}
S_{1}=\frac{d_{1}}{b^{2}},\quad S_{2}=\frac{d_{2}}{b^{3}},  \label{1.42}
\end{equation}%
turns into a conservative one, with the conservation of the phase volume.

Let us write down the explicit form of (\ref{1.38}), eliminating the
parameters $S_{1}$,$S_{2}$ with the help of Eqs. (\ref{1.42}):%
\begin{equation*}
\dot{P}=d_{2}\frac{P^{2}}{2}-d_{1}P-Y,
\end{equation*}%
\begin{equation}
\dot{Y}=b^{2}P+d_{1}Y-\frac{d_{2}}{b}\frac{Y^{2}}{2}.  \label{1.43}
\end{equation}

\FRAME{ftbpFU}{4.0421in}{5.9093in}{0pt}{\Qcb{The limit cycle in the system
(1.38).}}{}{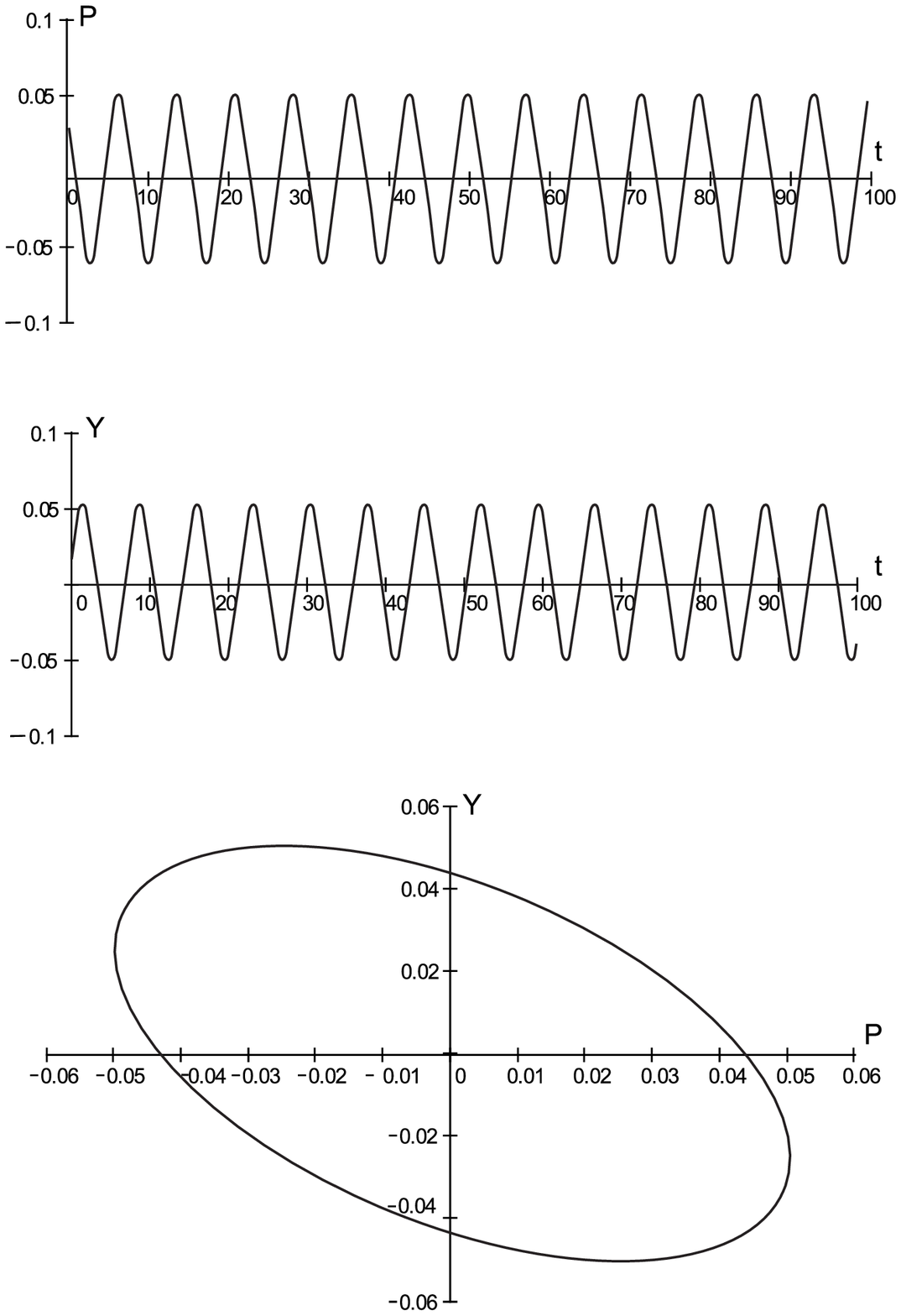}{\raisebox{-5.9093in}{\includegraphics[height=5.9093in]{fig_1-4.ps}}}

It is important for us to find the first integral of the system (\ref{1.43}%
). To this end, we transform the variables $P$ and $Y$ with the help of the
linear substitution%
\begin{equation}
Y_{1}=P-\frac{Y}{b},\quad Y_{2}=\frac{1}{\xi }\left( P+\frac{Y}{b}\right) +%
\frac{1}{2a},  \label{1.44}
\end{equation}%
where $\xi ^{2}=\frac{b-d_{1}}{b+d_{1}}$, $a=\frac{d_{2}}{4\omega _{1}}$, $%
\omega ^{2}=b^{2}-d_{1}^{2}$.

Then, equations (\ref{1.43}) can be represented in the form%
\begin{equation*}
\dot{Y}_{1}=aY_{1}^{2}+a\xi ^{2}Y_{2}^{2}-\left( \xi ^{2}+1\right) Y_{2}+%
\frac{\xi ^{2}+2}{4a},
\end{equation*}%
\begin{equation}
\dot{Y}_{2}=2aY_{1}Y_{2}.  \label{1.45}
\end{equation}

The system (\ref{1.45}) can be reduced to a total differential equation:%
\begin{equation}
\left( aY_{1}^{2}+a\xi ^{2}Y_{2}^{2}-\left( \xi ^{2}+1\right) Y_{2}+\frac{%
\xi ^{2}+2}{4a}\right) dY_{2}=2aY_{1}Y_{2}dY_{1}.  \label{1.46}
\end{equation}

By introducing the integrating factor $Y_{2}^{-2}$ into (\ref{1.45}), we
obtain the following equation:%
\begin{equation}
\left( a\xi ^{2}-\frac{\xi ^{2}+1}{Y_{2}}+\frac{\xi ^{2}+2}{4aY_{2}^{2}}%
\right) dY_{2}=2ad\left( \frac{Y_{1}^{2}}{Y_{2}}\right) .  \label{1.47}
\end{equation}

The integration of Eq. (\ref{1.45}) yields%
\begin{equation}
Y_{1}^{2}=\xi ^{2}Y_{2}^{2}-\frac{\xi ^{2}+1}{a}Y_{2}\ln \left\vert
Y_{2}\right\vert +KY_{2}-\frac{\xi ^{2}+2}{4a},  \label{1.48}
\end{equation}%
where $K$ is an arbitrary constant determined from the initial conditions.

Thus, expressions (\ref{1.44}) and (\ref{1.48}) determine a functional
interrelation between the price and the volume of the commodity, which is a
permanent balance between the income of the producer and his expenses at any
moment of time.

Returning to the system (\ref{1.38}), we should be reminded of the fact that
the coefficients $d_{1}$, $d_{2}$ are the demand elasticity and its
derivative with respect to the price, whereas $S_{1}$, $S_{2}$ are price
demand elasticity and its derivative with respect to the volume of the
commodity, i.e., 
\begin{equation*}
d_{1}=\eta _{D},\quad d_{2}=\eta _{D}^{\prime },\quad S_{1}=\eta _{S},\quad
S_{2}=\eta _{S}^{\prime }.
\end{equation*}

(It is understood that the elasticities and their derivatives are evaluated
at the trivial state of equilibrium.)

Then, conditions (\ref{1.42}) take a somewhat different form:%
\begin{equation*}
\eta _{D}=b^{2}\eta _{S},
\end{equation*}%
\begin{equation}
\eta _{D}^{\prime }=b^{3}\eta _{S}^{\prime }.  \label{1.49}
\end{equation}

The first expression in (\ref{1.49}) implies a transition from a stable
state of equilibrium to an unstable one with the formation (or annihilation)
of a limit cycle, whereas the second expression in (\ref{1.49}) determines
the boundary between the types of stability of self-oscillation regimes.

\chapter{Periodic regimes in nonlinear models of the multiplier-accelerator}

\section{A multiplier-accelerator model with finite duration of an
investment lag}

In this Chapter, we consider an example of a certain kind of
multiplier-accelerator models that illustrates at a phenomenological level
different aspects of the problems of business cycles. As the starting point,
we use the investment model of Goodwin \cite{[1]} that possesses a nonlinear
element built in the system of a multiplier-accelerator interaction. The
dynamics of this model is characterized by the lag of two types: from the
point of view of investment demand, there is a finite-duration lag of the
action of the accelerator, whereas there is a continuously distributed lag
on the part of supply.

As the main variable, the model involves profit or an output $Y=Y\left(
t\right) $. An excess demand is formalized by the equation%
\begin{equation}
D=C+I+G,  \label{2.1}
\end{equation}%
where $D$ is a cumulative demand comprising a consumer demand $C$, an
investment demand $I$, and independent expenses $G$. All the above-mentioned
terms are the actual costs. It is assumed that the consumption $C$ is
directly proportional to the profit and the lag is absent, i.e., $C=cY$,
where the quantity $0<c<1$ characterizes a marginal propensity to consume.
We also assume that the actual investments outlay is carried out with a
certain fixed lag of the duration of $\theta $ units of time after an
investment decision $F\left( t\right) $ is made, i.e., $I\left( t\right)
=F\left( t-\theta \right) $. Exactly here, the action of the accelerator
manifests itself as a functional relation between the volume of investment
decisions $F\left( t\right) $ and an instantaneous velocity of the change in
the profit (output volume) $\frac{dY\left( t\right) }{dt}$. In the most
general form, this relation can be represented as follows:%
\begin{equation*}
F\left( t\right) =\varphi \left( \frac{dY\left( t\right) }{dt}\right) ,
\end{equation*}%
where $\varphi $ is a certain nonlinear function possessing the property of
saturation.

In other words, for small changes in the profit, the classical linear
accelerator comes into play, whereas for a further increase in the profit
(output volume), the function $F$ reaches its upper bound $F_{\max }$
determined by the resource and capacity limits of the structure of the
production.

When the volume of the production output substantially decreases, the
quantity $F$ tends to its lower bound $F_{\min }$ that depends, generally
speaking, on the amortization quota of the fixed capital. For the sake of
convenience of further mathematical transformations, within the framework of
the model under consideration, we restrict ourselves to a Taylor expansion
of the function $\varphi \left( x\right) $ up to the third order:%
\begin{equation*}
\varphi \left( x\right) \approx \varphi _{1}x+\varphi _{2}\frac{x^{2}}{2}%
+\varphi _{3}\frac{x^{3}}{6},
\end{equation*}%
where $\varphi _{i}$, $i=\overline{1,3}$, are corresponding derivatives of
the function $\varphi \left( x\right) $.

Given that we have already described the main components of the cumulative
demand function, we obtain the following equation:%
\begin{equation}
D\left( t\right) =cY\left( t\right) +\varphi \left( \frac{dY\left( t\right) 
}{dt}\right) +G.  \label{2.2}
\end{equation}

Concerning the independent expenses $G$, it should be noted that they can be
considered fixed, i.e., $G=const$.

Consider next the situation on the part of supply. The main assumption is
that value of the output volume $Y$ lags behind the value of the cumulative
demand $D$. The lag is considered to be continuously distributed, and it can
be represented by a linear first-order differential equation:%
\begin{equation}
\varepsilon \frac{dY}{dt}=D-Y,  \label{2.3}
\end{equation}%
where $\varepsilon >0$ is a constant of the time lag characterizing the
dynamics of adjustment between the demand and the supply.

A balance synthesis of the demand and the supply, by (\ref{2.2}) and (\ref%
{2.3}), yields a differential equation with a retarded argument:%
\begin{equation}
\varepsilon \frac{dY\left( t\right) }{dt}-\varphi \left( \frac{dY\left(
t-\theta \right) }{dt}\right) +\left( 1-c\right) Y-G=0.  \label{2.4}
\end{equation}

Equation (\ref{2.4}) represents the general \textit{multiplier-accelerator
model of Goodwin} with nonlinear interactions.

With regard to the form of (\ref{2.4}), it is nothing but a mixed
differential-difference equation.

As is obvious, equation (\ref{2.4}) has a singular solution $Y^{\ast }=\frac{%
G}{1-c}$ that represents an equilibrium level of profit (output volume)
resulting from the action of the static multiplier. In what follows, it is
reasonable to introduce the variable $y\left( t\right) =\frac{Y\left(
t\right) -G}{1-c}$ that represents a deviation of the profit from its
equilibrium value.

In terms of the new variable $y\left( t\right) $, equation (\ref{2.4}) takes
the form%
\begin{equation}
\varepsilon s\frac{dy\left( t\right) }{dt}-\varphi \left( s\frac{dy\left(
t-\theta \right) }{dt}\right) +s^{2}y\left( t\right) =0,  \label{2.5}
\end{equation}%
where $s=1-c$ is a marginal propensity to save.

The given model, represented by Eq. (\ref{2.5}), determines the dynamics of
profit (output volume) in terms of deviations from the above-mentioned
equilibrium level $Y^{\ast }=\frac{G}{1-c}$.

For the sake of a further analysis of dynamic properties of the considered
process of changes in profit, it is necessary to carry out a sequence of
mathematical transformations and simplifications that will allow us to
reduced the mixed first-order differential-difference equation to an
ordinary differential equation of higher order \cite{[35]}.

Let us introduce new time $\tau =t-\theta $ instead of $t$. Then, equation (%
\ref{2.5}) can be represented as%
\begin{equation}
\varepsilon s\frac{dy\left( \tau +\theta \right) }{d\tau }-\varphi \left( s%
\frac{dy\left( \tau \right) }{d\tau }\right) +s^{2}y\left( \tau +\theta
\right) =0.  \label{2.6}
\end{equation}

Our next step consists in expanding the left-hand side of Eq. (\ref{2.6})
into a power series in $\theta $, while retaining terms containing the first
power of $\theta $. We obtain:%
\begin{equation*}
\varepsilon s\frac{dy\left( \tau \right) }{d\tau }+\varepsilon s\theta \frac{%
d^{2}y\left( \tau \right) }{d\tau ^{2}}+s^{2}\left( y\left( \tau \right)
+\theta \frac{dy\left( \tau \right) }{d\tau }\right) -\varphi \left( s\frac{%
dy\left( \tau \right) }{d\tau }\right) =0,
\end{equation*}%
or%
\begin{equation}
\varepsilon s\frac{d^{2}y}{d\tau ^{2}}+\left( \varepsilon +\theta s\right) 
\frac{dy}{d\tau }-\frac{1}{s}\varphi \left( s\frac{dy}{d\tau }\right) +y=0.
\label{2.7}
\end{equation}

In Eq. (\ref{2.7}), we substitute the explicit form of the function $\varphi 
$ as a cubic polynomial. As a result of transformations, we obtain an
ordinary second-order differential equation:%
\begin{equation}
\frac{d^{2}y}{d\tau ^{2}}+\left( \frac{\varepsilon +\theta s-\varphi _{1}}{%
\varepsilon \theta }\right) \frac{dy}{d\tau }+\frac{s}{\varepsilon \theta }y-%
\frac{s\varphi _{2}}{2\varepsilon \theta }\left( \frac{dy}{d\tau }\right)
^{2}-\frac{s^{2}\varphi _{3}}{6\varepsilon \theta }\left( \frac{dy}{d\tau }%
\right) ^{3}=0.  \label{2.8}
\end{equation}

Making the change of variables $y=y_{1}$, $\frac{dy}{d\tau }=y_{2}$, we
represent (\ref{2.8}) in the form of a system of two differential equations:%
\begin{equation*}
\frac{dy_{1}}{d\tau }=y_{2},
\end{equation*}%
\begin{equation}
\frac{dy_{2}}{d\tau }=-\frac{s}{\varepsilon \theta }y_{1}+\frac{\varphi
_{1}-\varepsilon -\theta s}{\varepsilon \theta }y_{2}+\frac{s\varphi _{2}}{%
2\varepsilon \theta }y_{2}^{2}+\frac{s^{2}\varphi _{3}}{6\varepsilon \theta }%
y_{2}^{3}.  \label{2.9}
\end{equation}

It is natural to begin the qualitative study of the dynamic system (\ref{2.9}%
) from an investigation into the states of equilibrium. As is obvious,
equations (\ref{2.9}) possess only the trivial state of equilibrium $%
y_{1}^{\ast }=0$, $y_{2}^{\ast }=0$. To classify the type of this singular
point, it is necessary to find characteristic numbers of the linear part of (%
\ref{2.9}) determined by the quadratic equation%
\begin{equation}
\lambda ^{2}-\frac{\varphi _{1}-\varepsilon -\theta s}{\varepsilon \theta }%
\lambda +\frac{s}{\varepsilon \theta }=0.  \label{2.10}
\end{equation}

From the explicit form of (\ref{2.10}), one can infer that the singular
point of the system (\ref{2.9}) may be either a stable (unstable) node or a
stable (unstable) focus. Of primary interest for us is the situation when a
complex focus changes its stability, which may be accompanied by the
formation of a limit cycle giving rise to a corresponding self-oscillation
regime. In this case, we represent the solution of (\ref{2.10}) in the form%
\begin{equation}
\lambda _{1,2}=\frac{\mu }{2}\mp i\omega ,  \label{2.11}
\end{equation}%
where $\omega ^{2}=\frac{s}{\varepsilon \theta }$, $i^{2}=-1$, and $\mu =%
\frac{\varphi _{1}-\varepsilon -\theta s}{\varepsilon \theta }$ is a small
quantity. In other words, for $\mu =0$, accordingly, the eigenvalues are
purely imaginary: $\lambda _{1,2}=\mp i\omega $.

By differentiation expression (\ref{2.10}) with respect to the parameter $%
\mu $, we obtain: $\lambda ^{\prime }=\frac{d\lambda }{d\mu }=\frac{1}{2}$.
This means that the eigenvalues $\lambda _{1,2}$ cross the imaginary axis
with a nonzero velocity. Therefore, we can argue that the conditions of
Hopf's bifurcation theorem are fulfilled, and the system (\ref{2.9}) allows
for the formation of a limit cycle from the complex focus.

It would be in order here to draw attention to the reason for the occurrence
of instability in the multiplier-accelerator model. As it seems, exactly the
accelerator "blows up" the damped oscillations induced by the multiplier and
generates a structural self-sustained oscillation motion
(self-oscillations). We have already seen that instability occurs in a given
economic system when one of its parameters changes. It is most natural to
consider as a variable parameter the coefficient of the linear accelerator $%
\varphi _{1}$ whose critical value $\varphi _{1}^{\ast }=\varepsilon +\theta
s$ changes the direction of damping in the system (\ref{2.9}). Thus, the
coefficient $\varphi _{1}$ plays the role of a bifurcation parameter. The
criterion of stability can now be written down in the form $\varphi
_{1}<\varphi _{1}^{\ast }$. A linear analysis shows that when $\varphi _{1}$%
, while increasing, passes through the value $\varphi _{1}^{\ast }$, a loss
of stability of the focus is caused by a pair of complex conjugate
eigenvalues of the matrix of the coefficients of the linear part of the
system (\ref{2.9}).

The loss of stability at $\varphi _{1}=\varphi _{1}^{\ast }$ occurs under
the conditions of Hopf's theorem that states that, in addition to a
stationary solution, there appear periodic solutions.

However, Hopf's theorem itself does not provide information on whether these
periodic solutions describe regimes that can be actually observed as steady
ones. Periodic solutions may prove to be unstable and, accordingly,
unobservable without the use of special procedures. Therefore, the next goal
of our study of emerging periodic solutions in the system (\ref{2.9}) is to
derive explicit formulas describing their stability, amplitude, and period.

To achieve the above-mentioned goal, we shall use the techniques presented
in \cite{[37]}. For the sake of convenience of the application of the
proposed methods, we shall retain original notation.

In order to reduce the system (\ref{2.9}) to the normal Poincar\'{e} form,
we make the change of variables $y_{1}=x_{1}$, $y_{2}=-\omega x_{2}$. As a
result, for $\mu =0$, we obtain%
\begin{equation*}
\dot{x}_{1}=-\omega x_{2},
\end{equation*}%
\begin{equation}
\dot{x}_{2}=\omega x_{1}-\frac{\omega ^{3}\varphi _{2}}{2}x_{2}^{2}+\frac{%
\omega ^{4}s\varphi _{3}}{6}x_{2}^{3}.  \label{2.12}
\end{equation}

Let us represent the system of the two ordinary differential equations (\ref%
{2.12}) in the form of a complex differential equation with respect to the
variable $z=x_{1}+x_{2}$:%
\begin{equation*}
\dot{z}=i\omega z+g_{20}\frac{z^{2}}{2}+g_{11}z\bar{z}+g_{02}\frac{\bar{z}%
^{2}}{2}+g_{30}\frac{z^{3}}{6}+g_{21}\frac{z^{2}\bar{z}}{2}
\end{equation*}%
\begin{equation}
+g_{12}\frac{z\bar{z}^{2}}{2}+g_{03}\frac{\bar{z}^{3}}{6},  \label{2.13}
\end{equation}%
where $\bar{z}$ is the complex conjugate of $z$, and the inverse change of
variables yields $x_{1}=\func{Re}z$, $x_{2}=\func{Im}z$.

For further evaluation, we need only the coefficients $g_{20}$, $g_{11}$, $%
g_{02}$, and $g_{21}$. Their explicit forms are%
\begin{equation}
g_{20}=-g_{11}=g_{02}=\frac{i\omega ^{3}\varphi _{2}}{4},\quad g_{21}=\frac{%
\omega ^{4}\varphi _{3}}{4}.  \label{2.14}
\end{equation}

Given expressions (\ref{2.14}), we obtain the following:

1) the value of the first Lyapunov quantity%
\begin{equation}
\func{Re}c_{1}\left( 0\right) =\frac{s^{3}\varphi _{3}}{16\varepsilon
^{2}\theta ^{2}};  \label{2.15}
\end{equation}

2) the amplitude of small oscillations%
\begin{equation}
\rho =\sqrt{\frac{8\varepsilon \theta \left( \varphi _{1}-\varepsilon
-\theta s\right) }{-s^{3}\varphi _{3}}};  \label{2.16}
\end{equation}

3) an approximate value of the period of oscillations%
\begin{equation}
T\approx \frac{2\pi }{\omega }\left( 1+\frac{s^{2}\varphi _{2}^{2}}{%
24\varepsilon ^{2}\theta ^{2}}\rho ^{2}\right) .  \label{2.17}
\end{equation}

The periodic solution itself, up to the choice of the initial phase, in
terms of the original variable, is written down in the form%
\begin{equation}
Y\left( \tau \right) =\frac{G}{s}+\rho \cos \left( \frac{2\pi \tau }{T}%
\right) +\frac{\rho ^{2}s\varphi _{2}}{12\varepsilon \theta }\left( \cos
\left( \frac{4\pi \tau }{T}\right) +3\right) .  \label{2.18}
\end{equation}

It is important for us to know the sign of the coefficients $\varphi _{3}$
that determines the sign of the first Lyapunov quantity in (\ref{2.15}). For 
$\varphi _{3}>0$, $\func{Re}c_{1}\left( 0\right) >0$, and, accordingly, the
limit cycle is unstable; that is, rigid excitation of self-oscillations
takes place, accompanied by the phenomenon of hysteresis. However, the
assumption of the positivity of $\varphi _{3}$ is unrealistic, because the
condition of achieving the limit saturation $F_{\max }$ ($F_{\min }$) under
an increase (decrease) in the velocity of changes of profit $Y$ will not be
satisfied. Therefore, we should set $\varphi _{3}<0$ ($\func{Re}c_{1}\left(
0\right) <0$). In this case, a stable limit cycle is generated with a soft
excitation regime of self-oscillations.

While analyzing the explicit form of the approximate solution for $Y\left(
\tau \right) $ in expression (\ref{2.18}), we should note the contribution
of the coefficient $\varphi _{2}$. A nonzero value of $\varphi _{2}$
introduces certain asymmetry into the structure of the resulting
oscillations. Obviously, they are nonharmonic even for small values of the
amplitude. Besides, the coefficient $\varphi _{2}$ induces an increase in
the period of oscillations with a growth in their amplitude.

An economic meaning of the asymmetry of the cycle consists in a difference
between the duration of periods of expansion and that of periods of decline,
which, on the whole, is a characteristic feature of nonlinear models of
economic dynamics.

Somewhat earlier, we considered in detail the influence of the accelerator
parameter $\varphi _{1}$ on the degeneracy of the linear part of the system (%
\ref{2.9}) that directly induced a bifurcation of limit-cycle generation
from an equilibrium state of the complex-focus type and the establishment of
the regime of self-sustaining oscillations \cite{[12]}.

Summarizing, we want to emphasize that, in the present study of behavioral
properties of the nonlinear multiplier-accelerator model of Goodwin, we have
ascertained the mechanism of the occurrence of a cycle; we have determined
the type of its stability, and we have given meaningful interpretation of
the influence of all the parameters of the nonlinear accelerator on
peculiarities of the self-oscillation regime.

\section{A model of the multiplier-accelerator with a continuously
distributed lag}

In the previous Section, we have studied in detail an example of one
nonlinear model of the multiplier-accelerator characterized by the
occurrence of a stable limit cycle in the neighborhood of an equilibrium
state. In that case, an essential element in the construction of the
accelerator has been a finite-dimension lag in the functional relation
between the measure of the volume of investment decisions and the
instantaneous velocity of the change in profit (output volume). As basic
assumptions in the synthesis of the initial model, we have employed
nonlinear dependence of the accelerator on the derivative of profit, as well
as linear dependence of the consumption function on the value of profit.

In the present consideration, the multiplier-accelerator model will be
represented in a different form. First of all, we assume that the action of
the accelerator is now expressed in terms of a continuously distributed lag 
\cite{[14]}.

According to R. Allen \cite{[1]}, if the investment function $I\left(
t\right) $ represents the actually induced investment at the moment of time $%
t$, caused by changes in the output volume $Y\left( t\right) $, the lag is
described by the differential equation%
\begin{equation}
\frac{dI}{dt}=\beta \left( \varphi \left( \frac{dY}{dt}\right) -I\right) ,
\label{2.19}
\end{equation}%
where $\varphi $ is a nonlinear accelerator function, and the parameter $%
\beta >0$ characterizes the rate of changes in the investment function $%
I\left( t\right) $. Concerning $\varphi \left( \frac{dY}{dt}\right) $, it
should be noted that its behavior for small changes in the profit is close
to linear one, whereas with a further increase in $\frac{dY}{dt}$ the
accelerator function grows slower and may even become non-monotonic. We
shall assume that the mechanism of the action of the accelerator is
satisfactorily described by the cubic parabola%
\begin{equation*}
\varphi \left( u\right) =\varphi _{1}u+\varphi _{2}\frac{u^{2}}{2}+\varphi
_{3}\frac{u^{3}}{6},
\end{equation*}%
where $\varphi _{i}$, $i=\overline{1,3}$, are corresponding derivatives of $%
\varphi \left( u\right) $.

The next step consists in the introduction of a lag into the model of the
multiplier. In an analysis that follows, we shall also employ a continuous
representation for the description of the action of the multiplier effect by
means of a corresponding differential equation.

Let us assume that on the part of cumulative demand, a lag is absent. The
planned consumption is given by $C=C\left( Y\right) $, and independent
expenses are determined by the quantity $G.$

Then, the cumulative demand can be represented in the form of the equation%
\begin{equation}
D=C\left( Y\right) +I+G.  \label{2.20}
\end{equation}

Here, we give up the hypothesis that the consumption function is linear and
assume that $C\left( Y\right) $ is a substantially nonlinear function of the
output volume $Y$.

In the most general case, we restrict ourselves to representing $C\left(
Y\right) $ in the form of a third-order polynomial, i.e.,%
\begin{equation*}
C\left( Y\right) =C_{1}Y+C_{2}\frac{Y^{2}}{2}+C_{3}\frac{Y^{3}}{6}.
\end{equation*}

The coefficients $C_{i}$, $i=\overline{1,3}$ have the same meaning as in the
case of the accelerator function. Moreover, $C_{1}>0$, whereas $C_{1}$ and $%
C_{2}$ may have opposite signs.

The parameter of independent expenses $G$ is considered to be constant.

Consider the situation on the part of supply. Here, a response of the output
volume $Y$ to the cumulative demand is considered to be non-instantaneous,
inertial, i.e., there exists a continuously distributed lag in the form of a
corresponding differential equation:%
\begin{equation}
\frac{dY}{dt}=\alpha \left( D-Y\right) ,  \label{2.21}
\end{equation}%
where $\frac{1}{\alpha }$ is a time constant of the multiplier

Equations (\ref{2.19})-(\ref{2.21}) completely determine the system of two
nonlinear differential equations that describe an interaction between the
multiplier and the accelerator. This system is represented as follows:%
\begin{equation*}
\frac{dY}{dt}=\alpha \left( C\left( Y\right) -Y+I+G\right) ,
\end{equation*}%
\begin{equation}
\frac{dI}{dt}=\beta \left( \varphi \left( \frac{dY}{dt}\right) -I\right) .
\label{2.22}
\end{equation}

Given that $\varphi \left( 0\right) =0$, the system (\ref{2.22}) has
singular solutions determined by the system of algebraic equations%
\begin{equation*}
C\left( Y\right) -Y+G=0,
\end{equation*}%
\begin{equation}
I=0.  \label{2.23}
\end{equation}

The first equation of (\ref{2.23}), by the explicit form of $C\left(
Y\right) $, may have up to three roots that represent the coordinates of an
equilibrium state of the system (\ref{2.22}).

We represent the system of two differential equations (\ref{2.22}) in the
form of a single second-order differential equation:%
\begin{equation}
\frac{d^{2}Y}{dt^{2}}=\alpha \left( \left( C^{\prime }\left( Y\right) -1-%
\frac{\beta }{\alpha }\right) \frac{dY}{dt}+\beta \varphi \left( \frac{dY}{dt%
}\right) +\beta \left( C\left( Y\right) -Y+G\right) \right) ,  \label{2.24}
\end{equation}%
where $C^{\prime }\left( Y\right) =\frac{dC\left( Y\right) }{dY}$.

With the help of (\ref{2.24}), we have succeeded in eliminating dependence
on the variable $I\left( t\right) $. In what follows, we shall operate a
vector field composed of the variables $Y\left( t\right) $ and $\frac{%
dY\left( t\right) }{dt}$. Introducing the coordinates $y_{1}=Y\left(
t\right) $ and $y_{2}=\frac{dY\left( t\right) }{dt}$, it is reasonable to
transform Eq. (\ref{2.24}) into a system of two first-order differential
equations:%
\begin{equation*}
\dot{y}_{1}=y_{2},
\end{equation*}%
\begin{equation}
\dot{y}_{2}=\alpha \left( \left( C^{\prime }\left( y_{1}\right) -1-\frac{%
\beta }{\alpha }\right) y_{2}+\beta \left( \varphi \left( y_{2}\right)
+C\left( y_{1}\right) -y_{1}+G\right) \right) .  \label{2.25}
\end{equation}

As is obvious, the system (\ref{2.25}) possesses the same states of
equilibrium as the system (\ref{2.24}) does.

Let the coordinates of a single point be given as $\left( y_{1}^{\ast
},y_{2}^{\ast }=0\right) $, where $y_{1}^{\ast }$ is the solution of the
equation $C\left( y_{1}\right) -y_{1}+G=0$. In order to analyze behavioral
properties of the system (\ref{2.25}) in the neighborhood of the given state
of equilibrium, we introduce new variables $u_{1}=y_{1}-y_{1}^{\ast }$ and $%
u_{2}=y_{2}$.

Given the explicit form of the nonlinear consumption function $C\left(
Y\right) $, as a result of some transformations, we arrive at the system of
differential equations%
\begin{equation*}
\dot{u}_{1}=u_{2},
\end{equation*}%
\begin{equation*}
\dot{u}_{2}=-\alpha \beta Su_{1}+\left( \alpha \beta \varphi _{1}-\beta
-\alpha S\right) u_{2}+\alpha \beta \left( C_{3}y_{1}^{\ast }+C_{2}\right) 
\frac{u_{1}^{2}}{2}
\end{equation*}%
\begin{equation*}
+\alpha \left( C_{3}y_{1}^{\ast }+C_{2}\right) u_{1}u_{2}+\alpha \beta
\varphi _{2}\frac{u_{2}^{2}}{2}+\alpha \beta C_{3}\frac{u_{1}^{3}}{6}
\end{equation*}%
\begin{equation}
+\alpha C_{3}\frac{u_{1}^{2}u_{2}}{2}+\alpha \beta \varphi _{3}\frac{%
u_{2}^{3}}{6},  \label{2.26}
\end{equation}%
where $\ S=1-C_{1}-C_{2}y_{1}^{\ast }-C_{3}\frac{\left( y_{1}^{\ast }\right)
^{2}}{2}$.

Let us consider a particular case of the system (\ref{2.26}) under the
conditions $\varphi _{2}=0$, $C_{3}y_{1}^{\ast }+C_{2}=0$. Under the
above-mentioned restrictions, the quadratic terms in (\ref{2.26}) vanish;
that is, there is symmetry with respect to the change $u_{1}\leftrightarrow
-u_{1}$, $u_{2}\leftrightarrow -u_{2}$. Furthermore, under the assumption
that%
\begin{equation*}
G=-\frac{C_{2}\left( C_{2}^{2}+3C_{3}\left( 1-C_{1}\right) \right) }{%
3C_{3}^{2}},
\end{equation*}%
the equation for singular points $y_{1}$ is factorized as follows:%
\begin{equation}
\left( C_{3}y_{1}+C_{2}\right) \left( \frac{y_{1}^{2}}{6}+\frac{C_{2}}{3C_{3}%
}y_{1}-\frac{C_{2}^{2}+3C_{3}\left( 1-C_{1}\right) }{3C_{3}^{2}}\right) =0.
\label{2.27}
\end{equation}

Accordingly, $y_{1}^{\ast }=-\frac{C_{2}}{C_{3}}$ is the coordinate of the
state of equilibrium in whose neighborhood the behavior of the system (\ref%
{2.26}) is being studied. To ensure the positivity of $y_{1}^{\ast }$, we
assume that $C_{2}<0$, $C_{3}>0$, and the parameter$\ S=1-C_{1}+\frac{%
C_{2}^{2}}{2C_{3}}>0$.

After the above-mentioned simplifications, the system (\ref{2.26}) takes the
form%
\begin{equation*}
\dot{u}_{1}=u_{2},
\end{equation*}%
\begin{equation*}
\dot{u}_{2}=-\alpha \beta Su_{1}+\left( \alpha \beta \varphi _{1}-\beta
-\alpha S\right) u_{2}+\alpha \beta C_{3}\frac{u_{1}^{3}}{6}
\end{equation*}%
\begin{equation}
+\alpha C_{3}\frac{u_{1}^{2}u_{2}}{2}+\alpha \beta \varphi _{3}\frac{%
u_{2}^{3}}{6}.  \label{2.28}
\end{equation}

We shall be concerned with qualitative properties of the system (\ref{2.28})
in the neighborhood of the trivial state of equilibrium $u_{1}^{\ast }=0$, $%
u_{2}^{\ast }=0$. To determine the type of equilibrium, it is necessary to
ascertain spectral properties of the linear part of (\ref{2.28}) with the
characteristic equation%
\begin{equation}
\lambda ^{2}-\mu _{1}\lambda +\alpha \beta S=0,  \label{2.29}
\end{equation}%
where $\mu _{1}=\alpha \beta \varphi _{1}-\alpha S-\beta $.

The explicit form of the quadratic equation (\ref{2.29}) is analogous to
that of the characteristic polynomial in \cite{[12]}. Therefore, when
analyzing the situation with the occurrence of a periodic regime in the
system (\ref{2.29}) resulting from the change of stability of the singular
point of the type of a complex focus%
\begin{equation*}
\lambda _{1,2}=\frac{\mu _{1}}{2}\pm i\omega ,\quad \omega ^{2}=\alpha \beta
S,
\end{equation*}%
we can arrive at the conclusion that the reason for this effect is
transition of the linear parameter of the accelerator through a certain
critical value $\varphi _{1}^{C}=\frac{\alpha S+\beta }{\alpha \beta }$.

Note that the derivative of the eigenvalue $\lambda $ with respect to the
parameter $\mu _{1}$ is nonzero:%
\begin{equation*}
\frac{d\lambda }{d\mu _{1}}=\frac{1}{2}\neq 0.
\end{equation*}

In this case, we may argue that the conditions of Hopf's theorem are
satisfied, and a limit cycle around the trivial state of equilibrium is
generated in the system (\ref{2.28}) from a complex focus.

The fact that a self-oscillation regime is present is rather remarkable in
itself; however, it does not provide much information. Hopf's bifurcation
theorem does not give any answer to the question about the uniqueness of the
limit cycle and the character of its stability.

To resolve the posed problems, we reduce (\ref{2.28}) to the normal Poincar%
\'{e} form for $\mu _{1}=0$, using the change of variables $u_{1}=x_{1}$, $%
u_{2}=-\omega x_{2}$:%
\begin{equation*}
\dot{x}_{1}=-\omega x_{2},
\end{equation*}%
\begin{equation}
\dot{x}_{2}=\omega x_{2}-\frac{\alpha \beta C_{3}}{\omega }\frac{x_{1}^{3}}{6%
}+\alpha C_{3}\frac{x_{1}^{2}x_{2}}{2}+\alpha \beta \varphi _{3}\omega ^{2}%
\frac{x_{2}^{3}}{6}.  \label{2.30}
\end{equation}

The system (\ref{2.30}) can be reduced to the complex differential equation%
\begin{equation}
\dot{Z}=i\omega Z+g_{30}\frac{Z^{3}}{6}+g_{21}\frac{Z^{2}\bar{Z}}{2}+g_{12}%
\frac{Z\bar{Z}^{2}}{2}+g_{03}\frac{\bar{Z}^{3}}{6},  \label{2.31}
\end{equation}%
where $Z=x_{1}+ix_{2}$, $\bar{Z}=x_{1}-ix_{2}$;%
\begin{equation*}
g_{30}=\frac{\alpha \left( 3C_{3}-\beta \omega ^{2}\varphi _{3}-\frac{i\beta
C_{3}}{\omega }\right) }{8};\quad g_{21}=\frac{\alpha \left( C_{3}+\beta
\omega ^{2}\varphi _{3}-\frac{i\beta C_{3}}{\omega }\right) }{8};
\end{equation*}%
\begin{equation*}
g_{12}=\frac{\alpha \left( -C_{3}-\beta \omega ^{2}\varphi _{3}-\frac{i\beta
C_{3}}{\omega }\right) }{8};\quad g_{03}=\frac{\alpha \left( -3C_{3}+\beta
\omega ^{2}\varphi _{3}-\frac{i\beta C_{3}}{\omega }\right) }{8}.
\end{equation*}

Making use of the explicit expressions for the coefficients $g_{ij}$, it is
not difficult to determine the first Lyapunov quantity:%
\begin{equation}
l_{1}\left( 0\right) =\func{Re}\frac{g_{21}}{2}=\frac{\alpha \left(
C_{3}+\beta \omega ^{2}\varphi _{3}\right) }{16}.  \label{2.32}
\end{equation}

As before \cite{[5]}, we assume that, as a result of the effect of
investment saturation, $\varphi _{3}<0$, whereas the coefficient $C_{3}>0$.
Therefore, a sign change in expression (\ref{2.32}) is possible, which is a
manifestation of different types of stability of the limit cycles. From (\ref%
{2.32}), the stability of the limit cycle for $C_{3}+\beta \omega
^{2}\varphi _{3}<0$ ($l_{1}\left( 0\right) <0$) follows directly, whereas
for the opposite sign of the inequalities an unstable self-oscillation
regime with a catastrophic loss of stability takes place.

The case when the first Lyapunov quantity is small and alternates the sign,
i.e.,%
\begin{equation}
l_{1}=\mu _{2},  \label{2.33}
\end{equation}%
is of much greater interest.

As is well-known \cite{[6]}, the behavior of dynamic systems in the vicinity
of the parameter values satisfying the equality $C_{3}+\beta \omega
^{2}\varphi _{3}=0$, such that the first Lyapunov quantity $l_{1}$ vanishes,
substantially depend on the sign of the second Lyapunov quantity $l_{2}$.
Depending on the first and the second Lyapunov quantities, as well as on the
sign of the real part of the roots of the characteristic equation $\mu _{1}$%
, in small neighborhood of the state of equilibrium on the phase plane, one
or two limit cycles may exist with all possible combinations of stability
and instability: namely, one stable or unstable limit cycle, or two limit
cycles (a stable one inside an unstable one or vice versa).

The second Lyapunov quantity is determined by the expression%
\begin{equation}
l_{2}=-\frac{1}{12\omega }\func{Im}g_{30}g_{12}.  \label{2.34}
\end{equation}

After substitution in (\ref{2.34}) of the actual values of the parameters,
we obtain:%
\begin{equation*}
l_{2}=\frac{\beta \alpha ^{2}C_{3}}{6\omega ^{2}}\left( C_{3}-\beta \omega
^{2}\varphi _{3}\right) ,
\end{equation*}%
or, by the validity of $C_{3}-\beta \omega ^{2}\varphi _{3}=0$,%
\begin{equation}
l_{2}=\frac{\beta \alpha ^{2}C_{3}}{3\omega ^{2}}.  \label{2.35}
\end{equation}

As is obvious, the quantity $l_{2}$ does not vanish for any values of the
parameters and is strictly positive, i.e., $l_{2}>0$.

If we make a conversion from the complex-valued variables to polar
coordinates, we get two independent equations for the amplitude and the
phase of the cycles:%
\begin{equation*}
\dot{\rho}=\rho \left( \mu _{1}+\mu _{2}\rho ^{2}+l_{2}\rho ^{4}\right) ,
\end{equation*}%
\begin{equation}
\dot{\psi}=\omega .  \label{2.36}
\end{equation}

The states of equilibrium for the first equation of (\ref{2.36}) satisfy the
biquadratic equation%
\begin{equation}
\mu _{1}+\mu _{2}\rho ^{2}+l_{2}\rho ^{4}=0.  \label{2.38}
\end{equation}

Equation (\ref{2.38}) may have either none or one, or two positive solutions
(cycles).\FRAME{ftbpFU}{3.039in}{2.38in}{0pt}{\Qcb{The bifurcation diagram.}%
}{}{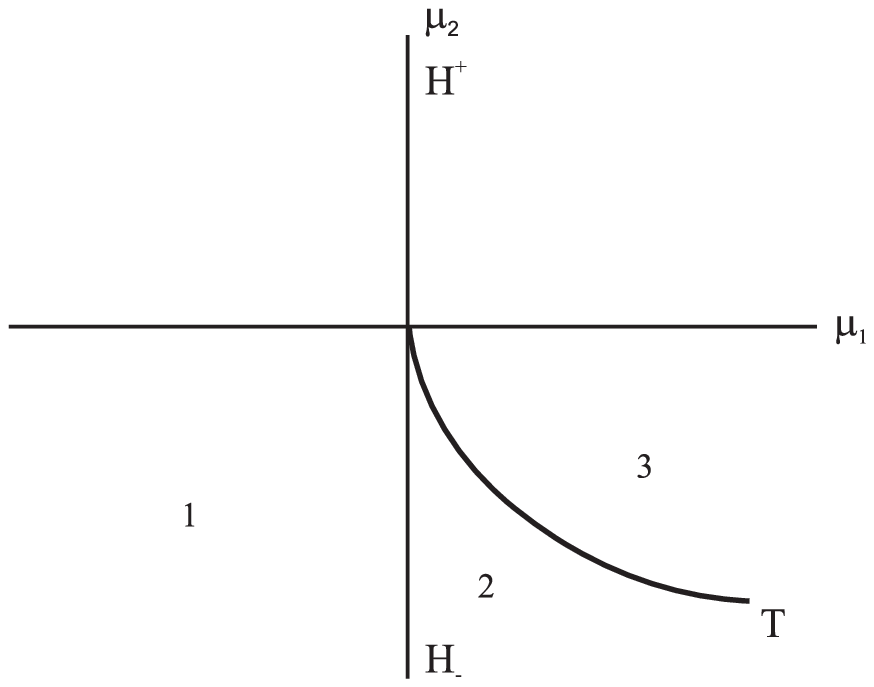}{\raisebox{-2.38in}{\includegraphics[height=2.38in]{fig_2-1.ps}}}

In Fig. 2.1, we present the corresponding bifurcation diagram. The line $%
H=\left\{ \left( \mu _{1},\mu _{2}\right) :\mu _{1}=0\right\} $ relates to
the usual Hopf's bifurcation. The state of equilibrium is stable for $\mu
_{1}<0$, and it is unstable for $\mu _{1}>0$. If we move along the line $\mu
_{1}=0$ to the points where $\mu _{2}<0$, the complex second-order focus on
the phase plane will generate an unstable (coarse) limit cycle, whereas the
focus itself becomes non-coarse and stable. Should we enter, while crossing $%
H^{-}$, region 2, the stable complex focus generates a stable limit cycle.
In region 2, both the cycle, the stable one and the unstable one, coexist
simultaneously; the merge and disappear on the line $T=\left\{ \left( \mu
_{1},\mu _{2}\right) :\mu _{2}^{2}=4l_{2}\mu _{1},\,\mu _{2}<0\right\} $.

The line $T$ characterizes the bifurcation of the double cycle. Further in
region 3, limit cycles are absent.

In Fig. 2.2, the region of the coexistence of the two limit cycles is shown.

\FRAME{ftbpFU}{3.0381in}{2.437in}{0pt}{\Qcb{The stable and unstable limit
cycles.}}{}{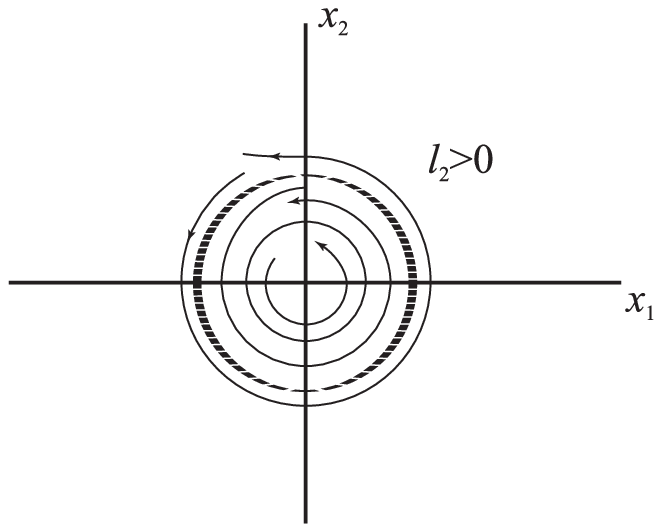}{\raisebox{-2.437in}{\includegraphics[height=2.437in]{fig_2-2.ps}}}

In conclusion of this investigation, we would like to point out the
following: the presence of two limit cycles in the initial dynamic system of
the multiplier-accelerator is stipulated not only by the nonlinearity of the
accelerator function, but also by substantial nonlinearity of the
consumption function, because exactly a relation between nonlinear
coefficients of these functions ensures the bifurcation of the double cycle.

\section{Cyclic regimes in a nonlinear model of the multiplier-accelerator
with two degrees of freedom}

Consider a model of the multiplier-accelerator with spatial inhomogeneity.
Such a model reflects peculiarities of interregional trade in the presence
of an import-export multiplier, which agrees with the already studied
multiplier of local expenses, as well as with the general concept of Keynes
and Samuelson \cite{[30],[31]}. Let the quantity of an imported commodity
depend on the local profit $Y$, which is now a function of time $t$ and of a
generalized spatial coordinate $r$. Assuming, as the first approximation,
that the action is local, we suppose that the commodity is imported from the
nearest neighborhood of a considered point, whereas the exported product is
produced under the influence of the same propensity to import in the
neighborhood of this point. Then, the net trade surplus is determined by the
product of a constant propensity to import and a profit margin in
export-import operations. In other words, the above-said can be represented
in the form of the following expression:%
\begin{equation*}
X-M=m\frac{\partial ^{2}Y}{\partial r^{2}},
\end{equation*}%
where $X$ is an export volume;

$M$ is an import volume;

$m$ is a constant propensity to import.

The spatial model of the multiplier-accelerator is represented in the form
of a single second-order differential equation with a nonlinear investment
function:%
\begin{equation}
\frac{\partial ^{2}Y}{\partial t^{2}}+SY-m\frac{\partial ^{2}Y}{\partial
r^{2}}=\left( \nu -1-S\right) \frac{\partial Y}{\partial t}-\frac{\nu }{3}%
\left( \frac{\partial Y}{\partial t}\right) ^{3},  \label{2.39}
\end{equation}%
where $S$ is a marginal propensity to save;

$\nu $ is the coefficient of the accelerator.

In our consideration, we assume that all the parameters of the model, i.e., $%
S$, $\nu $, and $m$, are constant positive quantities that do not depend
either on time or on the spatial coordinate.

The model described by Eq. (\ref{2.39}) is a rather complicated mathematical
object that exhibits a variety of forms of spatial-temporal organization.
Therefore, in what follows we shall focus on dynamic processes, having
preliminarily subdivided the space into two parts interrelated by the
regional trade. This will allow us to study such phenomena as frequency
matching and quasi-periodic motion. In other words, we may encounter a new
form of the attractor, namely, \textit{an invariant torus}.

The partial differential equation (\ref{2.39}) is represented in the form of
two coupled ordinary differential equations of the second order:%
\begin{equation*}
\ddot{Y}_{1}+\left( S_{1}+m_{1}\right) Y_{1}-m_{2}Y_{2}=\left( \nu
_{1}-1-S_{1}\right) \dot{Y}_{1}-\frac{\nu _{1}}{3}\dot{Y}_{1}^{3},
\end{equation*}%
\begin{equation}
\ddot{Y}_{2}+\left( S_{2}+m_{2}\right) Y_{2}-m_{1}Y_{1}=\left( \nu
_{2}-1-S_{2}\right) \dot{Y}_{2}-\frac{\nu _{2}}{3}\dot{Y}_{2}^{3}.
\label{2.40}
\end{equation}

Here, we have assumed that the parameters of the accelerator, the rates of
accumulation and of import are different for each region of the subdivision.

Previously, for the model of the multiplier-accelerator in one region, we
have found periodic regimes with the emergence of corresponding
self-oscillations, and we have studied the character of their stability. As
it seems, for the model with two regions, cyclic motion is also possible.
Moreover, quasi-periodic motion with two matched frequencies is possible as
well.

Let us represent the system (\ref{2.40}) in the traditional form of a system
of first-order differential equations. This new system is four-dimensional.

Let $x_{1}=Y_{1}$, $x_{2}=\dot{Y}_{1}$, $x_{3}=Y_{2}$, and $x_{4}=\dot{Y}%
_{2} $. As a result, the system (\ref{2.40}) takes the form%
\begin{equation*}
\dot{x}_{1}=x_{2},
\end{equation*}%
\begin{equation*}
\dot{x}_{2}=-\left( S_{1}+m_{1}\right) x_{1}+\left( \nu _{1}-1-S_{1}\right)
x_{2}+m_{2}x_{3}-\nu _{1}\frac{x_{2}^{3}}{3},
\end{equation*}%
\begin{equation*}
\dot{x}_{3}=x_{4},
\end{equation*}%
\begin{equation}
\dot{x}_{4}=m_{1}x_{1}-\left( S_{2}+m_{2}\right) x_{3}+\left( \nu
_{2}-1-S_{2}\right) x_{4}-\nu _{2}\frac{x_{4}^{3}}{3}.  \label{2.41}
\end{equation}

Obviously, the system (\ref{2.41}) has trivial equilibrium $x_{i}^{\ast }=0$%
, $i=\overline{1,4}$. The matrix of the linear part of (\ref{2.41}),
corresponding to this singular point, has the form%
\begin{equation}
A=\left\{ 
\begin{array}{cccc}
0 & 1 & 0 & 0 \\ 
-\left( S_{1}+m_{1}\right) & \nu _{1}-1-S_{1} & m_{2} & 0 \\ 
0 & 0 & 0 & 1 \\ 
m_{1} & 0 & -\left( S_{2}+m_{2}\right) & \nu _{2}-1-S_{2}%
\end{array}%
\right\} .  \label{2.42}
\end{equation}%
The matrix (\ref{2.42}) has the characteristic polynomial%
\begin{equation*}
\left( \lambda ^{2}-\left( \nu _{1}-1-S_{1}\right) \lambda
+S_{1}+m_{1}\right) \left( \lambda ^{2}\right.
\end{equation*}%
\begin{equation}
\left. -\left( \nu _{2}-1-S_{2}\right) \lambda +S_{2}+m_{2}\right)
-m_{1}m_{2}=0.  \label{2.43}
\end{equation}

Equation (\ref{2.43}) is a fourth-order equation; hence, it has four roots.
Of primary interest for us is the situation when (\ref{2.43}) has two pairs
of complex conjugate roots with small parameters in their real parts, i.e.,%
\begin{equation}
\lambda _{1,2}=\frac{\alpha _{1}}{2}\pm i\omega _{1},\quad \lambda _{3,4}=%
\frac{\alpha _{2}}{2}\pm i\omega _{2}.  \label{2.44}
\end{equation}

Under the assumption that $\nu _{i}=1+S_{i}+\alpha _{i}$, $i=1,2$, there
exist, for $\alpha _{i}=0$, critical values of the parameters of the
accelerator $\nu _{i}^{c}=1+S_{i}$ that are responsible for a possible
formation of limit cycles.

For $\alpha _{1}=\alpha _{2}=0$, equation (\ref{2.43}) reduces to to the
biquadratic equation%
\begin{equation*}
\lambda ^{4}+\left( S_{1}+m_{1}+S_{2}+m_{2}\right) \lambda ^{2}+\left(
S_{1}+m_{1}\right) \left( S_{2}+m_{2}\right) -m_{1}m_{2}=0,
\end{equation*}%
which, for $\lambda ^{2}=-\omega ^{2}$, yields an equation for the
frequencies:%
\begin{equation}
\omega ^{4}-\left( S_{1}+m_{1}+S_{2}+m_{2}\right) \omega ^{2}+\left(
S_{1}+m_{1}\right) \left( S_{2}+m_{2}\right) -m_{1}m_{2}=0.  \label{2.45}
\end{equation}

As the free term in (\ref{2.45}) is positive, equation (\ref{2.45}) has two
positive roots that determine the squared frequencies:%
\begin{equation*}
\omega _{1,2}^{2}=\frac{\left( S_{1}+m_{1}+S_{2}+m_{2}\right) \pm \sqrt{D}}{2%
},
\end{equation*}%
\begin{equation*}
D=\left( S_{1}+m_{1}+S_{2}+m_{2}\right) ^{2}-4\left( S_{1}+m_{1}\right)
\left( S_{2}+m_{2}\right) -m_{1}m_{2}>0.
\end{equation*}

For definiteness, we assume that $\omega _{1}>\omega _{2}$. If we compare
the values of the so-called eigenfrequencies $\bar{\omega}_{i}=\sqrt{%
S_{i}+m_{i}}$, $i=1,2$, it is not difficult to prove that $\omega _{1}>\bar{%
\omega}_{1}$, whereas $\omega _{2}<\bar{\omega}_{2}$. It means that the
common frequency of the matched oscillations is either higher than the
maximum natural frequency or lower than the minimum natural frequency. On
the other hand, in the case of free oscillations, the matched system will
never be able to oscillate with an intermediate, compared to natural
frequencies, frequency. From an economic point of view, this fact means that
a connection between two regions by means of trade relations either speeds
up or slows down a cycle of business activity in both the regions.

For further consideration of the properties of the four-dimensional flux
that possesses a state of equilibrium with two pairs of purely imaginary
eigenvalues, it is necessary to construct the normal form for the system of
ordinary differential equations (\ref{2.41}). This can be done by means of a
sequence of linear transformations of the initial variables as follows:%
\begin{equation*}
x_{1}=m_{2}\left( y_{1}+y_{3}\right) ,
\end{equation*}%
\begin{equation*}
x_{2}=-m_{2}\left( \omega _{1}y_{2}+\omega _{2}y_{4}\right) ,
\end{equation*}%
\begin{equation*}
x_{3}=\left( S_{1}+m_{1}-\omega _{1}^{2}\right) y_{1}+\left(
S_{1}+m_{1}-\omega _{2}^{2}\right) y_{3},
\end{equation*}%
\begin{equation}
x_{4}=-\omega _{1}\left( S_{1}+m_{1}-\omega _{1}^{2}\right) y_{2}-\omega
_{2}\left( S_{1}+m_{1}-\omega _{2}^{2}\right) y_{4}.  \label{2.46}
\end{equation}

The transformation (\ref{2.46}) converts the system of four ordinary
differential equations in real variables into a system of two complex
differential equations that takes the following form in terms of the polar
coordinates $Z_{j}=\rho _{j}e^{i\varphi _{j}}$:%
\begin{equation*}
\dot{\rho}_{1}=\alpha _{1}\rho _{1}+a_{11}\rho _{1}^{3}+a_{12}\rho _{1}\rho
_{2}^{2},
\end{equation*}%
\begin{equation*}
\dot{\rho}_{2}=\alpha _{2}\rho _{2}+a_{21}\rho _{1}^{2}\rho _{2}+a_{22}\rho
_{2}^{3},
\end{equation*}%
\begin{equation*}
\dot{\varphi}_{1}=\omega _{1}+O\left( \left\vert \rho \right\vert
^{2}\right) ,
\end{equation*}%
\begin{equation}
\dot{\varphi}_{2}=\omega _{2}+O\left( \left\vert \rho \right\vert
^{2}\right) .  \label{2.47}
\end{equation}

Here, $\left\vert \rho \right\vert ^{2}=\rho _{1}^{2}+\rho _{2}^{2}$; $%
\alpha _{1}$, $\alpha _{2}$ are small sign-alternating parameters; the
coefficients $Z_{j}=\rho _{j}e^{i\varphi _{j}}$, $j=\overline{1,2}$, are
functions of the initial parameters of the system.

We can learn a lot about the dynamics of the system (\ref{2.46}) from the
consideration of a plane vector field obtained by discarding the angular
coordinates, following the methods proposed in \cite{[42]}.

In order to reduce the number of parameters, we scale the variables $\rho
_{1}$ and $\rho _{2}$. Setting $\bar{\rho}_{1}=\rho _{1}\sqrt{\left\vert
a_{11}\right\vert }$ and $\bar{\rho}_{2}=\rho _{2}\sqrt{\left\vert
a_{22}\right\vert }\,$, dropping for notation convenience in what follows
the bar over $\rho _{1}$ and $\rho _{2}$, and, if necessary, scaling the
time variable, we obtain:%
\begin{equation*}
\dot{\rho}_{1}=\rho _{1}\left( \alpha _{1}+\rho _{1}^{2}+b\rho
_{2}^{2}\right) ,
\end{equation*}%
\begin{equation}
\dot{\rho}_{2}=\rho _{2}\left( \alpha _{2}+c\rho _{1}^{2}+d\rho
_{2}^{2}\right) ,  \label{2.48}
\end{equation}%
\begin{equation*}
d=\pm 1,\quad b=\frac{a_{12}}{\left\vert a_{22}\right\vert },\quad c=\frac{%
a_{21}}{\left\vert a_{11}\right\vert }.
\end{equation*}

The system (\ref{2.48}) is characterized by twelve topologically different
situations, presented in the following table:

\begin{tabular}{|l|l|l|l|l|l|l|l|l|l|l|l|l|}
\hline
Case & 1 & 2 & 3 & 4 & 5 & 6 & 7 & 8 & 9 & 10 & 11 & 12 \\ \hline
$d$ & + & + & + & + & + & + & - & - & - & - & - & - \\ \hline
$b$ & + & + & + & - & - & - & + & + & + & - & - & - \\ \hline
$c$ & + & + & - & + & - & - & + & - & - & + & + & - \\ \hline
$d-bc$ & + & - & + & + & + & - & - & + & - & + & - & - \\ \hline
\end{tabular}

This classification is based on an analysis of secondary "pitchfork"
bifurcations from nontrivial states of equilibrium of the plane vector
field. Note that the singular point $\left( \rho _{1},\rho _{2}\right)
=\left( 0,0\right) $ is always a state of equilibrium; besides, up to three
states of equilibrium may exist in the positive quadrant:%
\begin{equation*}
\left( \rho _{1},\rho _{2}\right) =\left( \sqrt{-\alpha _{1}},0\right)
,\quad \text{for }\alpha _{1}<0;
\end{equation*}%
\begin{equation}
\left( \rho _{1},\rho _{2}\right) =\left( 0,\sqrt{\frac{-\alpha _{2}}{d}}%
\right) ,\quad \text{for }\alpha _{2}d<0;  \label{2.49}
\end{equation}%
\begin{equation*}
\left( \rho _{1},\rho _{2}\right) =\left( \sqrt{\frac{d\alpha _{2}-b\alpha
_{1}}{Q}},\sqrt{\frac{c\alpha _{1}-\alpha _{2}}{Q}}\right) ,\quad \text{for }%
\frac{d\alpha _{2}-b\alpha _{1}}{Q},\frac{c\alpha _{1}-\alpha _{2}}{Q}<0,
\end{equation*}%
where $Q=d-bc$, $d=\pm 1$.

In general, the behavior of the system remains comparatively simple until
the occurrence of secondary Hopf bifurcations from the fixed point $\left(
\rho _{1}^{\ast },\rho _{2}^{\ast }\right) =\left( \sqrt{\frac{d\alpha
_{2}-b\alpha _{1}}{Q}},\sqrt{\frac{c\alpha _{1}-\alpha _{2}}{Q}}\right) $.
In order to detect such bifurcations, we linearize the dynamic equations in
a small neighborhood of this singular point. As a result, we obtain the
following matrix:%
\begin{equation*}
T=\left( 
\begin{array}{cc}
\alpha _{1}+3\left( \rho _{1}^{\ast }\right) ^{2}+b\left( \rho _{2}^{\ast
}\right) ^{2} & 2b\rho _{1}^{\ast }\rho _{2}^{\ast } \\ 
2c\rho _{1}^{\ast }\rho _{2}^{\ast } & \alpha _{2}+c\left( \rho _{1}^{\ast
}\right) ^{2}+3d\left( \rho _{2}^{\ast }\right) ^{2}%
\end{array}%
\right) ,
\end{equation*}%
whose trace is%
\begin{equation*}
\func{tr}T=\frac{2}{Q}\left( \alpha _{1}d\left( c-1\right) +\alpha
_{2}\left( b-d\right) \right) ,
\end{equation*}%
and the determinant is%
\begin{equation*}
\det T=\frac{4}{Q}\left( \left( b\alpha _{2}-d\alpha _{1}\right) \left(
c\alpha _{1}-\alpha _{2}\right) \right) .
\end{equation*}

Taking into account the conditions of the existence of the singular point (%
\ref{2.49}), we infer that the secondary Hopf bifurcation may occur only on
the straight line%
\begin{equation}
\alpha _{2}=\frac{d\left( 1-c\right) }{b-d}\alpha _{1},  \label{2.50}
\end{equation}%
and, at that, $Q>0$.

From this fact, it follows immediately that the secondary bifurcation does
not occur in cases 2, 6, 7, 9, 11, 12. It is also possible to show that this
bifurcation does not occur in cases 1, 3, 4, 5, because, for its
realization, it is important that the angular coefficient of the straight
line (\ref{2.50}) should lie in between the angular coefficients of the
straight lines that correspond to the "pitchforks", i.e.,%
\begin{equation}
\alpha _{2}=c\alpha _{1},\quad \alpha _{2}=\frac{d}{b}\alpha _{1},
\label{2.51}
\end{equation}%
which is equivalent to%
\begin{equation*}
c<\frac{d\left( 1-c\right) }{b-d}<\frac{d}{b}
\end{equation*}%
in a corresponding sector of the plane $\left( \alpha _{1},\alpha
_{2}\right) $.

As can be shown by means of simple evaluation, in each case, this
requirement does not ensure the condition $Q>0$.

Consider case 8, in which a Hopf bifurcation may occur. Some bifurcation
sets and phase portraits for this case are represented in Fig. 2.3.

\FRAME{ftbpFU}{4.0421in}{3.5293in}{0pt}{\Qcb{A partial bifurcation set of
the secondary Hopf bifurcation.}}{}{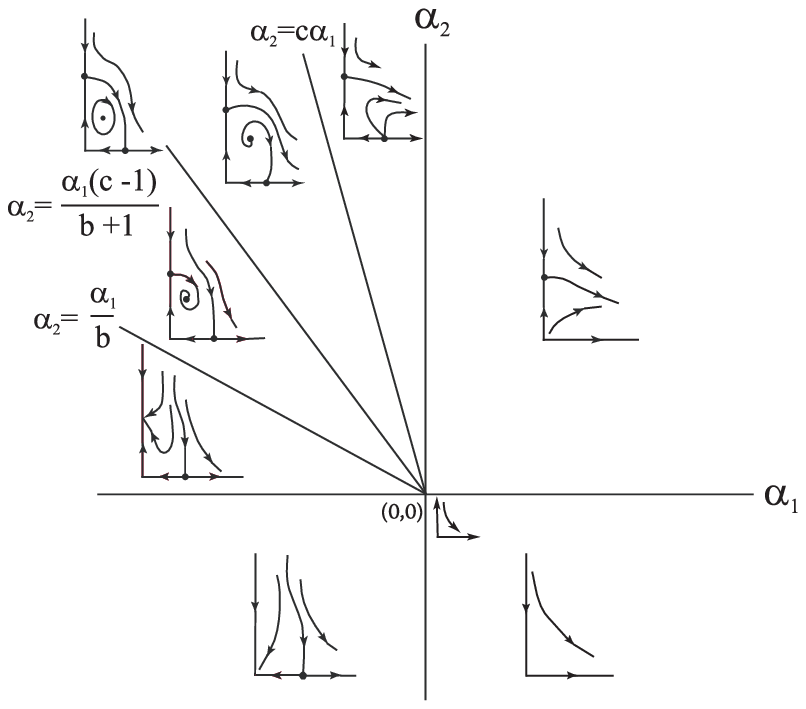}{\raisebox{-3.5293in}{\includegraphics[height=3.5293in]{fig_2-3.ps}}}

On the Hopf-bifurcation line (\ref{2.50}), the system%
\begin{equation*}
\dot{\rho}_{1}=\rho _{1}\left( \alpha _{1}+\rho _{1}^{2}+b\rho
_{2}^{2}\right) ,
\end{equation*}%
\begin{equation}
\dot{\rho}_{2}=\rho _{2}\left( \frac{c-1}{b+1}\alpha _{1}+c\rho
_{1}^{2}-\rho _{2}^{2}\right)  \label{2.52}
\end{equation}%
is integrable, whereas the function%
\begin{equation}
R\left( \rho _{1},\rho _{2}\right) =\rho _{1}^{\theta }\rho _{2}^{\beta
}\left( \alpha _{1}+\rho _{1}^{2}+\gamma \rho _{2}^{2}\right) ,  \label{2.53}
\end{equation}%
with $\theta =\frac{2\left( 1-c\right) }{Q}$, $\beta =\frac{2\left(
1+b\right) }{Q}$, and $\gamma =\frac{1+b}{1-c}$, is constant along the
solutions. In case 8, we have: $b>0>c$, $Q=-1-bc>0$, and $\alpha
_{1}=-\alpha <0$; therefore, level lines of this function have the form
shown in Fig. 2.4.

\FRAME{ftbpFU}{2.0358in}{2.3419in}{0pt}{\Qcb{The level line $R\left( \protect%
\rho _{1},\protect\rho _{2}\right) $ for case 8.}}{}{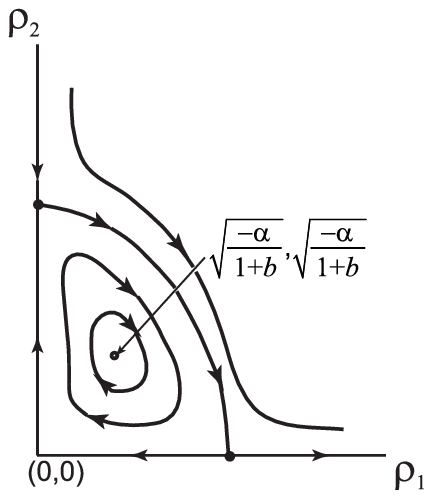}
{\raisebox{-2.3419in}{\includegraphics[height=2.3419in]{fig_2-4.ps}}}

As the system (\ref{2.52}) is integrable, the secondary Hopf bifurcation is
degenerate. Therefore, to study the topology of this bifurcation in full
detail, it is necessary to consider in (\ref{2.50}) terms containing higher
powers of the initial variables.

Unfortunately, we do not possess necessary information, because nonlinearity
in the accelerator model is restricted to cubic terms, whereas an exhaustive
analysis requires terms of the fifth order. Nonetheless, it is possible to
discuss some conclusions that can be drawn on the basis of the obtained
results about the total four-dimensional flux of the system (\ref{2.47}). It
is necessary to bear in mind that here we have two rotations $\dot{\varphi}%
_{1}=\omega _{1}$ and $\dot{\varphi}_{2}=\omega _{2}$. Moreover, the ratio
of the frequencies $\omega _{1}:\omega _{2}$ that should be restored for
final conclusions. As is easy to see, the initial four-dimensional system (%
\ref{2.47}) possesses four types of attractors corresponding to the fixed
points of the plane system (\ref{2.48}):

1) the trivial fixed point $\left( \rho _{1}=0,\rho _{2}=0\right) $;

2) a periodic orbit with the period $\approx \frac{2\pi }{\omega _{1}}\left(
\rho _{1}=\rho _{1}^{\ast },\rho _{2}=0\right) $;

3) a periodic orbit with the period $\approx \frac{2\pi }{\omega _{2}}\left(
\rho _{1}=0,\rho _{2}=\rho _{2}^{\ast }\right) $;

4) an invariant two-dimensional torus with the periods 
\begin{equation*}
\approx \frac{2\pi }{\omega _{1}},\quad \approx \frac{2\pi }{\omega _{2}}%
\left( \rho _{1}=0,\rho _{2}=\rho _{2}^{\ast }\right) ;
\end{equation*}

5) an invariant three-dimensional torus with the periods 
\begin{equation*}
\approx \frac{2\pi }{\omega _{1}},\quad \approx \frac{2\pi }{\omega _{2}}%
,\quad O\left( \frac{1}{\alpha }\right) .
\end{equation*}

The last (long) period on the tree-dimensional torus is associated with a
secondary Hopf bifurcation in the plane system (\ref{2.48}). Here, we can
foresee the presence of subtle resonance effects and phenomena that remain
outside the scope of our consideration. Thus, we expect to find a narrow
"wedge" around the line secondary Hopf bifurcation where chaotic dynamics,
including transversal homoclinic orbits and "horseshoes", takes place.
Consequently, we can argue that the multiplier-accelerator model of business
cycles with a nonlinear investment function, extended to the case of
interregional trade, and whose motion is induced by a linear multiplier of
import, may initiate random motion.

This fact implies that processes of economic forecasting are problematic. As
it seems, short-term forecasts are the most efficient ones, because
exponential divergence of close trajectories does not occur over small
periods of time.

\chapter{Self-organization in Keynesian models}

As a science, macroeconomics was born owing to J. M. Keynes and, in the
first place, to his outstanding work "The General Theory of Employment,
Interest and Money" where, for the first time, the problems of
macroeconomics appeared as the main subject of the research. Keynes's
viewpoints, revolutionary at that time, had substantially formed as a result
of an analysis of the reasons for the occurrence of the Great Depression.
The global crisis of 1929-1933 prompted Keynes, as well as many other
economists after him, to study seriously the economy as an integral system 
\cite{[27]}.

In this Chapter, we shall pay attention to some problems of economic
dynamics, the models of which are based on the postulates of Keynes's
theory. As such, the following models will be successively considered:

1) the model of the growth of the gross domestic product;

2) the LS-LM Keynes model;

3) the Kaldor model.

\section{The dynamics of GDP growth}

In the construction of long-term programs of social-economic development and
in macroeconomic modeling, the most important, in terms of criteria, factor
is the volume of GDP per capita. Following the post-Keynesian tradition,
consider, as the basis of the model of GDP growth, the main macroeconomic
identity for the volume of the total income:%
\begin{equation}
Y=C+I+G+N_{E},  \label{3.1}
\end{equation}%
where $Y$ is the GDP volume in value terms.

Let us analyze the terms of expression (\ref{3.1}) from the point of view of
their economic meaning.

1) $C=C\left( y\right) $ is the consumption volume that depends on GDP. We
shall assume that $C\left( y\right) $ is a nonlinear function that can be
represented by quadratic dependence:%
\begin{equation*}
C\left( y\right) =C_{0}+C_{1}y+C_{2}y^{2},
\end{equation*}%
where $C_{0}$ is autonomous consumption; $C_{1}$ is the limit propensity to
consume for small values of $y$; $C_{2}$ characterizes extremal properties
of the consumption curve. If $C_{2}<0$, then $C\left( y\right) $ has a
maximum, which agrees with Keynes's postulate that the propensity to consume
declines with an increase in income. For $C_{2}<0$, the function $C\left(
y\right) $ has a minimum, according to the Modigliani-Duesenberry hypothesis
of relative income \cite{[32]}.

2) $I=I\left( Y,\frac{dY}{dt}\right) $ is the investment function. We assume
that investment react only to the rate of changes in GDP, i.e., it carries
out the function of a simple accelerator:%
\begin{equation*}
I=\upsilon \frac{dY}{dt},
\end{equation*}%
where $\upsilon $ is the marginal capital coefficient.

3) $G$ is the volume of government expenditure. In a simplified version, we
assume that the quantity $G$ is independent of GDP and is constant in time.

4) $N_{E}$ is the volume of pure export characterizing the factor of
external economic activities.

By analogy with $G$, we also assume that $\ N_{E}=\func{const}$.

Taking into account all the above assumptions, we can represent the model (%
\ref{3.1}) in the form of an ordinary differential equation:%
\begin{equation}
\upsilon \frac{dY}{dt}=-R+\left( 1-C_{1}\right) Y-C_{2}Y^{2},  \label{3.2}
\end{equation}%
where $R=C_{0}+G+N_{E}=\func{const}$.

Singular solutions to Eq. (\ref{3.2}) that correspond to states of static
equilibrium can be found from the condition%
\begin{equation*}
\frac{dY}{dt}=0,
\end{equation*}%
or%
\begin{equation}
C_{2}Y^{2}-\left( 1-C_{1}\right) Y+R=0.  \label{3.3}
\end{equation}

The quadratic equation (\ref{3.3}) has the following representation for the
roots:%
\begin{equation}
Y_{1,2}^{\ast }=\frac{1-C_{1}\pm \sqrt{A}}{2C_{2}},\quad A=-4C_{2}R+\left(
1-C_{1}\right) ^{2}.  \label{3.4}
\end{equation}

If $C_{1}<1$, which is usually assumed, and $A>0$, then $Y_{1,2}^{\ast }$
are different positive numbers. In other words, equation (\ref{3.2}) has two
states of equilibrium, and the coefficients $C_{2}$ and$\ R$ have the same
sign at that. In the case $A=0$, there exists a double state of equilibrium: 
$Y_{1,2}^{\ast }=\frac{1-C_{1}}{2C_{2}}$.

For $A<0$, equation (\ref{3.3}) has no real solutions, i.e., equation (\ref%
{3.2}) does not possess states of equilibrium. If the coefficients $C_{2}$
and$\ R$ have opposite signs, one of the roots is a positive number, whereas
the other one is negative. As the negative solution has no economic meaning
in this case, one speaks about a single state of equilibrium.

For the convenience of a further analysis of the properties of the
differential equation (\ref{3.2}), we introduce new variables $X=Y-\frac{%
1-C_{1}}{2C_{2}}$, $\tau =\frac{t}{\upsilon }$. As a result of the change of
variables, equation (\ref{3.2}) takes the form%
\begin{equation}
\frac{dX}{d\tau }=\frac{A}{4C_{2}}-C_{2}X^{2}.  \label{3.5}
\end{equation}

Expression (\ref{3.5}) is a typical differential equation with separable
variables. The integral curve that passes through the point $\tau =0$, $%
X\left( 0\right) =X_{0}$ for $C_{2}>0$ is given by the following equations:%
\begin{equation}
X\left( \tau \right) =\frac{X_{0}}{1+C_{2}X_{0}\tau },\quad \text{if }A=0;
\label{3.6}
\end{equation}%
\begin{equation}
X\left( \tau \right) =\frac{4X_{0}C_{2}\sqrt{A}+A\tanh \left( \sqrt{A}\tau
\right) }{4C_{2}\sqrt{A}+4X_{0}C_{2}^{2}\tanh \left( \sqrt{A}\tau \right) }%
,\quad \text{if }A>0;  \label{3.7}
\end{equation}%
\begin{equation}
X\left( \tau \right) =\frac{4X_{0}C_{2}\sqrt{-A}+A\tan \left( \sqrt{-A}\tau
\right) }{4C_{2}\sqrt{-A}+4X_{0}C_{2}^{2}\tan \left( \sqrt{-A}\tau \right) }%
,\quad \text{if }A<0.  \label{3.8}
\end{equation}

It is not difficult to notice that, for different values of the quantity $A$%
, the solutions to the differential equation (\ref{3.5}) differ
substantially with regard to their properties, i.e., even for small
variations of $A$ in the neighborhood of zero, a qualitative change in the
scenario of the evolution of $X\left( \tau \right) $ takes place. Therefore,
the development of the situation can be diagnosed by means of qualitative
methods, without resort to complicated and expensive calculations. In
qualitative forecasting, special attention should be paid to those factors
that can change the dynamics of GDP growth either in negative or positive
direction.

In other words, if the process under consideration is in a zone of stable
development, the rest of qualitative information (such as the actual
trajectory of development) becomes less important. As a rule, in
macroeconomic modeling, mistakes and errors of approximation in the
parameters of the model are possible; accounting for perturbations of
exogenous character is also rather difficult. In this regard, it is
reasonable to draw conclusions not about a single trajectory of development,
but rather about a region of space of possible trajectories. The evaluation
of such regions is the subject of qualitative theory of differential
equations.

Let us normalize the variable $X\left( \tau \right) $ in such a way that
will allow us to reduce the number of parameters in the differential
equation (\ref{3.5}) to a single one.

For $X\left( \tau \right) =-\frac{U\left( \tau \right) }{C_{2}}$, equation (%
\ref{3.5}) takes the form%
\begin{equation}
\frac{dU}{d\tau }=f\left( U,A\right) ,\quad f\left( U,A\right) =U^{2}-\frac{A%
}{4}.  \label{3.9}
\end{equation}

Let $U^{\ast }=U^{\ast }\left( A\right) $ be a state of equilibrium of Eq. (%
\ref{3.9}) for certain for certain fixed values of the parameter $A$, and
let $\lambda \left( A\right) =f^{\prime }\left( U^{\ast },A\right) $.

For $\lambda <0$, the state of equilibrium $U^{\ast }$ is stable, whereas it
is unstable for $\lambda >0$. For small variations of the parameter $A$, the
behavior of the trajectory (\ref{3.9}) in the neighborhood of the state of
equilibrium with $\lambda \neq 0$ does not change qualitatively.

Thus, the inequality $\lambda \neq 0$ is the condition of non-degeneracy
that singles out a coarse case.

In the neighborhood of the state of equilibrium, the coarse system is
modeled by the linearized equation (\ref{3.9}):%
\begin{equation}
\frac{dU}{d\tau }=\lambda U,\quad \lambda \neq 0.  \label{3.10}
\end{equation}

A different situation occurs when, for certain values of the parameter $A$,
the eigenvalue $\lambda $ vanishes in the vicinity of the state of
equilibrium:%
\begin{equation*}
\lambda \left( A\right) =f^{\prime }\left( U^{\ast },A\right) =0,
\end{equation*}%
and the condition of non-degeneracy, $f"_{UU}\neq 0$, is fulfilled. In this
case, $U^{\ast }$ is a double root of the equation $f\left( U^{\ast }\right)
=0$.

A model equation for this bifurcation depends on a single parameter and has
the form (\ref{3.9}). Then, for $A>0$, the system (\ref{3.9}) possesses two
states of equilibrium: namely, a stable one and an unstable one. (The latter
is the boundary of the attraction region of the stable state of
equilibrium.) For $A=0$, they merge into a "semi-stable" state of
equilibrium and a non-coarse system appears. For $A<0$, the states of
equilibrium disappear: see Fig. 3.1.

\FRAME{ftbpFU}{4.0421in}{1.9337in}{0pt}{\Qcb{The bifurcation diagram for the
case of "double equilibrium".}}{}{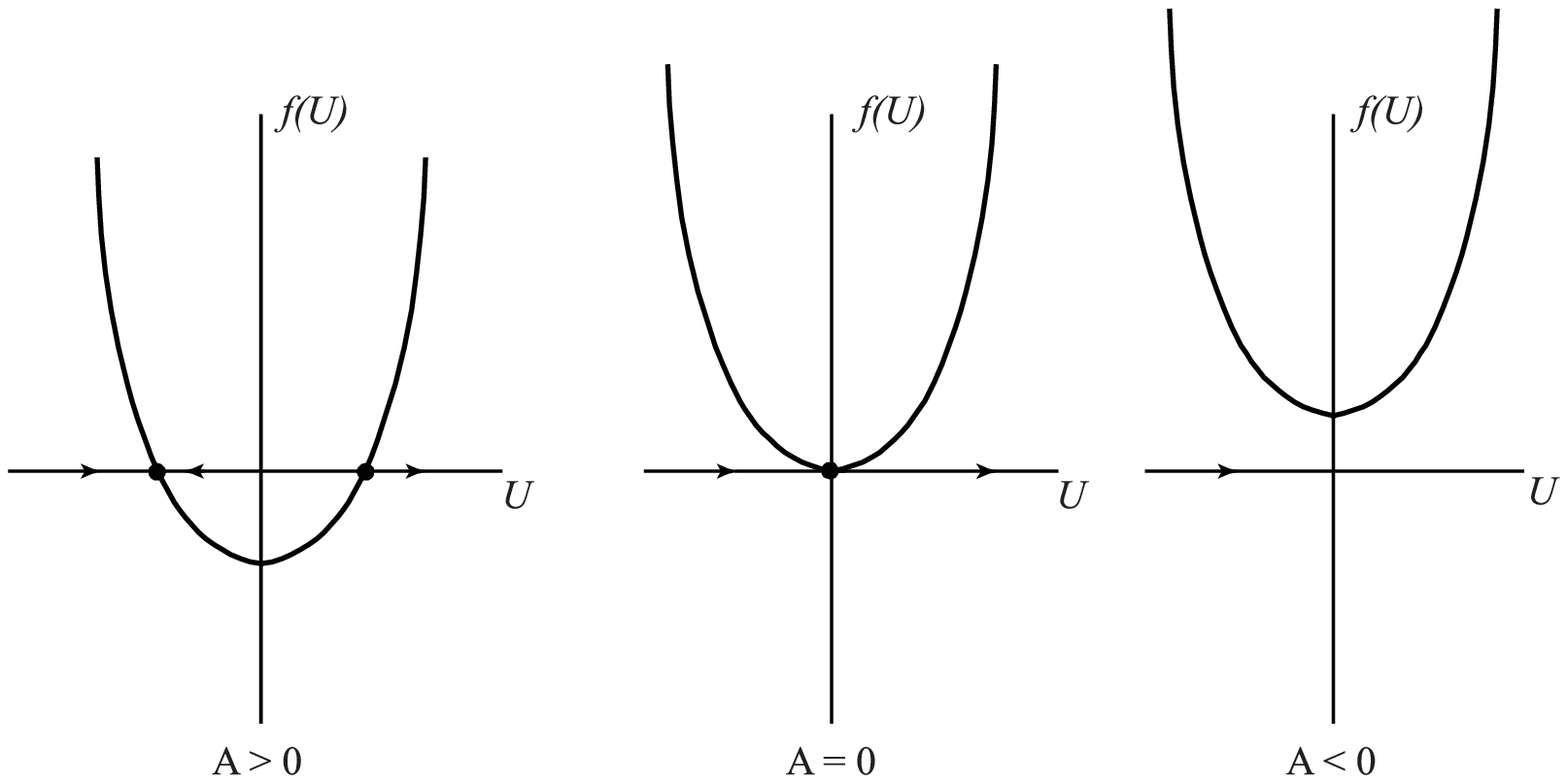}{\raisebox{-1.9337in}{\includegraphics[height=1.9337in]{fig_3-1.ps}}}

Let us monitor the stability of the state of equilibrium. Thus, when the
parameter approaches the bifurcation value $A=0$, the attraction region
shrinks on one side, and, after the disappearance of equilibrium, all the
solutions leave the considered phase space. In economic applications, such a
phenomenon is called "a break-down of equilibrium".

Note that, for the bifurcation value of the parameter $A=0$, the projective
mapping $f\left( U,A\right) =0$ onto the parameter space has a "fold"-type
singularity. In other words, there occurs a loss of stability, i.e., a
catastrophe of the "fold" type.

An important quantitative parameter of the zone of stable development, in
the model of economic development, is represented by the parameter $A$,
which is a measure of the distance between the states of equilibrium $Y_{1}$
and $Y_{2}$ (or $X_{1}$ and $X_{2}$):%
\begin{equation*}
A=C_{2}^{2}\left( Y_{1}-Y_{2}\right) ^{2}=C_{2}^{2}\left( X_{1}-X_{2}\right)
^{2}.
\end{equation*}

In the case, when $A$ is sufficiently far from zero, one of the states of
equilibrium is stable, whereas the second one is unstable. Such a situation
is rather typical of, for example, the study of logistic economic growth.
When $A$ tends to zero, the two states of equilibrium merge into a single
one, and one can observe a state of dynamic uncertainty in the sense of the
stability of economic growth trajectory, with a possible "ejection" of the
corresponding trajectory into the region of depressive dynamics. Returning
to the basic macroeconomic equation (\ref{3.2}), we can formulate the
condition of stability: the derivative of the consumption function with
respect to the GDP volume at the point of equilibrium should be larger that
unity:%
\begin{equation}
C_{Y}^{\prime }\left( Y^{\ast }\right) >1.  \label{3.11}
\end{equation}

In this case, according to the Keynes absolute-income hypothesis, the
consumption function is linear:%
\begin{equation*}
C\left( Y\right) =C_{0}+C_{1}Y.
\end{equation*}

Then, from (\ref{3.11}), it follows that exponentially unstable growth is
observed, because earlier we have assumed that the condition $C_{1}<1$ is
fulfilled.

In other words, the model of economic growth with a linear consumption
function can be stable only under the condition that a marginal propensity
to consume is larger than unity. As is obvious, such a condition cannot be
fulfilled in practice for an arbitrary GDP volume. This, in turn, induces a
discussion about differences in the behavior of the consumption function in
short-term and long-term periods.

The consumption function we have mentioned before and which rests on the
relative-income hypothesis is devoid of such shortcomings.

Thus, the problem of studying the stability of economic growth, described by
Eqs. (\ref{3.1}) and (\ref{3.2}), under the above-mentioned substantial
simplifying assumption, has been reduced to structural properties of the
consumption function. Quite naturally, there always exist certain parameters
that, for some reason, do not fit into our formal analysis. This fact
hampers the application of structural methods of macroeconomic forecasting,
such as, e.g., simulative modeling. For this reason, in the present work, we
employ the methodology of ascertaining, analyzing and forecasting economic
processes and phenomena of self-organization by means of qualitative methods
of the examination of dynamic models. By ascertaining substantially
nonlinear processes of self-organization, we are able to single out the
controlled factors as well as those factors that are beyond out influence 
\cite{[34]}.

As a result of our investigation, we have clearly demonstrated that the
macroeconomic model of growth may have several states of static equilibrium.
Accordingly, in dynamics, there exist evolutionary trajectories that
correspond to the states of equilibrium of the system. In the case when the
structure of the economy is coarse, small perturbations of the environment
are unable to "dislodge" it from its trajectory of growth. In the situation
of a non-coarse system, even small jumps of the parameters (or a "shock")
may cause a sudden transition to a different state of equilibrium, and the
trajectories of development undergo qualitative changes. On the other hand,
a merger of states of equilibrium is accompanied by a bifurcation with a
catastrophic loss of stability. In this work, for a quadratic consumption
function, we have studied effects in the vicinity of a double state of
equilibrium, with the occurrence of a "fold"-type catastrophe. Taking into
account higher-order nonlinearity makes possible to observe a bifurcation of
threefold degeneracy of the state of equilibrium, with an "assembly"-type
catastrophe, of fourfold degeneracy of the state of equilibrium, with an
"swallow-tail"-type catastrophe, etc.

A qualitative study of properties of the macroeconomic model should precede
the elaboration of efficient methods of the management of the economic
system on the basis of self-organization principles. The use of
self-organization allows one to optimize shortages of the functioning of the
economic system and to prevent the control parameters from entering a zone
of undesirable bifurcations and catastrophes \cite{[22]}.

\section{The LS-LM Keynes model}

Consider the dynamic economic system proposed by J. M. Keynes in his work
"The General Theory of Employment, Interest and Money" \cite{[20]}. This
model represents the conditions of mutual balance in the goods and money
market and is called, accordingly, the \textit{LS-LM} model. In its simplest
form, a business cycle is described by the system of ordinary differential
equations%
\begin{equation*}
\tau _{Y}\frac{dY}{dt}=I\left( Y,R\right) -S\left( Y,R\right) ,
\end{equation*}%
\begin{equation}
\tau _{R}\frac{dR}{dt}=L\left( Y,R\right) -M.  \label{3.12}
\end{equation}

According to W. Zang \cite{[19]}, all the parameters and variables here are
positive, and their meaning is the following: $Y=Y\left( t\right) $ is the
volume of the national income; $R=R\left( t\right) $ is the interest rate; $%
I=I\left( Y,R\right) $ is the investment demand function that increases with
respect to the volume of the national income, i.e., $\frac{\partial I}{%
\partial Y}=I_{Y}>0$, and decreases with respect to the interest rate, i.e., 
$\frac{\partial I}{\partial R}=-I_{R}<0$; $S=S\left( Y,R\right) $ is the
savings function that increases with respect to both the variables, i.e., $%
\frac{\partial S}{\partial Y}=S_{Y}>0$, $\frac{\partial S}{\partial R}%
=S_{R}>0$; $L=L\left( Y,R\right) $ is the total demand for money that
increases with respect to the income, i.e., $\frac{\partial L}{\partial Y}%
=L_{Y}>0$, and decreases with respect to the interest rate, i.e., $\frac{%
\partial L}{\partial R}=-L_{R}<0$; $M$ is the constant supply of money; $%
\tau _{Y}$ and $\tau _{R}$ are corresponding time constants.

The system (\ref{3.12}) illustrates the action of a simple mechanism: an
excess demand for investment, compared to the volume of savings, leads to an
increase in the national income, and vice versa; in the case when the total
demand for money resources is higher than the available supply, the interest
rate increases.

Concerning the investment demand function, it should be noted that the
values of $I\left( Y,R\right) $ stand in a direct relationship to the volume
of the national income and in an inverse relationship to the interest rate.
On the other hand, this means that an increase in the income or in the
interest rate will stimulate the population to enhance the savings; and
under the condition of growth in the output (income) volume, the demand for
money $L\left( Y,R\right) $ increases.

For the system (\ref{3.12}), we assume the existence of, as a minimum, one
positive singular solution $Y_{0}$, $R_{0}$, which represents the state of
static equilibrium of the \textit{LS-LM} model. For algebraic evaluation of
the combination of the values of the national income and of the interest
rate, it is necessary to solve the following system of equations:%
\begin{equation*}
I\left( Y_{0},R_{0}\right) =S\left( Y_{0},R_{0}\right) ,
\end{equation*}%
\begin{equation}
L\left( Y_{0},R_{0}\right) =M.  \label{3.13}
\end{equation}

It is sufficient to confine the analysis of dynamic properties of the system
(\ref{3.12}) to a local domain of the two-dimensional space of the initial
variables $Y\left( t\right) $ and $R\left( t\right) $ near the state of
equilibrium $Y_{0}$, $R_{0}$.

To this end, we introduce new variables $\bar{Y}\left( t\right) =Y\left(
t\right) -Y_{0}$ and $\bar{R}\left( t\right) =R\left( t\right) -R_{0}$, with
the meaning of deviations from the equilibrium values of the income and of
the interest rate, and expand the right-hand sides of the system (\ref{3.12}%
) in a Taylor series at the point of equilibrium, retaining the first and
the second powers of the corresponding variables. For convenience, we drop
the bar over the variables and, without loss of generality, set $\tau
_{R}=\tau _{Y}=1$, $F\left( Y,R\right) =I\left( Y,R\right) -S\left(
Y,R\right) $.

As a result, the system (\ref{3.12}) can be represented in the form%
\begin{equation*}
\frac{dY}{dt}=F_{Y}Y-F_{R}R+F_{YY}\frac{Y^{2}}{2}+F_{YR}YR+F_{RR}\frac{R^{2}%
}{2}+O\left( \left\vert Y\right\vert ^{2},\left\vert R\right\vert
^{2}\right) ^{\frac{3}{2}},
\end{equation*}%
\begin{equation}
\frac{dR}{dt}=L_{Y}Y-L_{R}R+L_{YY}\frac{Y^{2}}{2}+L_{YR}YR+L_{RR}\frac{R^{2}%
}{2}+O\left( \left\vert Y\right\vert ^{2},\left\vert R\right\vert
^{2}\right) ^{\frac{3}{2}},  \label{3.14}
\end{equation}%
where the coefficients of the quadratic terms are the second-order
derivatives with respect to the corresponding variables at the point of
equilibrium $Y_{0}$, $R_{0}$, and $F_{Y}=I_{Y}-S_{R}>0$, $%
F_{R}=I_{Y}+S_{R}>0 $.

The matrix of the linear part of (\ref{3.14}) at the point of equilibrium
has the following representation:%
\begin{equation*}
A=\left( 
\begin{array}{cc}
F_{Y} & -F_{R} \\ 
L_{Y} & -L_{R}%
\end{array}%
\right) ,
\end{equation*}%
with the characteristic polynomial%
\begin{equation}
\lambda ^{2}-\func{tr}A\cdot \lambda +\det A=0,  \label{3.15}
\end{equation}%
where $\func{tr}A=F_{Y}-L_{R}$ is the trace of the matrix $A$, and $\det
A=F_{R}L_{Y}-F_{Y}L_{R}$ is the determinant of the matrix $A$.

In the case when $\func{tr}A<0$ and $\det A>0$, we can argue that the system
(\ref{3.14}) is stable in the linear approximation. We shall consider in
more detail the situation in a small neighborhood of the boundary of the
region of linear stability $\func{tr}A=\mu $, where $\mu $ is a small
sign-alternating quantity. This means that the quantities $F_{Y}$ and $L_{R}$
are sufficiently close to each other, and when they are exactly equal to
each other, the divergence of the vector field of the variables $Y\left(
t\right) $, $R\left( t\right) $ passes through zero.

As $\det A>0$, we can set $\omega ^{2}\left( \mu \right) =\det
A=F_{R}L_{Y}-F_{Y}L_{R}$ and $F_{Y}=L_{R}+\mu $.

Equation (\ref{3.14}) then takes the following form:%
\begin{equation}
\lambda ^{2}-\mu \lambda +\omega _{0}^{2}-\mu L_{R}=0,  \label{3.16}
\end{equation}%
\begin{equation*}
\omega _{0}^{2}=F_{R}L_{Y}-L_{R}^{2}.
\end{equation*}

Differentiating (\ref{3.16}) with respect to the parameter $\mu $, at $\mu
=0 $, we obtain:%
\begin{equation}
\frac{d\lambda }{d\mu }=\frac{1}{2}-i\frac{L_{R}}{2\omega _{0}}.
\label{3.17}
\end{equation}

From (\ref{3.16}), it follows that the real part of the derivative of the
eigenvalue does not vanish.

In other words, the eigenvalues cross the imaginary axis with a non-zero
velocity.

Thus, the conditions of Hopf's bifurcation theorem are fulfilled, and, in
the system (\ref{3.14}), when the stability of the complex-focus-type state
of equilibrium changes, the formation (or annihilation) of a bifurcation of
the limit cycle takes place, accompanied by the generation of a
corresponding self-oscillation regime.

As a bifurcation parameter, we have chosen $F_{Y}=L_{R}$, which is
equivalent to the condition%
\begin{equation*}
I_{Y}-S_{Y}-L_{R}=0.
\end{equation*}

To analyze this bifurcation, we construct, using the change of variables $%
Y=F_{R}x_{1}$, $R=L_{R}x_{1}+\omega _{0}x_{2}$, the normal form of the
system (\ref{3.14}) at $\mu =0$.

As a result of transformations with the change of the time scale $\tau
=\omega t$, we obtain the system%
\begin{equation*}
\dot{x}_{1}=-x_{2}+a_{20}\frac{x_{1}^{2}}{2}+a_{11}x_{1}x_{2}+a_{02}\frac{%
x_{2}^{2}}{2},
\end{equation*}%
\begin{equation}
\dot{x}_{2}=x_{1}+b_{20}\frac{x_{1}^{2}}{2}+b_{11}x_{1}x_{2}+b_{02}\frac{%
x_{2}^{2}}{2},  \label{3.18}
\end{equation}%
where%
\begin{equation*}
a_{20}=\frac{F_{R}^{2}F_{YY}+2F_{R}L_{R}F_{YR}+L_{R}^{2}F_{RR}}{\omega
_{0}F_{R}};
\end{equation*}%
\begin{equation*}
a_{11}=F_{YR}+\frac{L_{R}F_{RR}}{F_{R}};\quad a_{02}=\frac{\omega _{0}F_{RR}%
}{F_{R}};
\end{equation*}%
\begin{equation*}
b_{20}=\frac{F_{R}^{2}\left( F_{R}L_{YY}-F_{YY}\right) +2F_{R}L_{R}\left(
F_{R}L_{YR}-F_{YR}\right) +L_{R}^{2}\left( F_{R}L_{RR}-F_{RR}\right) }{%
\omega _{0}F_{R}};
\end{equation*}%
\begin{equation*}
b_{11}=\frac{F_{R}^{2}L_{YR}-F_{R}F_{YR}+L_{R}\left(
F_{R}L_{RR}-F_{RR}\right) }{\omega _{0}F_{R}};\quad b_{02}=L_{RR}-\frac{%
F_{RR}}{F_{R}}.
\end{equation*}

The system of two ordinary differential equations (\ref{3.18}) is the normal
Poincar\'{e} form; it can be directly applied for the evaluation of the main
characteristics of the forming limit cycles, such as the amplitude, the
frequency, and the period of oscillations; it can also be applied for the
determination of the stability of periodic solutions. The relevant formulas
are given work by V. Zang \cite{[19]}; it is shown therein that the limit
cycle can be either stable or unstable, depending on the values of the
typical parameters.

It would be in order here to emphasize that the above-mentioned work deals
with the situation when the forming limit cycle is unique and has a quite
definite type of stability. However, at the same time, the most important
question in the studies of the Hopf bifurcation concerns the maximum number
of limit cycles that can be generated from the state of equilibrium (the
fixed point) under parametric excitation of the given system. This problem
is completely resolved only for the quadratic case of polynomial systems by
N. N. Bautin \cite{[6]}: it is shown that the maximum number of limit cycles
that can be generated in the quadratic system from a focus-type singular
point is equal to three.

To determine maximum multiplicity of the limit cycle in the system of
differential equations (\ref{3.18}), it is necessary to evaluate the first
three Lyapunov focus quantities.

Let us represent the system (\ref{3.18}) in the form of a single complex
differential equation in the variable $Z=x_{1}+ix_{2}$, for $\mu \neq 0$:%
\begin{equation}
\dot{Z}=\left( i+\mu \right) z+g_{20}\frac{Z^{2}}{2}+g_{11}Z\bar{Z}+g_{20}%
\frac{\bar{Z}^{2}}{2},  \label{3.19}
\end{equation}%
where 
\begin{equation*}
g_{jk}=g_{jk}\left( a_{jk},b_{jk}\right) ,\quad j,k=\overline{0,2},\quad
j+k=2.
\end{equation*}

The singular point turns from a focus into a center under the following
conditions:

\begin{equation*}
\text{1) }\mu =g_{11}=0;
\end{equation*}%
\begin{equation*}
\text{2) }\mu =g_{20}+\bar{g}_{11}=0;
\end{equation*}%
\begin{equation*}
\text{3) }\mu =\func{Im}\left( g_{20}g_{11}\right) =\func{Im}\left( \bar{g}%
_{11}^{3}g_{02}\right) =\func{Im}\left( g_{20}^{3}g_{02}\right) =0;
\end{equation*}%
\begin{equation}
\text{4) }\mu =g_{20}-4\bar{g}_{11}=\left\vert g_{02}\right\vert
-2\left\vert g_{11}\right\vert =0.  \label{3.20}
\end{equation}

Expressions (\ref{3.20}) constitute the conditions for the existence of the
first integral, or the Hamiltonian, of the system (\ref{3.18}). In such a
system, there exists an infinite set of periodic trajectories that
continuously depend on the initial conditions. Quite naturally, isolated
closed trajectories (limit cycles) cannot exist under the Hamiltonian
conditions.

In the paper by H. Zoladek \cite{[48]}, the following formulas for the
evaluation of the three Lyapunov quantities for Eq. (\ref{3.19}) are given:%
\begin{equation*}
l_{1}=-\frac{1}{2}\func{Im}\left( g_{20}g_{11}\right) ,
\end{equation*}%
\begin{equation}
l_{2}=-\frac{1}{12}\func{Im}\left( \left( g_{20}-4\bar{g}_{11}\right) \left(
g_{20}+\bar{g}_{11}\right) \bar{g}_{11}g_{02}\right) ,  \label{3.21}
\end{equation}%
\begin{equation*}
l_{3}=-\frac{5}{64}\func{Im}\left( \left( 4\left\vert g_{11}\right\vert
^{2}-\left\vert g_{02}\right\vert ^{2}\right) \left( g_{20}+\bar{g}%
_{11}\right) \bar{g}_{11}^{2}g_{02}\right) .
\end{equation*}

Thus, using (\ref{3.21}), it is not difficult to establish the cyclicity of
the singular point:

1) no cycles are present if 
\begin{equation}
\mu \neq 0;  \label{3.22}
\end{equation}

2) a single cycle is present if%
\begin{equation}
\mu =0,\quad \func{Im}\left( g_{20}g_{11}\right) \neq 0;  \label{3.23}
\end{equation}

3) two limit cycle coexist if%
\begin{equation}
\mu =\func{Im}\left( g_{20}g_{11}\right) =0,\quad g_{20}\neq 4\bar{g}_{11};
\label{3.24}
\end{equation}

3) three limit cycles are observed if%
\begin{equation}
\mu =g_{20}-4\bar{g}_{11}=0.  \label{3.25}
\end{equation}

By virtue of the relation between (\ref{3.18}) and (\ref{3.19}), i.e.,%
\begin{equation*}
g_{20}=\frac{1}{4}\left( a_{20}-a_{02}+2b_{11}+i\left(
b_{20}-b_{02}-2a_{11}\right) \right) ,
\end{equation*}%
\begin{equation*}
\bar{g}_{11}=\frac{1}{4}\left( a_{20}+a_{02}-i\left( b_{20}+b_{02}\right)
\right) ,
\end{equation*}%
we obtain a parametric restriction on the existence of three limit cycles.

According to the condition (\ref{3.25}),%
\begin{equation*}
a_{20}-a_{02}+2b_{11}=4\left( a_{20}+a_{02}\right) ,
\end{equation*}%
\begin{equation*}
b_{20}-b_{02}-2a_{11}=-4\left( b_{20}+b_{02}\right) ,
\end{equation*}%
or%
\begin{equation*}
2b_{11}=3a_{20}+5a_{02},
\end{equation*}%
\begin{equation}
2a_{11}=5b_{20}+3b_{02}.  \label{3.26}
\end{equation}

Relations (\ref{3.26}) constitute algebraic conditions of the existence of
three limit cycles in a system of the general form (\ref{3.18}). They have
been obtained by direct evaluation of the corresponding Lyapunov quantities
for the six-parameter quadratic system without the use of any canonical
models of the type of the system of Bautin, Andronova \textit{et al}. \cite%
{[6]}, which employs a five-parameter form of the representation of
considered models.

In the presence of parametric perturbations, we write down the following
differential equation characterizing the dynamics of the amplitude of
oscillations with three small parameters $\beta _{j}$, $j=\overline{1,3}$:%
\begin{equation}
\dot{\rho}=\beta _{1}\rho +\beta _{2}\rho ^{2}+\beta _{3}\rho ^{3}+l_{3}\rho
^{4}.  \label{3.27}
\end{equation}

Here, $\beta _{1}=\mu $, $\beta _{2}=l_{1}$, $\beta _{3}=l_{2}$, and $%
l_{3}\neq 0$.

The bifurcation diagram of the system for the case $l_{3}<0$ is presented in
Fig. 3.2 \cite{[4]}.

\FRAME{ftbpFU}{4.0421in}{1.7979in}{0pt}{\Qcb{Relative disposition of
bifurcation surfaces in the neighborhood of the bifurcation $l_{2}=0$.}}{}{%
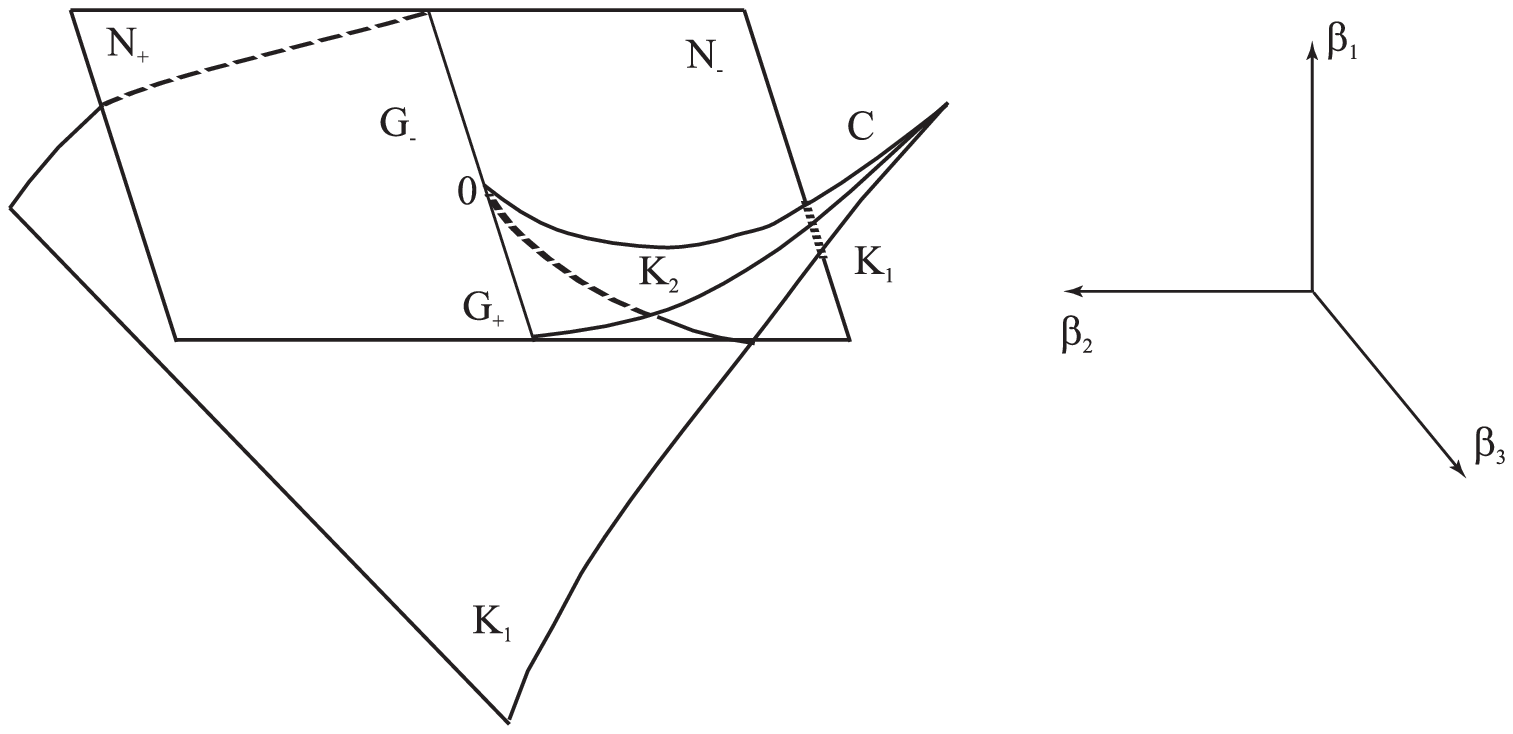}{\raisebox{-1.7979in}{\includegraphics[height=1.7979in]{fig_3-2.ps}}}

The plane $N$ corresponds to bifurcation formation of a limit cycle from the
fixed point $O$ (the state of equilibrium). On the half-plane $N_{-}$ the
loss of stability of the focus is "soft", whereas it is "rigid" on $N_{+}$.
The curved surface $K$ corresponds to a non-local bifurcation of
co-dimension one "double cycle". The part $K_{1}$ of this surface
corresponds to a stable, from the outside, multiple cycle. The other part, $%
K_{2}$, corresponds to an unstable, from the outside, multiple cycle. On the
surface $K$, there is a rib of return $C$, i.e., a common line for the
above-mentioned parts of the surface of the multiple cycle. On the line $C$,
there forms a non-local bifurcation of co-dimension two, i.e., a bifurcation
of a merger of three cycles. The line of the intersection of the half-plane $%
N_{-}$ and the surface $K_{1}$ corresponds to a bifurcation of co-dimension
"one plus one" accompanied simultaneously by a change of the stability of
the focus and a merger of a remote pair of cycles.

In Fig. 3.3, we present the bifurcation diagram in the neighborhood of $0$
(zero) for $l_{3}<0$.

\FRAME{ftbpFU}{4.0439in}{2.636in}{0pt}{\Qcb{The bifurcation diagram for the
bifurcation $l_{2}=0$ with $l_{3}<0$.}}{}{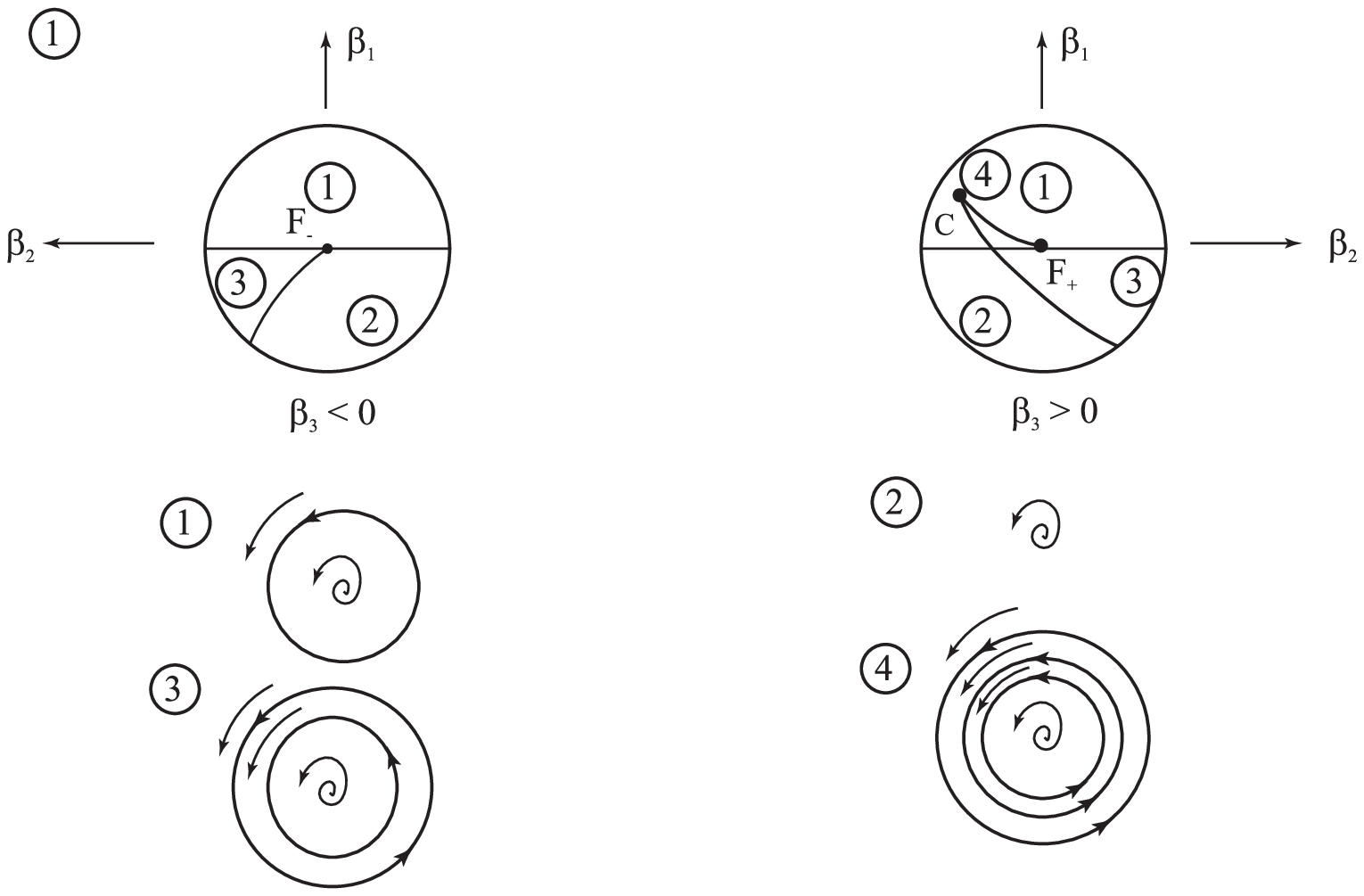}{\raisebox{-2.636in}{\includegraphics[height=2.636in]{fig_3-3.ps}}}

Returning to the description of the initial Keynes model (\ref{3.1}), we
want to emphasize that we have used a set of restrictions on the parameters
of the linear part of the system (\ref{3.14}) that has allowed us to satisfy
the requirements of Hopf's theorem on the existence of the limit cycle. At
the same time, we have not stipulated any restrictive conditions on the
coefficients of the quadratic terms, and, therefore, we may argue that, in
the most general case, the maximum possible number of limit cycles in the
neighborhood of the state of equilibrium $Y_{0}$, $R_{0}$ is equal to three.

Up to this point, we have studied the behavior of the system (\ref{3.14}) on
one boundary of the stability region determined by the trace of the matrices
of dynamics. In what follows, we shall assume that both the trace and the
determinant of this matrix are small sign-alternating quantities. Let $\det
A=-\mu _{1}$ and $\func{tr}A=\mu _{2}$.

In this case, the characteristic equation (\ref{3.15}) takes the form%
\begin{equation}
\lambda ^{2}-\mu _{2}\lambda -\mu _{1}=0.  \label{3.28}
\end{equation}

For $\mu _{1}=\mu _{2}=0$, the eigenvalues are multiple and equal to zero: $%
\lambda _{1,2}=0$. Such degeneracy in the linear part of (\ref{3.14}) may
lead to the formation of the Bogdanov-Takens bifurcation \cite{[7]}; a study
of this bifurcation is the subject of what follows.

As is obvious, in this case, one should choose two bifurcation parameter
determined by the expressions for the trace and the determinant of the
matrix $A$:%
\begin{equation*}
F_{Y}-L_{R}=\mu _{2},
\end{equation*}%
\begin{equation}
F_{R}L_{Y}-F_{Y}L_{R}=-\mu _{1}.  \label{3.29}
\end{equation}

Let the functions, found from condition (\ref{3.29}), serve as the
bifurcation parameters:%
\begin{equation*}
F_{Y}=L_{R}+\mu _{2},
\end{equation*}%
\begin{equation}
F_{R}=\frac{L_{R}^{2}+L_{R}\mu _{2}-\mu _{1}}{L_{Y}}.  \label{3.30}
\end{equation}

The next important stage is the construction of a corresponding normal form
for the considered bifurcation of co-dimension two. Making the change of
variables $Y=L_{R}y_{1}+y_{2}$, $R=L_{Y}y_{1}$, we obtain the following
system:%
\begin{equation*}
\dot{y}_{1}=y_{2}+m_{20}y_{1}^{2}+m_{11}y_{1}y_{2}+m_{02}y_{2}^{2},
\end{equation*}%
\begin{equation}
\dot{y}_{2}=\mu _{1}y_{1}+\mu
_{2}y_{2}+n_{20}y_{1}^{2}+n_{11}y_{1}y_{2}+n_{02}y_{2}^{2},  \label{3.31}
\end{equation}%
where%
\begin{equation*}
m_{20}=\frac{L_{R}^{2}L_{YY}}{2L_{Y}}+L_{R}L_{YR}+\frac{L_{Y}L_{YR}}{2},
\end{equation*}%
\begin{equation*}
m_{11}=\frac{L_{R}L_{YR}}{L_{Y}}+L_{YR},\quad m_{02}=\frac{L_{Y}L_{YY}}{2},
\end{equation*}%
\begin{equation*}
n_{20}=\frac{1}{2}\left( L_{R}^{2}\left( F_{YY}-\frac{L_{R}}{L_{Y}}%
L_{YY}\right) +2L_{Y}L_{R}\left( F_{YR}-\frac{L_{R}}{L_{Y}}L_{YR}\right)
\right.
\end{equation*}%
\begin{equation*}
\left. +L_{Y}^{2}\left( F_{RR}-\frac{L_{R}}{L_{Y}}L_{RR}\right) \right) ,
\end{equation*}%
\begin{equation*}
n_{11}=L_{R}\left( F_{YY}-\frac{L_{R}}{L_{Y}}L_{RR}\right) +L_{Y}\left(
F_{YR}-\frac{L_{R}}{L_{Y}}L_{YR}\right) ,
\end{equation*}%
\begin{equation*}
n_{02}=\frac{L_{R}^{2}}{2}\left( F_{RR}-\frac{L_{R}}{L_{Y}}L_{RR}\right) .
\end{equation*}

Carrying out a nonlinear reduction of the variables $y_{1}$ and $y_{2}$,
i.e.,%
\begin{equation*}
y_{1}=u_{1}+\frac{m_{11}+n_{02}}{2}u_{1}^{2}+m_{02}u_{1}u_{2},
\end{equation*}%
\begin{equation*}
y_{2}=u_{2}-m_{02}u_{1}^{2}+n_{02}u_{1}u_{2},
\end{equation*}%
and dropping terms of order higher than two, we arrive at the following
system:%
\begin{equation*}
\dot{u}_{1}=u_{1},
\end{equation*}%
\begin{equation}
\dot{u}_{2}=\mu _{1}u_{1}+\mu _{2}u_{2}+n_{20}u_{1}^{2}+\left(
n_{11}+2m_{20}\right) u_{1}u_{2}.  \label{3.32}
\end{equation}

By means of the substitution $w_{1}=u_{1}+\delta $, $w_{2}=u_{2}$, we
eliminate from (\ref{3.32}) the linear, in the variable $u_{2}$, term:%
\begin{equation*}
\dot{w}_{1}=w_{2},
\end{equation*}%
\begin{equation}
\dot{w}_{2}=\theta _{1}+\theta _{2}w_{1}+n_{20}w_{1}^{2}+\left(
n_{11}+2m_{20}\right) w_{1}w_{2}.  \label{3.33}
\end{equation}%
where%
\begin{equation*}
\theta _{1}=\frac{n_{20}}{\left( n_{11}+2m_{20}\right) ^{2}}\mu _{2}^{2}-%
\frac{\mu _{1}\mu _{2}}{n_{11}+2m_{20}},
\end{equation*}%
\begin{equation*}
\theta _{2}=\mu _{1}-\frac{2n_{20}}{n_{11}+2m_{20}}\mu _{2}.
\end{equation*}

To complete the construction of the normal form of the system (\ref{3.31})-(%
\ref{3.33}), we need yet another scaling of the variables, $\xi _{1}=\frac{%
w_{1}}{n_{20}K^{2}}$, $\xi _{1}=\frac{\func{sign}\left( K\right) }{%
n_{20}K^{2}}w_{2}$, and of time, $t=\left\vert K\right\vert \tau $, $K=\frac{%
n_{11}+2m_{20}}{n_{20}}\neq 0$:%
\begin{equation*}
\dot{\xi}_{1}=\xi _{2},
\end{equation*}%
\begin{equation}
\dot{\xi}_{2}=\alpha _{1}+\alpha _{2}\xi _{1}+\xi _{1}^{2}+S\xi _{1}\xi _{2},
\label{3.34}
\end{equation}%
where%
\begin{equation*}
\alpha _{1}=\left( n_{11}+2m_{20}\right) K^{3}\theta _{1},\quad \alpha
_{2}=K^{2}\theta _{2},\quad S=\func{sign}\left( K\right) =\pm 1.
\end{equation*}

Eliminating $\theta _{1}$ and $\theta _{2}$, we express the small parameters 
$a_{1}$, $\alpha _{2}$ in terms of $\mu _{1}$, $\mu _{2}$:%
\begin{equation*}
\alpha _{1}=K^{2}\mu _{2}^{2}-K^{3}\mu _{2}\mu _{1},
\end{equation*}%
\begin{equation}
\alpha _{2}=K^{2}\mu _{1}-2K\mu _{2}.  \label{3.35}
\end{equation}

The bifurcation diagram for the case $S=-1$ is represented in Fig. 3.4.

\FRAME{ftbpFU}{3.039in}{2.8539in}{0pt}{\Qcb{The diagram of the
Bogdanov-Takens bifurcation in the system (3.34) for $S=-1$.}}{}{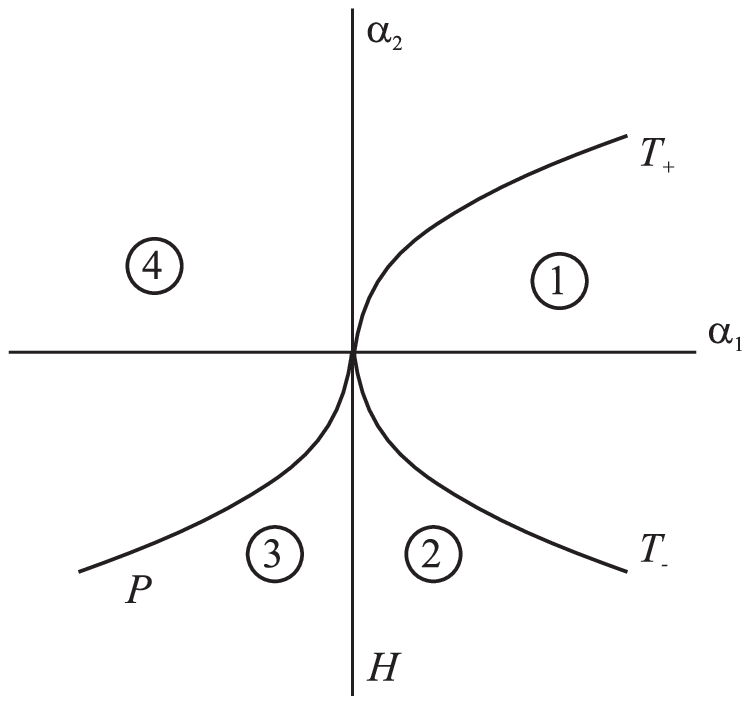%
}{\raisebox{-2.8539in}{\includegraphics[height=2.8539in]{fig_3-4.ps}}}

Our analysis of the system (\ref{3.34}) we begin with the evaluations of the
coordinates of the fixed point. As is obvious, since $\xi _{2}=0$, they are
positioned on the horizontal axis in the phase plane and satisfy the
quadratic equation%
\begin{equation}
\alpha _{1}+\alpha _{2}\xi _{1}+\xi _{1}^{2}=0.  \label{3.36}
\end{equation}

Equation (\ref{3.36}) may have from zero to two roots. The discriminant
parabola,%
\begin{equation}
T=\left\{ \left( \alpha _{1},\alpha _{2}\right) :4\alpha _{1}-\alpha
_{2}^{2}=0\right\} ,  \label{3.37}
\end{equation}%
is related to the "fold" bifurcation. Along this line, the system (\ref{3.34}%
) has equilibrium with the zero eigenvalue. If $\alpha _{2}\neq 0$, the
"fold" bifurcation is non-degenerate, and, on crossing the line $T$ from
left to right, two states of equilibrium are formed.

Denoting the left and the right states of equilibrium by $E_{1}$ and $E_{2}$%
, respectively, we get:%
\begin{equation*}
E_{1,2}=\left( \xi _{1,2}^{0},0\right) =\left( \frac{1}{2}\left( -\alpha
_{2}\pm \sqrt{\alpha _{2}^{2}-4\alpha _{1}}\right) ,0\right) .
\end{equation*}

The point $\alpha =0$ separates the two branches, $T_{-}$ and $T_{+}$, of
the "fold"-bifurcation line for $\alpha _{2}<0$ and $\alpha _{2}>0$,
respectively. Along the line $T_{-}$, a stable node $E_{1}$ coexists with a
saddle point $E_{2}$, and, in the vicinity of $T_{+}$, an unstable node $%
E_{1}$ coexists with a saddle $E_{2}$.

The vertical line $\alpha _{1}=0$ is the line on which the state of
equilibrium $E_{1}$ has a pair of eigenvalues of the zero sum: $\lambda
_{1}+\lambda _{2}=0$. The lower part%
\begin{equation}
H=\left\{ \left( \alpha _{1},\alpha _{2}\right) :\alpha _{1}=0,\,\alpha
_{2}<0\right\}  \label{3.38}
\end{equation}%
is related to a non-degenerate Hopf bifurcation, whereas the upper half-line
is a non-bifurcation line related to a neutral saddle. As a result of the
Hopf bifurcation, a stable limit cycle is generated, i.e., the first
Lyapunov quantity is negative.

The cycle exists in the neighborhood of $H$, for $\alpha _{1}<0$.

The state of equilibrium $E_{2}$ is still a saddle for all the values of the
parameters to the left of the line $T$. Here, there are no other local
bifurcations.

Let us pass around the point $\alpha _{1}=\alpha _{2}=0$ in circle of a
small radius counterclockwise. In region 1, both states of equilibrium and
cycles are absent. When passing from region 1 to region 2 through the part $%
T_{-}$ of the multiplicity line, two states of equilibrium are formed in the
system (\ref{3.34}): namely, a saddle $E_{2}$ and a stable node $E_{1}$.
Further, the node turns into a focus that loses stability as a result of a
Hopf bifurcation, which is accompanied by the formation of a stable limit
cycle on the neutrality line of the focus $H$. This limit cycle disappears
on the line $P=\left\{ \left( \alpha _{1},\alpha _{2}\right) :\alpha _{1}=-%
\frac{6}{25}\alpha _{2}^{2},\,\alpha _{2}<0\right\} $, being destroyed as a
result of a global bifurcation on the loop of the separatrix of the saddle.
(The period of motion along the cycle grows to infinity at that.) Finally,
when passing from region 4 to region 1 through the part $T_{+}$ of the
multiplicity line, \ the unstable node merges with the saddle, and both of
them disappear.

Using formulas (\ref{3.35}), we write down the equations for the bifurcation
lines in terms of the initial parameters $\mu _{1}$, $\mu _{2}$:

a) the line of the "fold" bifurcation is given by%
\begin{equation*}
T=\left\{ \left( \mu _{1},\mu _{2}\right) :\mu _{1}=0,\,\mu _{2}\neq
0\right\} ;
\end{equation*}

b) the line of the Hopf bifurcation is given by%
\begin{equation*}
H=\left\{ \left( \mu _{1},\mu _{2}\right) :\mu _{2}\left( \mu _{2}-K\mu
_{1}\right) =0,\,\mu _{2}>\frac{K}{2}\mu _{1}\right\} ;
\end{equation*}

c) the line of global bifurcation is given by%
\begin{equation*}
P=\left\{ \left( \mu _{1},\mu _{2}\right) :\left( 7\mu _{2}-K\mu _{1}\right)
\left( 7\mu _{2}-6K\mu _{1}\right) =0,\,\mu _{2}>\frac{K}{2}\mu _{1}\right\}
.
\end{equation*}

This point completes our study of the Bogdanov-Takens bifurcation of
co-dimension two, as applied to the system (\ref{3.14}). The case $S=+1$ can
be studied using the substitution $t\rightarrow -t$, $\xi _{2}\rightarrow
-\xi _{2}$. Here, a substantial difference lies in the instability of the
limit cycle. Besides, it is important to note that the limit cycle in the
vicinity of the point of the Bogdanov-Takens bifurcation has a frequency
which is proportional to the square root of the small parameter. This means
that a business cycle in the Keynes model may have a very large period,
which rather difficult to detect in numerical simulations of the initial
model.

\section{Bifurcations in the nonlinear Kaldor model}

The main assumption of this model, constructed in the Keynesian spirit, is
that the investment and savings functions are substantially nonlinear
functions of the income $Y$ and of the fixed assets rate $K$ \cite{[41]}. As
regards the investment function $I=I\left( Y,K\right) $, it is assumed that
the limit propensity to invest $I_{Y}=\frac{\partial I}{\partial Y}$ is
positive, although it is variable. This means that it takes the so-called
"normal" value for "normal" values of the income rate $Y$. For values of
income that are lower than a given "normal" interval, the limit propensity
to invest declines as a result of losses of income in the period of low
activity rate compared to the "normal" rate. It also decreases for values of 
$Y$ that are higher than the "normal" interval because of a positive effect
of the scale of expenditure and an increase in it. Thus, the investment
function is an S-shaped curve. Besides, Kaldor assumes that a higher capital
assets rate leads to a decrease in the marginal efficiency of the fixed
assets; that is, $\frac{\partial I}{\partial K}=-I_{K}<0$ ($I_{K}>0$).

The savings function is also nonlinear: $S=S\left( Y,K\right) $. The limit
propensity to save $S_{Y}=\frac{\partial S}{\partial Y}$ is positive and
less than unity, although it varies. This assumption may be justified as
follows: there exists a "normal" rate of the propensity to save that
corresponds to a "normal" interval of changes in the income. Below this
interval, the savings decrease towards consumption, whereas above this
interval, they increase. In other words, the savings function is an upturned
S-shaped curve.

Additionally, Kaldor assumes that $S_{K}=\frac{\partial S}{\partial K}$ is a
positive quantity; that is, the savings function is shifted upwards with an
increase in the capital rate. This assumption is questioned by a number of
researchers \cite{[40],[44]}, because Kaldor himself did not provide any
satisfactory justification for it. In our further consideration, we shall
assume that $S_{K}$ may change sign.

In Fig. 3.5, we present qualitative dependencies of the investment and
savings functions on the income rate, for a fixed value of the capital rate.

\FRAME{ftbpFU}{4.0421in}{2.7588in}{0pt}{\Qcb{The structure of the states of
equilibrium in the Kaldor model.}}{}{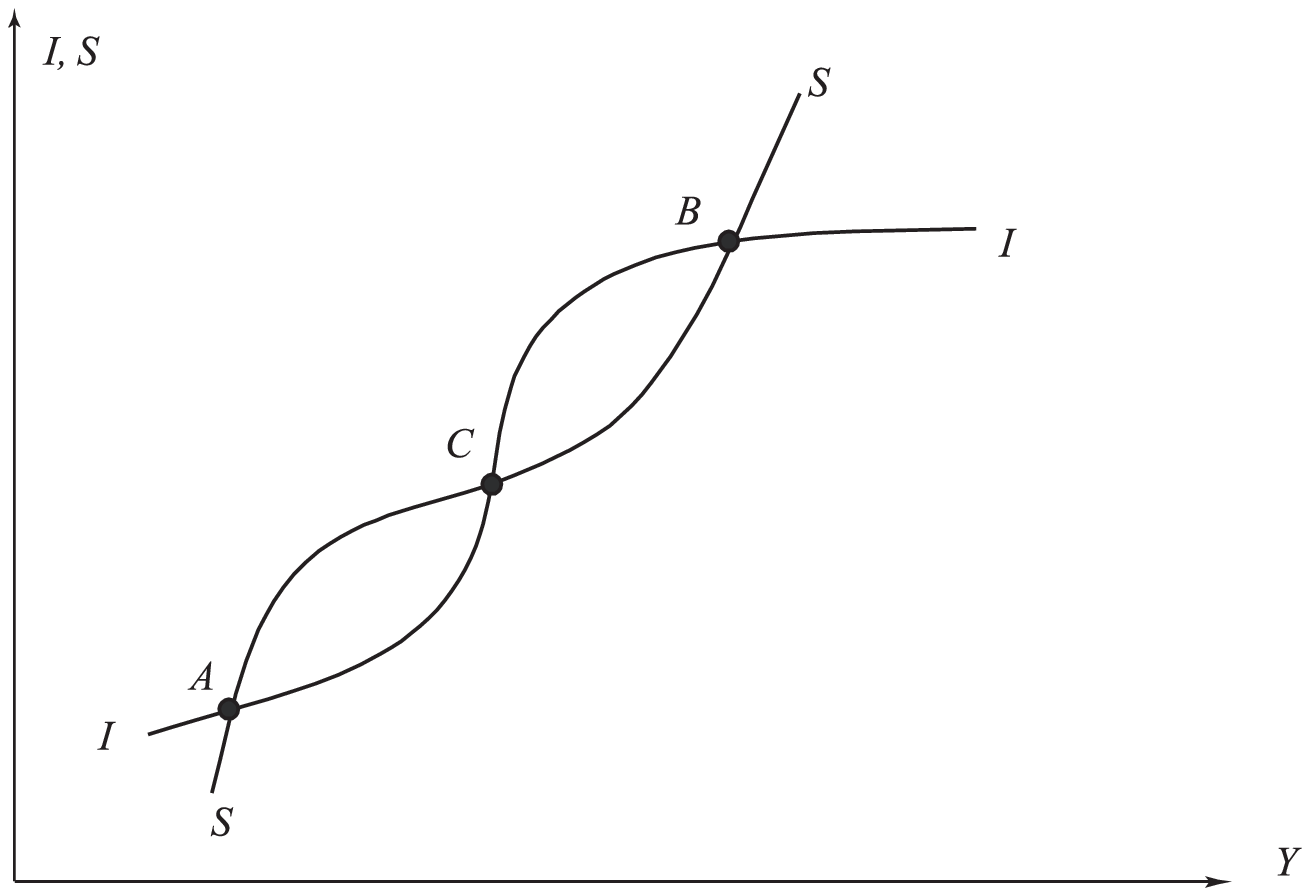}{\raisebox{-2.7588in}{\includegraphics[height=2.7588in]{fig_3-5.ps}}}

The dynamic Kaldor model is represented by the following equations:%
\begin{equation*}
\dot{Y}=\alpha \left[ \left( I\left( Y,K\right) -S\left( Y,K\right) \right) %
\right] ,
\end{equation*}%
\begin{equation}
\dot{K}=I_{0}\left( Y,K\right) ,  \label{3.39}
\end{equation}%
where $I_{0}\left( Y,K\right) $ is the realized investment, which, generally
speaking, is different from the planned one $I\left( Y,K\right) $; the
quantity $\alpha >0$ characterizes the rate of the change of the income in
the time domain.

The functions $I\left( Y,K\right) $ and $S\left( Y,K\right) $ are symmetric
with respect to the point of equilibrium. (Such an assumption is completely
justified.) We also assume that the dependence of the above-mentioned
functions on the variable $\ I$ is linear, and that the realized investment
coincides with the planned one, i.e., $I_{0}=I$. It is then convenient to
introduce new variables $\tilde{Y}=Y-Y_{0}$ and $\tilde{K}=K-K_{0}$ that
constitute deviations of the initial investment and capital rates from their
equilibrium values at the point $C$.

By symmetry, the investment and income functions are odd, and they can be
represented in the following form:%
\begin{equation*}
I\left( \tilde{Y},\tilde{K}\right) =-I_{K}\tilde{K}+I_{Y}\tilde{Y}+I_{3}%
\tilde{Y}^{3}+O\left( \tilde{Y}^{5}\right) ,
\end{equation*}%
\begin{equation}
S\left( \tilde{Y},\tilde{K}\right) =S_{K}\tilde{K}+S_{Y}\tilde{Y}+S_{3}%
\tilde{Y}^{3}+O\left( \tilde{Y}^{5}\right) .  \label{3.40}
\end{equation}

Here, $I_{3}$ and $S_{3}$ are the corresponding coefficients of the Taylor
series expansion of the initial function, and, besides, the condition $%
I_{Y}>S_{Y}$ holds. Thus, the system (\ref{3.39}) has the following explicit
form:%
\begin{equation*}
\overset{\cdot }{\tilde{Y}}=\alpha \left( I_{Y}-S_{Y}\right) \tilde{Y}%
-\alpha \left( I_{K}-S_{K}\right) \tilde{K}+\alpha \left( I_{3}-S_{3}\right) 
\tilde{Y}^{3},
\end{equation*}%
\begin{equation}
\overset{\cdot }{\tilde{K}}=I_{Y}\tilde{Y}-I_{K}\tilde{K}+I_{3}\tilde{Y}^{3}.
\label{3.41}
\end{equation}

The system (\ref{3.41}) has three states of equilibrium:

a) the trivial one, $\tilde{Y}_{0}=0$, $\tilde{K}_{0}=0$, which corresponds
to the point $C$ in Fig. 3.5;

b) nontrivial ones, $\tilde{Y}_{1,2}=\mp \sqrt{-\frac{S_{Y}I_{K}+S_{K}I_{Y}}{%
S_{K}I_{3}+S_{3}I_{K}}}$, $\tilde{K}_{1,2}=\frac{S_{K}I_{Y}+S_{Y}I_{K}}{%
S_{3}I_{K}+S_{K}I_{3}}\tilde{Y}_{1,2}$, which corresponds to the points $A$
and $B$ in Fig. 3.5.

The points $A$, $B$ and $C$ represent possible variants of static
equilibrium. In \cite{[41]}, it is argued that equilibrium at the point $C$
is unstable, whereas it is stable at the points $A$ and $B$. At the point $C$%
, the instability of equilibrium is due to the fact that, for $y_{A}<y<y_{C}$%
, savings exceed investments and a surplus appears in the goods market,
which provokes a further decline in the production. In the case $%
y_{C}<y<y_{B}$, since the volume of investments exceeds that of savings, a
deficiency of goods occurs, which stimulates a growth in the production.

As regards the stability of the points $A$ and $B$, we point out that a
deviation from $A$ or $B$ to the right leads to a goods excess and to a
decrease in their production, whereas a deviation to the left leads to a
deficiency and to a growth in the production. The state of the economic
system that corresponds to the point $A$ is characterized by a low volume of
investments that is insufficient for a complete reimbursement of the
worn-out capital. A decrease in the capital, after a certain time, will
raise the entrepreneurs' propensity to invest, and the demand for
investments will grow, which will lead to an increase in the investment
function $I\left( Y,K\right) $; equilibrium will be destroyed.

On the contrary, the point $B$ represents a state of equilibrium with high
economic activities. As a result of an achieved optimal capital volume, the
demand for investments starts to fall, the value of the function $I\left(
Y,K\right) $ begins to decrease, and the economy leaves the state of
equilibrium.

In the process of changes in market conditions, when the graphs of the
savings and investment functions move towards each other, the points $A$ and 
$C$ may merge. In the opposite case, the points $B$ and $C$ may merge. It is
important that these states of equilibrium, i.e., $A$, $B$ and $C$, become
unstable.

In this regard, it is reasonable to study in detail the situation, when all
the three states of equilibrium, $A$, $B$ and $C$, are sufficiently close to
each other and may merge into a single point. Our further consideration will
be devoted to a detailed analysis of properties of the dynamic system (\ref%
{3.41}), taking into account the above-mentioned assumption.

Consider the behavior of the system (\ref{3.41}) in the vicinity of the
trivial state of equilibrium. The linear part of (\ref{3.41}) has the
characteristic polynomial%
\begin{equation}
\lambda ^{2}+\left( I_{K}-\alpha \left( I_{Y}-S_{Y}\right) \right) \lambda
+\alpha \left( I_{Y}S_{K}+I_{K}S_{Y}\right) =0.  \label{3.42}
\end{equation}

Let us assume that the coefficients in (\ref{3.42}) are small quantities,
i.e., $\mu _{1}=-\alpha \left( I_{Y}S_{K}+I_{K}S_{Y}\right) $ and $\mu
_{2}=\alpha \left( I_{Y}-S_{Y}\right) -I_{K}$. In this case, the system (\ref%
{3.41}) may have a Bogdanov-Takens bifurcation, i.e., a so-called
"double-zero" bifurcation \cite{[46]}. As bifurcation parameters, we choose
the following ones: $I_{K}=\alpha \left( I_{Y}-S_{Y}\right) -\mu _{2}$ and $%
S_{K}=-\frac{\alpha S_{Y}}{I_{Y}}\left( I_{Y}-S_{Y}\right) -\frac{\mu _{1}}{%
\alpha I_{Y}}+\mu _{2}$.

As $\mu _{1}$ and $\mu _{2}$ are small quantities, it is obvious that $%
S_{K}<0$. Making the change of variables $\tilde{Y}=\left[ \alpha \left(
I_{Y}-S_{Y}\right) -\mu _{2}\right] U_{1}+U_{2}$ and $\tilde{K}=I_{Y}U_{1}$,
we reduce the system to the form of a nonlinear oscillator.

We have:%
\begin{equation*}
\dot{U}%
_{1}=U_{2}+a_{30}U_{1}^{3}+a_{21}U_{1}^{2}U_{2}+a_{12}U_{1}U_{2}^{2}+a_{03}U_{2}^{3},
\end{equation*}%
\begin{equation}
\dot{U}_{2}=\mu _{1}U_{1}+\mu
_{2}U_{2}+b_{30}U_{1}^{3}+b_{21}U_{1}^{2}U_{2}+b_{12}U_{1}U_{2}^{2}+b_{03}U_{2}^{3},
\label{3.43}
\end{equation}%
where%
\begin{equation*}
a_{30}=\frac{I_{3}}{I_{Y}}\left( \alpha \left( I_{Y}-S_{Y}\right) -\mu
_{2}\right) ^{3};\quad a_{21}=\frac{3I_{3}}{I_{Y}}\left( \alpha \left(
I_{Y}-S_{Y}\right) -\mu _{2}\right) ^{2};
\end{equation*}%
\begin{equation*}
a_{12}=\frac{3I_{3}}{I_{Y}}\left( \alpha \left( I_{Y}-S_{Y}\right) -\mu
_{2}\right) ;\quad a_{30}=\frac{I_{3}}{I_{Y}};
\end{equation*}%
\begin{equation*}
b_{ij}=\left[ \alpha \left( S_{Y}-\frac{S_{3}}{I_{3}}\right) +\mu _{2}\right]
a_{ij};\quad i+j=3;\quad j=\overline{0,3}.
\end{equation*}

To construct the normal Poincar\'{e} form of the system of ordinary
differential equations (\ref{3.43}), we perform a nonlinear reduction of the
variables of the state:%
\begin{equation*}
U_{1}=V_{1}+\left( \frac{a_{21}}{3}+\frac{b_{12}}{6}\right) V_{1}^{3}+\left( 
\frac{a_{12}+b_{03}}{2}\right) V_{1}^{2}V_{2}+a_{03}V_{1}V_{2}^{2},
\end{equation*}%
\begin{equation}
U_{2}=V_{2}-a_{30}V_{1}^{3}+\frac{b_{12}}{2}%
V_{1}^{2}V_{2}+b_{03}V_{1}V_{2}^{2}.  \label{3.44}
\end{equation}

As a result of transformations, taking into account expressions (\ref{3.44}%
), we obtain the following representation of the system (\ref{3.43}):%
\begin{equation*}
\dot{V}_{1}=V_{2},
\end{equation*}%
\begin{equation}
\dot{V}_{2}=\mu _{1}V_{1}+\mu _{2}V_{2}+MV_{1}^{3}+NV_{1}^{2}V_{2},
\label{3.45}
\end{equation}%
where $M=b_{30}$, $N=b_{21}+3a_{30}$.

For $\mu _{1}=0$, $\mu _{2}=0$, we have: $N=3\alpha ^{3}\left(
I_{Y}-S_{Y}\right) ^{2}\left( \frac{S_{Y}I_{3}}{I_{Y}}-S_{3}+I_{Y}-S_{Y}%
\right) $, $M=\alpha ^{4}\left( I_{Y}-S_{Y}\right) ^{3}\left( \frac{%
S_{Y}I_{3}}{I_{Y}}-S_{3}\right) $. It is reasonable to reduce the system (%
\ref{3.45}), which is already a normal Poincar\'{e} form, to a still simpler
form by means of the linear transformation%
\begin{equation*}
x_{1}=p\sqrt{\left\vert M\right\vert }V_{1},\quad x_{2}=p^{2}\sqrt{%
\left\vert M\right\vert }V_{2},\quad \tau =\frac{1}{p}t,
\end{equation*}%
where%
\begin{equation*}
p=\left\vert \frac{N}{M}\right\vert =\frac{3}{\alpha }\left\vert \frac{1}{%
I_{Y}-S_{Y}}-\frac{1}{S_{3Y}-S_{Y}\frac{I_{3}}{I_{Y}}}\right\vert .
\end{equation*}

Finally, we get:%
\begin{equation*}
\dot{x}_{1}=x_{2},
\end{equation*}%
\begin{equation}
\dot{x}_{2}=p^{2}\mu _{1}x_{1}+p\mu _{2}x_{2}+Sx_{1}^{3}-x_{1}^{2}x_{2},
\label{3.46}
\end{equation}%
where $S=\func{sign}\left( p\right) =\pm 1$.

As regards the system (\ref{3.46}), it is not difficult to notice that it is
invariant with respect to the substitution $x_{1}\rightarrow -x_{1}$ and $%
x_{2}\rightarrow -x_{2}$, and it always has trivial equilibrium $%
E_{0}=\left( 0,0\right) $. The two other possible states of equilibrium have
the coordinates $E_{1,2}=\left( \pm \sqrt{-S\mu _{1}},0\right) $; they exist
for $\mu _{1}<0$, if $S=1$, and for $\mu _{1}>0$, if $S=-1$.

Important, for the system (\ref{3.46}), is the fact that all the three
states of equilibrium merge into the single trivial one for $\mu _{1}=0$.

Let $S=1$. The bifurcation diagram is presented in Fig. 3.6.

\FRAME{ftbpFU}{3.039in}{2.3618in}{0pt}{\Qcb{The bifurcation diagram for $%
S=+1 $.}}{}{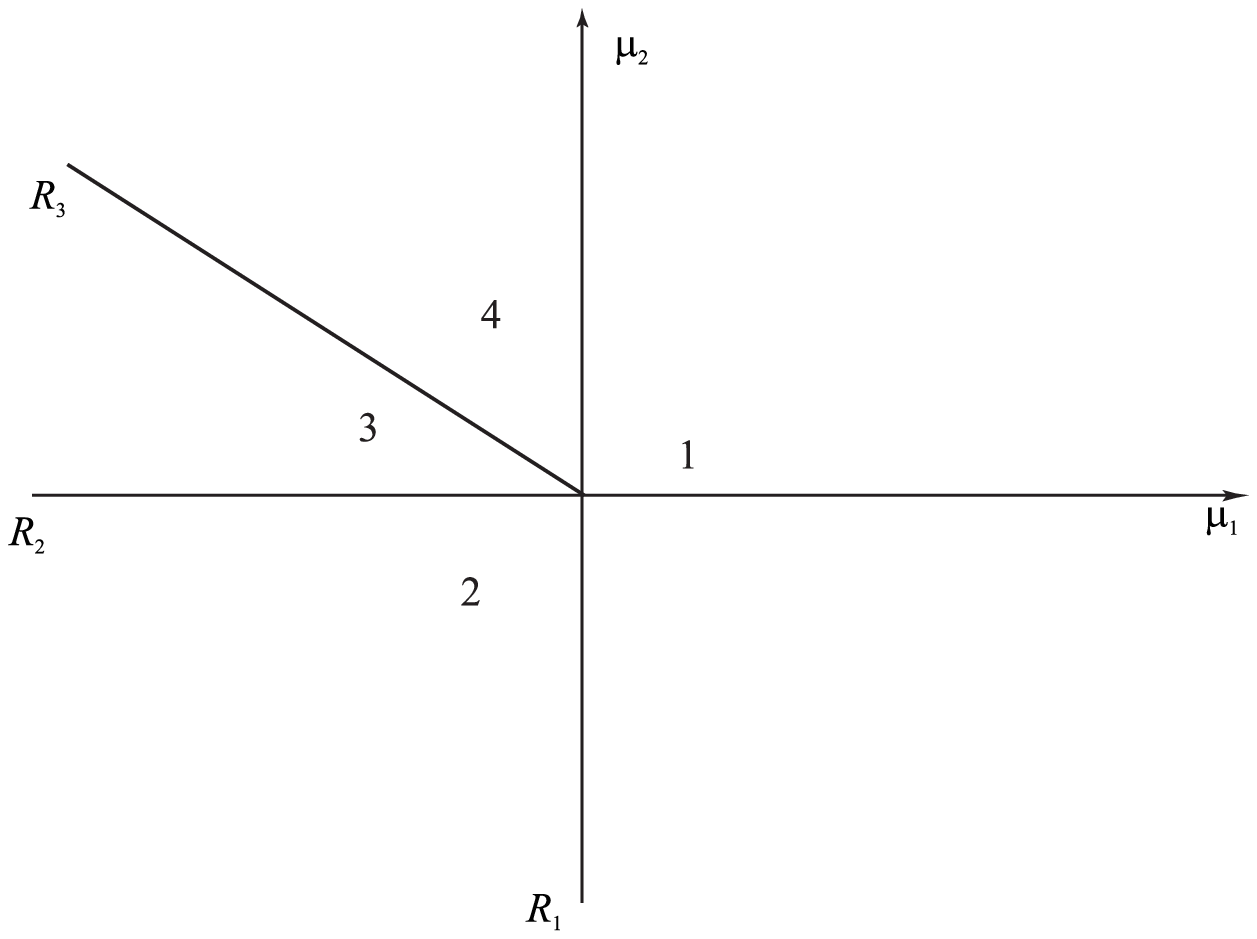}{\raisebox{-2.3618in}{\includegraphics[height=2.3618in]{fig_3-6.ps}}}

In region 1, there is the single trivial state of equilibrium $E_{0}$ that
is a saddle point. On crossing the lower branch of the line $R_{1}=\left\{
\left( \mu _{1},\mu _{2}\right) :\mu _{1}=0\right\} $, a "pitchfork" takes
place, accompanied by the appearance of a pair of symmetric saddles $E_{1,2}$%
, until the trivial equilibrium $E_{0}$ becomes a stable node. This node
turns into a focus in region 2 that, on crossing the half-line $%
R_{2}=\left\{ \left( \mu _{1},\mu _{2}\right) :\mu _{1}=0,\,\mu
_{2}<0\right\} $, undergoes a Hopf bifurcation accompanied by the generation
of a stable limit cycle.

On crossing the line $R_{3}=\left\{ \left( \mu _{1},\mu _{2}\right) :\mu
_{2}=-\frac{p}{5}\mu _{1},\,\mu _{1}<0\right\} $, a global heteroclinic
bifurcation, accompanied by the appearance of corresponding orbits that are
related to the saddles $E_{1,2}$, takes place, and, in region 4, a
heteroclinic cycle is formed. Further, all the three states of equilibrium
coexist up to the crossing of the upper part of the straight line $R_{1}$,
and a return to region 1 takes place.

Consider the case $S=-1$. The corresponding bifurcation diagram is depicted
in Fig. 3.7.

\FRAME{ftbpFU}{3.039in}{2.2857in}{0pt}{\Qcb{The bifurcation diagram for $%
S=-1 $.}}{}{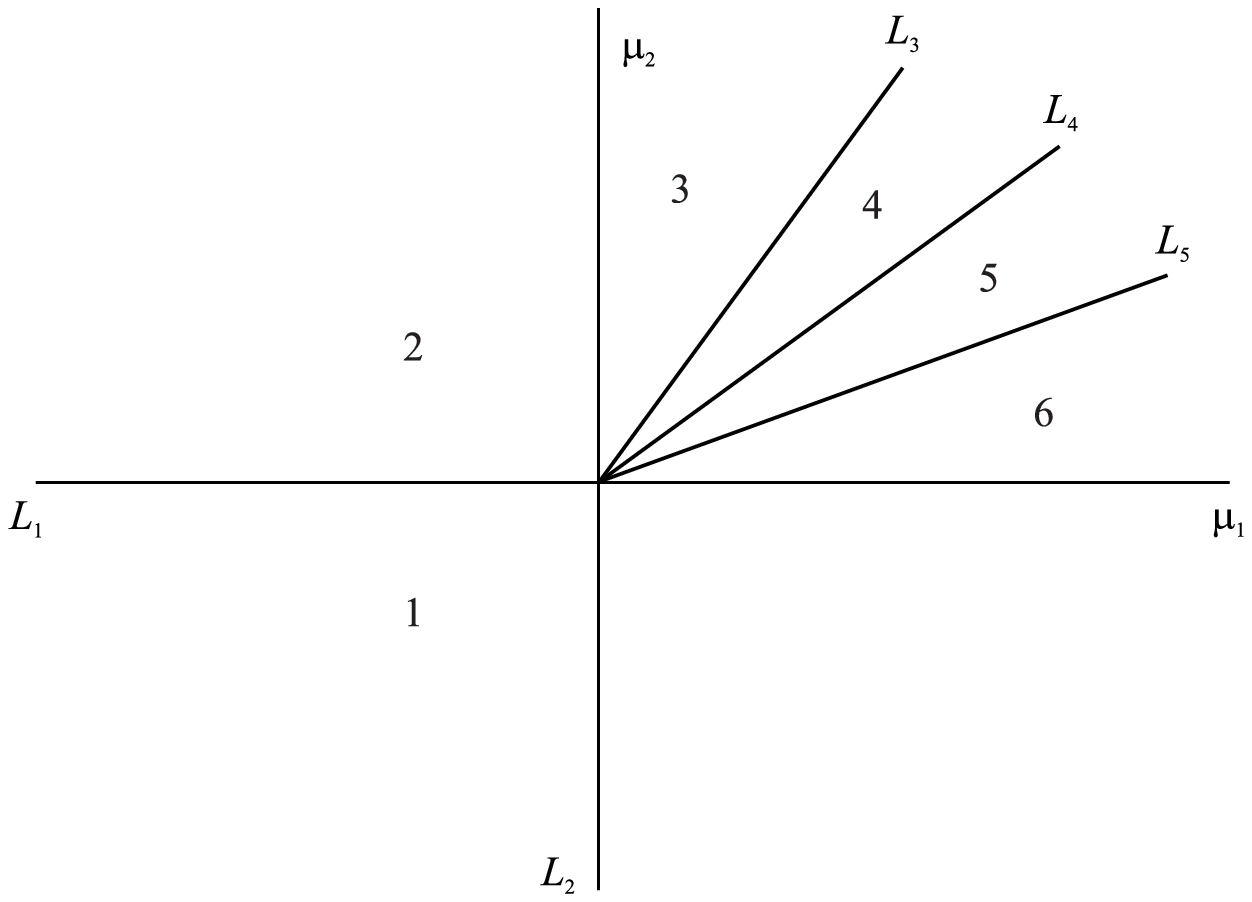}{\raisebox{-2.2857in}{\includegraphics[height=2.2857in]{fig_3-7.ps}}}

In region 1, there is the single trivial equilibrium $E_{0}$ that is a
stable node; further, it goes over to a focus. On the half-line $%
L_{1}=\left\{ \left( \mu _{1},\mu _{2}\right) :\mu _{2}=0,\,\mu
_{1}<0\right\} $, a Hopf bifurcation takes place, and a stable limit cycle
is generated. Two unstable nodes separate from the trivial equilibrium on
crossing the upper part of the line $L_{2}=\left\{ \left( \mu _{1},\mu
_{2}\right) :\mu _{1}=0,\,\mu _{2}>0\right\} \,$, when passing from region 2
region 3, as a result of a "pitchfork" bifurcation. In region 3, all the
three states of equilibrium are localized inside a "large" limit cycle. On
the half-line $L_{3}=\left\{ \left( \mu _{1},\mu _{2}\right) :\mu _{2}=p\mu
_{1},\,\mu _{1}>0\right\} $, the nontrivial focuses $E_{1,2}$ undergo a Hopf
bifurcation. This bifurcation leads to the appearance of two "small"
unstable limit cycles around the nontrivial states of equilibrium. The
points of equilibrium become stable. Thus, in region 4, there are three
limit cycles: an external "large" one and two internal "small" ones. Along
the line $L_{4}=\left\{ \left( \mu _{1},\mu _{2}\right) :\mu _{2}=\frac{4p}{5%
}\mu _{1},\,\mu _{1}>0\right\} $, the "small" cycles form a symmetric figure
that resembles the Bernoulli lemniscate, with the center at $E_{0}$, as a
result of the occurrence of a global homoclinic bifurcation. Along the line $%
L_{4}$, the saddle $E_{0}$ has two homoclinic orbits. These orbits can be
transformed from one into the other by means of symmetry transformations. On
crossing the line $L_{4}$, when passing from region 4 to region 5, not only
the "small" cycles are destroyed, but also an external "large" unstable
cycle is generated. Then, in region 5, two "large" cycles coexist: the
external one is stable, while the internal one is unstable. These two cycles
merge and disappear along the line $L_{5}=\left\{ \left( \mu _{1},\mu
_{2}\right) :\mu _{2}=k_{0}p\mu _{1},\,\mu _{1}>0\right\} $, where $%
k_{0}=0.752\ldots $. This is a saddle-node bifurcation of the limit cycle.
After the occurrence of this bifurcation, there are no limit cycles in the
system. In region 6, all the three states of equilibrium are present: the
trivial saddle $E_{0}$ and two stable nontrivial focuses (nodes) $E_{1,2}$.
The nontrivial states of equilibrium merge with the trivial one on the lower
part of the line $L_{2}=\left\{ \left( \mu _{1},\mu _{2}\right) :\mu
_{1}=0,\,\mu _{2}>0\right\} $ as a result of a "pitchfork" bifurcation, and
we return to region 1.

Thus, the behavioral properties of the system (\ref{3.46}) are determined by
comparison between the quantities $I_{Y}-S_{Y}$ and $S_{3}-\frac{S_{Y}I_{3}}{%
I_{Y}}$. If the condition $I_{Y}-S_{Y}>S_{3}-\frac{S_{Y}I_{3}}{I_{Y}}$ is
fulfilled, then $S=\func{sign}\left( p\right) =-1$. In the opposite case $%
I_{Y}-S_{Y}>S_{3}-\frac{S_{Y}I_{3}}{I_{Y}}$, we get: $S=+1$. Returning to
the question of the choice of the sign of the quantity $S_{K}$, we note that
states of equilibrium of the saddle type are possible in the initial dynamic
system (\ref{3.41}) only under the condition $S_{K}<-\frac{I_{K}S_{Y}}{I_{Y}}
$, i.e., for $S_{K}<0$.

Thus, in this chapter, we have demonstrated dynamic behavior of the Kaldor
model in its whole variety, in the case when three states of equilibrium
degenerate into a single one. We have also followed the hierarchy of
instabilities accompanied by a cascade of corresponding bifurcations \cite%
{[11]}.

\chapter{Dynamics of economic processes with a lag}

In many economic models, when constructing balance relations, one has to
take into account a lag responsible for various interactions. In Chapter 2,
we have already considered the Goodwin model involving a lag of two types:
on the part of the demand for investments there is a lag with fixed duration
of the action of the accelerator, whereas on the part of the supply there is
a continuously distributed lag. In this chapter, we shall demonstrate the
application of methods of the theory of nonlinear oscillations in the class
of differential equations with a lag and of integro-differential equations
that may have a continuously distributed lag.

\section{Instability of price dynamics in Fisher's model}

In modern economic literature, there is a sufficiently detailed qualitative
description of the mechanism of the formation of domestic prices of the
output based on an analysis of the dynamics of financial flows in
export-import operations. At the same time, special attention is paid to
such an important factor as an index of the trade balance whose surplus or
deficit determines the direction of price change.

Before proceeding with the consideration of a formalized mathematical model
based on classical Fisher's macroeconomic equation, it is necessary to put
forward a number of assumptions \cite{[41]}:

1) a free-trade scheme, without any influence of governments and
monopolistic structures, is considered;

2) the rate of the national income is considered to be sufficient, and the
price rate is determined on the basis of qualitative theory of the currency
of money;

3) in the course of the considered period, changes in the supply of money
stock are stipulated exclusively by a surplus (or deficit) of the balance of
trade;

4) since exchange rates are considered to be fixed, they can be set equal to
unity, which is equivalent to complete unification of international
operations;

5) transportation expenses, insurances and other costs are not taken into
account either for commodity or financial flows.

In what follows, we use the following notation: $Q$ is money supply; $V$ is
the rate of the currency of money; $Y$ is the national income rate (a
constant); $P$ is the index of the domestic price; $P_{M}$ is the index of
the export price (a constant); $M$ is the volume of import; $X$ is the
volume of export.

The basic equation of Fisher's model has the form%
\begin{equation}
QV=PY.  \label{4.1}
\end{equation}

The volume of export is a decreasing function of the domestic price:%
\begin{equation*}
X=X\left( P\right) ,\quad \frac{\partial X}{\partial P}<0.
\end{equation*}%
On the contrary, the volume of import is an increasing function of the
domestic price:%
\begin{equation*}
M=M\left( P\right) ,\quad \frac{\partial M}{\partial P}>0.
\end{equation*}

A relation of the form $P^{\ast }X\left( P^{\ast }\right) -P_{M}M\left(
P^{\ast }\right) =0$ characterizes a condition of trade balance equilibrium,
and it is assumed that this algebraic equation has positive solutions that
determine equilibrium values of the domestic price $P^{\ast }$.

A disturbance of equilibrium is accompanied by changes in the money supply
and is expressed by means of the following equation:%
\begin{equation}
\frac{dQ}{dt}=PX\left( P\right) -P_{M}M\left( P\right) .  \label{4.2}
\end{equation}

From expression (\ref{4.1}), it follows that a change in the money supply
leads to a change in the domestic price $P=P\left( t\right) $. We assume
that this change is not instantaneous, i.e., there is a time lag determined
by a constant positive quantity $\tau $. In this case, equation (\ref{4.1})
takes the form%
\begin{equation}
\frac{dP\left( t\right) }{dt}=\frac{V}{Y}\frac{dQ\left( t-\tau \right) }{dt}.
\label{4.3}
\end{equation}

From (\ref{4.2}) and (\ref{4.3}), we obtain a differential-difference
equation that describes the dynamics of the domestic price:%
\begin{equation}
\frac{dP\left( t\right) }{dt}=\frac{V}{Y}\left\{ P\left( t-\tau \right) X%
\left[ P\left( t-\tau \right) \right] -P_{M}M\left[ P\left( t-\tau \right) %
\right] \right\} .  \label{4.4}
\end{equation}

Let us assume that the export function $X\left( P\right) $ and import
function $X\left( P\right) $ are inherently nonlinear and can be expanded in
a Taylor series up to the third power in the neighborhood of the state of
equilibrium $P^{\ast }$, i.e.,%
\begin{equation*}
X\left( P\right) =X_{0}+X_{1}\left( P-P^{\ast }\right) +X_{2}\left(
P-P^{\ast }\right) ^{2}+X_{3}\left( P-P^{\ast }\right) ^{3}+O\left(
P^{4}\right) ,
\end{equation*}%
\begin{equation}
M\left( P\right) =M_{0}+M_{1}\left( P-P^{\ast }\right) +M_{2}\left(
P-P^{\ast }\right) ^{2}+M_{3}\left( P-P^{\ast }\right) ^{3}+O\left(
P^{4}\right) ,  \label{4.5}
\end{equation}%
where $X_{i}=\frac{\partial ^{i}X\left( P^{\ast }\right) }{i!\partial P^{i}}$%
, $M_{i}=\frac{\partial ^{i}M\left( P^{\ast }\right) }{i!\partial P^{i}}$,
with $i=\overline{0,3}$, are corresponding derivatives of the functions $%
X\left( P\right) $ and $M\left( P\right) $ at the point of equilibrium $%
P^{\ast }$. We introduce a new variable $\bar{P}\left( t\right) =P\left(
t\right) -P^{\ast }$ that has the meaning of a deviation of the domestic
price from its equilibrium value. In this case, taking into account (\ref%
{4.5}), equation (\ref{4.4}) reduces to the following:%
\begin{equation}
\frac{d\bar{P}}{d\bar{t}}=\frac{\tau V}{Y}\left[ G_{1}\bar{P}\left( \bar{t}%
-1\right) +G_{2}\bar{P}^{2}\left( \bar{t}-1\right) +G_{3}\bar{P}^{3}\left( 
\bar{t}-1\right) +O\left( \bar{P}^{4}\right) \right] ,  \label{4.6}
\end{equation}%
where $\bar{t}=\tau t$, $G_{i}=X_{i-1}+P^{\ast }X_{i}-P_{M}M_{i}$, $i=%
\overline{1,3}$.

It is reasonable to begin an analysis of the process described by (\ref{4.6}%
) with a study of the conditions of local stability, restricting ourselves
only to the linear part, i.e.,%
\begin{equation}
\frac{d\bar{P}}{d\bar{t}}=\frac{\tau V}{Y}G_{1}\bar{P}\left( \bar{t}%
-1\right) .  \label{4.7}
\end{equation}

The characteristic equation for (\ref{4.7}) is given by%
\begin{equation}
\lambda -\frac{\tau VG_{1}}{Y}e^{-\lambda }=0.  \label{4.8}
\end{equation}

Using a well-known result of the theory of stability of
differential-difference equations \cite{[43]}, as applied to (\ref{4.7}), we
obtain the necessary and sufficient conditions of linear stability:%
\begin{equation}
0<-\frac{\tau VG_{1}}{Y}<\frac{\pi }{2}.  \label{4.9}
\end{equation}

From the form of the left-hand part of the double inequality (\ref{4.9}) it
follows that the quantity $G_{1}$ is negative, whereas the right-hand part
of (\ref{4.9}) sets the upper bound for the absolute value of $G_{1}$.

Condition (\ref{4.9}) has a rather transparent and meaningful
interpretation. To demonstrate this interpretation, we perform a
transformation of the initial parameters of the considered model (\ref{4.6}).

Let 
\begin{equation*}
G_{1}=X_{0}\left[ 1-\eta _{X}-\eta _{M}\right] ,
\end{equation*}%
where%
\begin{equation*}
\eta _{X}=\frac{P^{\ast }X_{1}}{X_{0}},\quad \eta _{M}=\frac{P^{\ast }M_{1}}{%
M_{0}},
\end{equation*}%
under the condition $P^{\ast }X_{0}=P_{M}M_{0}$.

The quantities $\eta _{X}$ and $\eta _{M}$ are elasticities of the export
and import functions with respect to the price $P$. As $X_{0}>0$, the
condition $G_{1}<0$ is equivalent to $\eta _{X}+\eta _{M}>1$, which
corresponds to the so-called Marshall-Lerner conditions \cite{[41]}. At the
same time, condition (\ref{4.9}) reduces to%
\begin{equation}
0<\eta _{X}+\eta _{M}<1+\frac{Y\pi }{2X_{0}V\tau }.  \label{4.10}
\end{equation}

Thus, the economic interpretation of the conditions of local stability
consists in the fact that the sum of the elasticities not only must be
larger than unity, but it also must be smaller than an additional critical
value. In other words, instability in the considered economic model may
arise not only when the sum of the elasticities is sufficiently small, but
also in the case when its value is much larger.

Let us study behavioral properties of the initial dynamic system (\ref{4.6})
in a small neighborhood of the bounds of the inequality (\ref{4.10}). First,
we consider the situation when stability is lost at the lower bound. To this
end, we introduce a small parameter $\nu _{1}=1-\eta _{X}-\eta _{M}$.

In this case, when the sign of $\nu _{1}$ changes, the eigenvalue of the
linearized problem passes through zero, and a stationary value $P^{\ast }$
may either not exist or split into several stationary states. This means
that a bifurcation of stationary solutions takes place.

The differential-difference equation (\ref{4.6}) can be represented as
follows:%
\begin{equation}
\frac{d\bar{P}}{d\bar{t}}=A_{1}\bar{P}\left( \bar{t}-1\right) +A_{2}\bar{P}%
^{2}\left( \bar{t}-1\right) +A_{3}\bar{P}^{3}\left( \bar{t}-1\right) ,
\label{4.11}
\end{equation}%
where $A_{i}=\frac{\tau V}{Y}G_{i}$, $i=\overline{1,3}$.

One should bear in mind that the quantity $A_{1}$ is small, i.e., $%
G_{1}=X_{0}\nu _{1}$. Additionally, we assume that $A_{2}$ is also small if
we introduce a small quantity $\nu _{2}=\frac{G_{2}}{X_{0}}$. Using the
techniques of the central manifold method \cite{[46]}, one can prove that
the differential-difference equation (\ref{4.11}), under the condition that $%
A_{1}$, $A_{2}$ are small and the time lag is finite, is topologically
equivalent to the differential equation%
\begin{equation}
\frac{d\bar{P}}{d\bar{t}}=A_{1}\bar{P}\left( \bar{t}\right) +A_{2}\bar{P}%
^{2}\left( \bar{t}\right) +A_{3}\bar{P}^{3}\left( \bar{t}\right)
\label{4.12}
\end{equation}%
in the neighborhood of $\bar{P}=0$.

By means of the linear change of variables $\tilde{P}=\bar{P}+\frac{A_{2}}{%
3A_{3}}$, equation (\ref{4.12}) is represented as follows:%
\begin{equation}
\frac{d\tilde{P}}{d\bar{t}}=\alpha _{1}+\alpha _{2}\tilde{P}+\alpha _{3}%
\tilde{P}^{3},  \label{4.13}
\end{equation}%
where%
\begin{equation*}
\alpha _{1}=\frac{2A_{2}^{3}}{27A_{3}^{2}}-\frac{A_{1}A_{2}}{3A_{3}},\quad
\alpha _{2}=A_{1}-\frac{A_{2}^{2}}{3A_{3}}.
\end{equation*}

The transformation $\tilde{P}\left( \bar{t}\right) =\left\vert \beta
\right\vert W\left( \bar{t}\right) $ yields an explicit form of the normal
Poincar\'{e} form for the differential equation (\ref{4.13}):%
\begin{equation}
\frac{dW}{d\bar{t}}=\beta _{1}+\beta _{2}W+SW^{3},  \label{4.14}
\end{equation}%
where%
\begin{equation*}
\beta =\frac{1}{\sqrt{A_{3}}},\quad \beta _{1}=\frac{\alpha _{1}}{\left\vert
\beta \right\vert },\quad \beta _{2}=\alpha _{2},\quad S=\func{sign}\beta
=\pm 1.
\end{equation*}

For definiteness, we set $S=-1$.

Equation (\ref{4.14}) can have three states of equilibrium. A "fold"
bifurcation is determined by a curve $R$ on the plane $\beta _{1},\beta _{2}$%
, given by a projection of the line%
\begin{equation*}
\tilde{A}:\left\{ 
\begin{array}{c}
\beta _{1}+\beta _{2}V-V^{3}=0, \\ 
\beta _{2}-3V^{2}=0%
\end{array}%
\right.
\end{equation*}%
to the parameter plane. By eliminating $V$ from these equations, we obtain
the projection:%
\begin{equation*}
R=\left\{ \left( \beta _{1},\beta _{2}\right) :4\beta _{2}^{3}+27\beta
_{1}^{2}=0\right\} .
\end{equation*}

The curve $R$ is called a semicubical parabola, and it has two branches $%
R_{1}$, $R_{2}$ that meet tangentially at the "cusp-of-the-beak" point (a
cusp bifurcation), for $\beta _{1}=\beta _{2}=0$. The corresponding
bifurcation diagram is represented in Fig. 4.1.

\FRAME{ftbpFU}{3.039in}{2.4976in}{0pt}{\Qcb{The diagram of the
"cusp-of-the-beak" bifurcation.}}{}{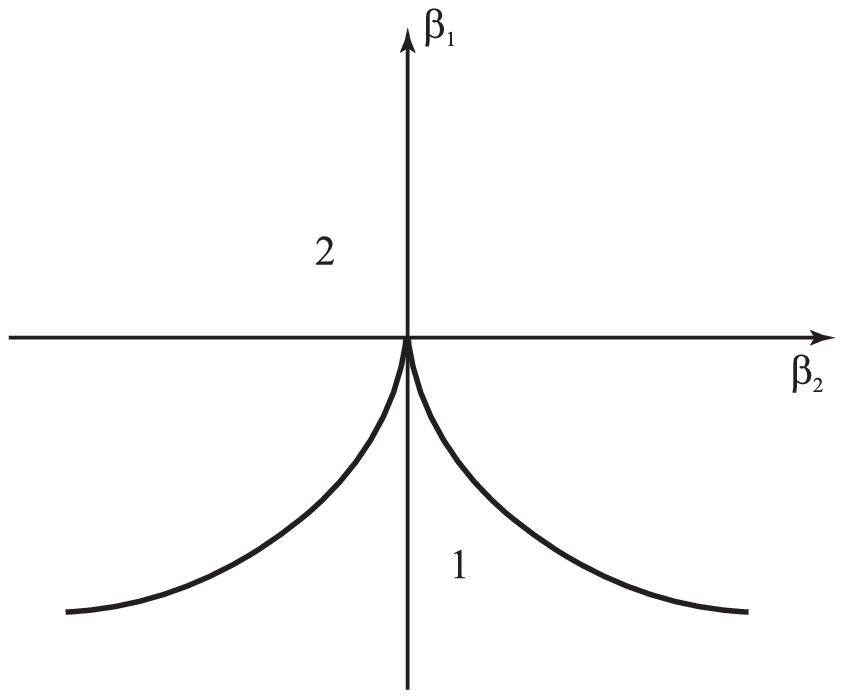}{\raisebox{-2.4976in}{\includegraphics[height=2.4976in]{fig_4-1.ps}}}

In region 1, in front of the boundary line, there are three states of
equilibrium: two stable states and one unstable state. In region 2, behind
the separation line, there is a single state of equilibrium which is stable.
A non-degenerate "fold" bifurcation takes place on crossing either $R_{1}$
or $R_{2}$ at any point of the parameter plane $\beta _{1},\beta _{2}$,
except for the origin. If the curve is crossed on passing from region 1 to
region 2, the right stable state of equilibrium merges with the unstable
one, and both of them disappear. Analogously, the left stable state of
equilibrium merges with the unstable one on the line $R_{2}$.

On approaching the "cusp-of-the-beak" point, in front of the region1, all
the three states of equilibrium merge into a single one as a triple root of
the right-hand side of the initial equation (\ref{4.14}). Of importance is
also the fact that, in the course of the transition from a stable regime to
an unstable one, the phenomenon of hysteresis is observed in (\ref{4.14}),
and a catastrophe occurs \cite{[46]}. The case $S=1$ can be considered
analogously.

Let us now clarify the situation when the sum of the elasticities $\eta
_{X}+\eta _{M}$ is close to the right bound of the inequality (\ref{4.10}).
We introduce into consideration a small parameter $\mu =-\left( A_{1}+\frac{%
\pi }{2}\right) $. Then, the differential-difference equation (\ref{4.6})
can be transformed into%
\begin{equation}
\frac{d\bar{P}}{d\bar{t}}=-\left( \mu +\frac{\pi }{2}\right) \bar{P}\left( 
\bar{t}-1\right) +A_{2}\bar{P}^{2}\left( \bar{t}-1\right) +A_{3}\bar{P}%
^{3}\left( \bar{t}-1\right) .  \label{4.15}
\end{equation}

The characteristic equation for the linear part of (\ref{4.15}) is given by%
\begin{equation}
\lambda +\left( \mu +\frac{\pi }{2}\right) e^{-\lambda }=0.  \label{4.16}
\end{equation}

We should find out when this equation has a pair of purely imaginary roots $%
\lambda =\pm i\omega $, $i^{2}=-1$, $\omega >0$.

If $\lambda =\pm i\omega $, then the conditions%
\begin{equation*}
\left( \mu +\frac{\pi }{2}\right) \cos \omega =0,\quad \omega -\left( \mu +%
\frac{\pi }{2}\right) \sin \omega =0
\end{equation*}%
hold.

From this fact, it follows that, for $\mu =0$, equation (\ref{4.16}) has a
pair of purely imaginary roots for $\omega =\frac{\pi }{2}$. It is not
difficult to show that (\ref{4.16}) has no roots with positive real parts.

As $\lambda $ is analytic with respect to $\mu $, the differentiation of (%
\ref{4.16}), at $\mu =0$, yields:%
\begin{equation*}
\frac{\partial \lambda }{\partial \mu }=\frac{\frac{\pi }{2}+i}{\frac{\pi
^{2}}{4}+1}.
\end{equation*}

Thus, all the conditions of Hopf's theorem on the existence of periodic
solutions are satisfied, because the real part of the derivative of the
eigenvalue with respect to the parameter does not vanish.

Based on the above-mentioned considerations, we shall demonstrate that Eq. (%
\ref{4.15}) has a family of periodic solutions $\bar{P}_{\varepsilon }\left( 
\bar{t}\right) $ ($\varepsilon >0$), where $\varepsilon $ is a measure of
the amplitude $\max_{\bar{t}}\left\vert \bar{P}_{\varepsilon }\left( \bar{t}%
\right) \right\vert $, and $\varepsilon $ is sufficiently small at that.

The problem reduces to a study of the bifurcation of the generation
(annihilation) of a cycle in the differential-difference equation (\ref{4.15}%
). To reduce this functional equation to a complex differential equation, we
use the method of the central manifold \cite{[37]}.

Equation (\ref{4.15}) contains a number of parameters; therefore, to
simplify further consideration, we make the substitution $\bar{P}%
_{\varepsilon }\left( \bar{t}\right) =-\frac{A_{1}}{A_{2}}u\left( \bar{t}%
\right) $.

For $\mu =0$, equation (\ref{4.15}) takes the form%
\begin{equation}
\frac{du\left( \bar{t}\right) }{d\bar{t}}=-\frac{\pi }{2}\left( u\left( \bar{%
t}-1\right) +u^{2}\left( \bar{t}-1\right) +\gamma u^{3}\left( \bar{t}%
-1\right) \right) ,  \label{4.17}
\end{equation}%
where $\gamma =\frac{X_{0}G_{3}}{G_{2}^{2}}$.

By the central manifold theorem, equation (\ref{4.17}) reduces to a
differential equation in a complex variable:%
\begin{equation}
\dot{Z}=\frac{i\pi }{2}Z+g_{20}\frac{Z^{2}}{2}+g_{11}Z\bar{Z}+g_{02}\frac{%
\bar{Z}^{2}}{2}+g_{21}\frac{Z^{2}\bar{Z}}{2}+\ldots ,  \label{4.18}
\end{equation}%
where%
\begin{equation*}
g_{20}=-g_{11}=g_{02}=\pi \bar{D},
\end{equation*}%
\begin{equation}
g_{21}=2\pi \left[ \left( \frac{2-11i}{5}-i\frac{3\gamma \pi }{4}\right) 
\bar{D}+\frac{7}{3}D\bar{D}+i\bar{D}^{2}\right] ,  \label{4.19}
\end{equation}%
\begin{equation*}
D=\frac{1+i\frac{\pi }{2}}{1+\frac{\pi ^{2}}{4}},\quad \bar{D}=\frac{1-i%
\frac{\pi }{2}}{1+\frac{\pi ^{2}}{4}}.
\end{equation*}

The existence of concrete values of the coefficients of the nonlinear part
of Eq. (\ref{4.18}) allows us to use the formulas of \cite{[37]} in order to
determine stability, the direction of generation, the period, and the
asymptotic form of periodic solutions of small amplitude of the limit cycle
that realizes the Andronov-Hopf bifurcation from the stationary state. Using
(\ref{4.19}), we obtain an explicit form of the first Lyapunov quantity:%
\begin{equation}
C_{1}\left( 0\right) =\frac{\pi }{1+\frac{\pi ^{2}}{4}}\left\{ \frac{2}{5}-%
\frac{\pi }{2}\left( \frac{11}{5}+\frac{3\gamma \pi }{4}\right) -i\left( 
\frac{\pi }{5}+\frac{11}{5}+\frac{3\gamma \pi }{4}\right) \right\} .
\label{4.20}
\end{equation}

The real part of (\ref{4.20}) is negative for%
\begin{equation}
\gamma >\gamma _{0}=\frac{16-44\pi }{15\pi ^{2}}=-0.826\ldots .  \label{4.21}
\end{equation}

This means that the limit cycle is stable if $\gamma >\gamma _{0}$, and it
is unstable if the condition (\ref{4.21}) is not fulfilled. For the stable
limit cycle, the following expressions for the major characteristics are
derived:

1) the amplitude is given by%
\begin{equation*}
\varepsilon =\left( \frac{20\mu }{15\gamma \pi ^{2}+44\pi -16}\right) ^{%
\frac{1}{2}};
\end{equation*}

2) the period is given by%
\begin{equation*}
T_{\varepsilon }=4\left( 1+\frac{2}{5\pi }\varepsilon ^{2}\right) ;
\end{equation*}

3) the asymptotic form of the periodic solution is given by%
\begin{equation*}
u_{\varepsilon }\left( \bar{t}\right) =2\varepsilon \cos \left( \frac{\pi 
\bar{t}}{2}\right) +2\varepsilon ^{2}\left( \frac{2}{5}\sin \left( \pi \bar{t%
}\right) -\frac{1}{5}\cos \left( \pi \bar{t}\right) -1\right) .
\end{equation*}

At the same time, the cycle is generated in the direction $\mu >0$, and the
emerging periodic solution is asymptotically stable. The corresponding
regime of the generation of self-oscillations is called soft. On the
contrary, if the condition $\gamma <\gamma _{0}$ is realized, an unstable
limit cycle takes place. The loss of stability with the generation of
self-oscillations occurs rigidly, i.e., a sharp transition (jump) to a new
stationary regime. In a realistic system, such a loss of stability results
in a catastrophe.

The most complicated behavior of the initial system (\ref{4.17}) is
exhibited in the situation when the parameter $\gamma $ is close to its
critical value $\gamma _{0}$, i.e., when the quantity $\xi =\gamma -\gamma
_{0}$ is small. In this case, one can observe in the considered system the
so-called Bautin bifurcation \cite{[37]} that is characterized by a
possibility of the coexistence of both the stable and unstable limit cycles.

To analyze qualitative properties of the above-mentioned bifurcation, we
should expand the right-hand side of (\ref{4.17}) in a Taylor series up to
fifth-order terms. After that, we should employ the central manifold method
to reduce the functional equation to a complex differential equation
involving nonlinear fifth-order terms, which is necessary for the evaluation
of the second Lyapunov quantity $C_{2}\left( 0\right) \neq 0$.

In this case, the complex differential equation has the form%
\begin{equation*}
\dot{Z}=Z\left( i\omega +\varepsilon _{1}+\varepsilon _{2}Z\bar{Z}+C_{2}Z^{2}%
\bar{Z}^{2}\right) ,
\end{equation*}%
where $\varepsilon _{1}=\varepsilon _{1}\left( \mu \right) $, $\varepsilon
_{2}=\varepsilon _{2}\left( \xi \right) $.

Depending on the sign of $\varepsilon _{1}$, $\varepsilon _{2}$ and $C_{2}$,
the following scenarios are possible:

1) $C_{2}<0$, $\varepsilon _{2}<0$. When $\varepsilon _{1}$ passes from
negative values to positive ones, the system softly achieves a stable
self-oscillation regime;

2) $C_{2}<0$, $\varepsilon _{2}>0$. When $\varepsilon _{1}$ passes from
negative values to positive ones, the system rigidly achieves a stable
periodic self-oscillation regime. It is generated before the loss of
stability by the state of equilibrium, together with an unstable oscillation
regime that settles on the state of equilibrium at the very moment when
stability is lost;

3) $C_{2}>0$, $\varepsilon _{2}<0$. The loss of stability is soft. However,
the generated cycle is quickly annihilated in the process of merging with an
unstable one, coming from a distance. After that, a new regime is rigidly
excited in the system;

4) $C_{2}>0$, $\varepsilon _{2}>0$. This is classical rigid excitation.

Consequently, whatever the sign of $C_{2}\left( 0\right) $, for
corresponding sign of $\varepsilon _{2}$, our analysis reveals a
qualitatively different, compared to the one-parameter case, phenomenon: for 
$C_{2}\left( 0\right) <0$ there exists a rigidly excited stationary regime,
whereas for $C_{2}\left( 0\right) >0$ a softly excited regime turns out to
be short-lived. In order to establish which of the two cases ($C_{2}\left(
0\right) <0$ or $C_{2}\left( 0\right) >0$) is realized in reality, one has
to perform scrupulous evaluation of the second Lyapunov quantity, which is
in itself a rather good exercise in symbolic transformations.

Thus, as a result of the study of dynamic properties of the
differential-difference equation (\ref{4.4}) in a small neighborhood of the
boundaries of the region of local stability, we can arrive at the conclusion
that there exists a bifurcation of co-dimension two (this fact is in itself
far from being trivial):

1) on the left boundary ($\eta _{X}+\eta _{M}=1$), a "cusp-of-the-beak"
bifurcation takes place;

2) on the right boundary ($\eta _{X}+\eta _{M}=1+\frac{Y\pi }{2X_{0}V\tau }$%
), a Bautin bifurcation of the limit cycle takes place.

In conclusion, we want to point out that the application of mathematical
methods to an analysis of concrete objects is associated with numerical
results and a corresponding meaningful interpretation. In this sense, the
role of qualitative theory of differential-difference equations is somewhat
different: it puts stress on a search for characteristic features of the
phenomenon as a whole, on qualitative forecasts of its behavior. The
objectives of the authors are the determination of irreducible topological
structures that form the phase portrait of the system. The applied part
consists in establishing a correspondence between these structures of the
phase space and the considered economic processes, together with carrying
out a bifurcation analysis. At the same time, we have to take into account
the properties of the realistic object that impose restrictions on both the
phase variables and the parameters of the initial equations \cite{[9]}.

\section{The cyclicity of innovation processes}

Presently, the humanity is concerned with a search for innovation ways of
stable development of civilization. A new paradigm of the 21st century,
i.e., the concept of stable development, has explicitly systematic,
synergetic character. By stable development we should understand a synthesis
of the necessities of stable economic, ecological and social evolution,
which is realized simultaneously on global, national and regional levels;
and, as is known, simultaneous cooperative action is the essence of the
synergetic effect.

In this work, by economic development we shall understand a substantially
nonlinear process, characterized by spasmodic transitions from one
stationary state to another. A fundamental basis of economic (innovation)
forecasting is formed by prediction theory of N. D. Kondratev \cite{[24]}
and innovation theory of Y. A. Shumpeter \cite{[47]}, further developed by
the modern Russian economist Yu. Yakovets \cite{[38]}.

The main theoretical prerequisites for the justification of qualitative
forecasting are the following:

1) a prediction of economic innovations is based on accounting for the
interaction of the laws of statics (that determines a multiple balance of
the functioning of the economic system), of cyclic dynamics (a combination
of the observed cycles of various duration), and of sociogenetics (the laws
of heredity, variability and selection in the dynamics of technological and
social-economic systems);

2) depending on the period and the intensity of influence on the economy,
evolutionary cycles differ substantially: there are medium-term, long-term
and super-long-term cycles. Innovation oscillations facilitate technological
crises accompanied by changes in the structure of innovations. At the
beginning of a super-long-term cycle, the most important seminal innovations
are initiated; once during a few centuries, they radically change the
structure of the economy by forming a new technological way of production.
Kondratev's half-century cycles are characterized by basic innovations that
determine the competitiveness of the production in the framework of a given
technological formation of the economy. Medium-term cycles (of 10-12 years
of duration) are caused by changes in the prevailing generations of
technology; they are realized in the cluster of basic innovations and in the
wave of improving ones. Thus, we may argue that innovation-technological
cycles of various periods serve as the basis of economic cycles of
corresponding duration: medium-term cycles (of 10-12 years of duration),
long-term (Kondratev's) cycles, and super-long-term (civilization) cycles 
\cite{[39]}.

A rather detailed description of the phases of the innovation cycle and
their application to an analysis of dynamic structure of the economy is
contained in the monograph \cite{[26]}. Conventionally, as the beginning of
the cycle, we shall take a decrease in the efficiency of the prevailing
generations of technics (technology) that leads to a decrease in the rate of
economic growth and a drop in the standard of living of a substantial part
of the population. This stimulates scientific and engineering activities in
the direction of obtaining new technological solutions.

However, conditions for their innovative application are generated only at
the end of the depression phase and in the phase of revival, when a renewal
of fixed capital takes place and the volume of investments grows, which
stimulates a demand for innovations. The rate of economic growth
accelerates, employment increases, and a growth in the standard of living of
the population is observed. By the end of the revival phase, these processes
reach their maximum. However, in the stability phase, against a background
of a considerable income rate, the rate of growth falls until a crisis
causes a sharp drop in the rate of growth and a decline in the gross
domestic product (GDP), which substantially reduces investments.

In the monograph by S. Yu. Glaziev \cite{[17]}, it is pointed out that,
simultaneously with the acceleration of economic development, the influence
of counteracting factors increases: this conclusion is drawn on the basis of
multiple statistical measurement of the dynamics of GDP for different
countries. As a result, economic growth either stabilizes or acquires cyclic
character. These arguments are illustrated by a mathematical model that
reflects feedback between the rate of economic growth and the rate of the
gross national product per capita. It is assumed that the influence of
counteracting factors increases with cumulative growth of GDP. In other
words, we assume that the rate of the growth of the cumulative volume of the
production of a new product depends on the average weighted cumulative
production volume in the past and not only on its volume at the moment. As a
result, the dynamics of the production volume can be described by the
integro-differential equation%
\begin{equation}
\dot{y}_{0}=y_{0}\left( r-\int_{-\infty }^{0}y_{0}\left( t+s\right) Q\left(
-s\right) ds\right) ,  \label{4.22}
\end{equation}%
where $y_{0}\left( t\right) $ is the cumulative volume of the production of
the new product;

$Q\left( t\right) $ is a function characterizing cumulative growth of the
production;

$r>0$ is a parameter that has the meaning of a technological limit of
production growth.

As a rule, as $Q\left( t\right) $, one employs a function with a
corresponding set of characteristic time lags. In terms of the theory of
automatic control, $Q\left( t\right) $ is an impulsive transition function
of a linear control system with an input signal $y_{0}\left( t\right) $. The
output of such a system is given by the quantity $u_{0}\left( t\right)
=\int_{-\infty }^{0}y_{0}\left( t+s\right) Q\left( -s\right) ds$, which is a
convolution of the functions $y_{0}\left( t\right) $ and $Q\left( t\right) $%
. In what follows, we shall use an operator representation of $Q\left(
t\right) $ in the form of a fractionally linear function of arbitrary order,
namely,%
\begin{equation}
Q_{n}\left( \lambda \right) =\frac{b_{n-1}\lambda ^{n-1}+\ldots
+b_{1}\lambda +b_{0}}{\lambda ^{n}+a_{n-1}\lambda ^{n-1}+\ldots
+a_{1}\lambda +a_{0}},  \label{4.23}
\end{equation}%
under the normalization condition%
\begin{equation*}
\int_{-\infty }^{0}Q\left( -s\right) ds=1.
\end{equation*}%
The nonlinear integro-differential equation (\ref{4.22}) has two states of
equilibrium:

1) $y_{0}^{\ast }=0$;

2) $y_{0}^{\ast }=r$.

The state of equilibrium $y_{0}^{\ast }=0$ is of no practical interest for
our analysis, because it implies complete absence of the GDP. The second
singular point, $y_{0}^{\ast }=r$, characterizing the limit of GDP growth,
is a major economic factor; hence, our objective will be a study of dynamic
properties of the process described by the integro-differential equation (%
\ref{4.22}) in the neighborhood of this state of equilibrium.

Let us rewrite Eq. (\ref{4.22}) in terms of a new variable $y=y_{0}-r$ that
has the meaning of a deviation of the production volume from its equilibrium
value:%
\begin{equation}
\dot{y}=-r\int_{-\infty }^{0}y\left( t+s\right) Q_{n}\left( -s\right)
d-y\int_{-\infty }^{0}y\left( t+s\right) Q_{n}\left( -s\right) ds.
\label{4.24}
\end{equation}

The characteristic equation of the linear part of (\ref{4.22}) has the form%
\begin{equation}
\lambda +rQ_{n}\left( \lambda \right) =0,  \label{4.25}
\end{equation}%
where $Q_{n}\left( \lambda \right) $ is defined by expression (\ref{4.23}).

Let us consider the simplest case $n=1$, $Q_{1}=\frac{b_{0}}{\lambda +a_{0}}$%
. To satisfy the normalization condition, it is necessary that $%
b_{0},a_{0}>0 $. Then, equation (\ref{4.25}) is represented in the form of
the quadratic equation%
\begin{equation*}
\lambda ^{2}+a_{0}\lambda +rb_{0}=0.
\end{equation*}

Note that, for $a_{0}^{2}\geq 4rb_{0}$, the state of equilibrium $y^{\ast
}=0 $ is a stable node. However, for $a_{0}^{2}<4rb_{0}$, in the vicinity of
the equilibrium point, stable oscillations with the attenuation rate $a_{0}$
are observed. This type of equilibrium is called a \textit{stable focus}. In
this system, a transition from a node to a focus has no bifurcation
character, which means that the system is stable.

In the sense of the diversity of dynamic behavioral properties, the
situation with $n=2$ is more interesting. In this case,%
\begin{equation*}
Q_{2}\left( \lambda \right) =\frac{b_{1}\lambda +b_{0}}{\lambda
^{2}+a_{1}\lambda +a_{0}},
\end{equation*}%
and the spectral equation (\ref{4.25}) takes the form of the cubic equation%
\begin{equation}
\lambda ^{3}+a_{1}\lambda ^{2}+\left( a_{0}+rb_{1}\right) \lambda +rb_{0}=0.
\label{4.26}
\end{equation}

If the coefficients of Eq. (\ref{4.26}) satisfy the relation%
\begin{equation}
a_{1}\left( a_{0}+rb_{1}\right) =rb_{0},  \label{4.27}
\end{equation}%
we obtain the solution $\lambda _{1,2}=\pm i\omega $, $\lambda _{3}=-a_{1}$,
where $\omega ^{2}=\frac{a_{0}b_{0}}{b_{0}-a_{1}b_{1}}$, $r=\frac{a_{1}a_{0}%
}{b_{0}-a_{1}b_{1}}$, $i^{2}=1$. As the parameters $\omega $ and $r$ are
positive, for the coefficients we have: $a_{1},a_{0},b_{0}>0$, and $%
b_{0}>a_{1}b_{1}$ at that.

The presence of a pair of purely imaginary eigenvalues in the spectrum of
the linear part of the integro-differential equation (\ref{4.24}) implies a
possibility of the excitation of a self-oscillation regime, i.e., an
occurrence of a limit cycle as a result of an Andronov-Hopf bifurcation.

In order to transform the integro-differential equation (\ref{4.24}) into a
system of nonlinear differential equations, we make the change of variables $%
x_{1}=y$, $x_{2}=u_{0}-r$, $x_{3}=\dot{u}_{0}$. As a result, we arrive at
the following system of ordinary third-order differential equations:%
\begin{equation*}
\dot{x}_{1}=-rx_{2}-x_{1}x_{2},
\end{equation*}%
\begin{equation}
\dot{x}_{2}=x_{3},  \label{4.28}
\end{equation}%
\begin{equation*}
\dot{x}_{3}=b_{0}x_{1}-\left( a_{0}+b_{1}r\right)
x_{2}-a_{1}x_{3}-b_{1}x_{1}x_{2}.
\end{equation*}

Now, we shall demonstrate the application of Hopf's bifurcation theorem to
the system of autonomous differential equations (\ref{4.28}).

First of all, we should decide on the choice of the bifurcation parameter
whose critical value allows for the existence of purely imaginary
eigenvalues, i.e., of those whose real part vanishes ($\func{Re}\lambda
_{1,2}=0$), whereas the imaginary part is nonzero ($\func{Im}\lambda
_{1,2}=\pm \omega $). It is convenient to take $r$ as the bifurcation
parameter and to study the properties of (\ref{4.28}) in a small
neighborhood of the critical value $r_{0}=\frac{a_{1}a_{0}}{b_{0}-a_{1}b_{1}}
$. That is to say, we consider the value $r=r_{0}+\mu $, where $\mu $ is a
small quantity. In this case, the eigenfrequency of the system (\ref{4.28})
in the linear approximation is defined by the expression%
\begin{equation}
\omega ^{2}=\frac{a_{0}b_{0}}{b_{0}-a_{1}b_{1}}=\frac{b_{0}r_{0}}{a_{1}}%
\quad \left( \mu =0\right) .  \label{4.29}
\end{equation}

In the new notation, the characteristic equation (\ref{4.26}) takes the form%
\begin{equation}
\lambda ^{3}+a_{1}\lambda ^{2}+\left( \omega ^{2}+b_{1}\mu \right) \lambda
+b_{0}\mu +a_{1}\omega ^{2}=0.  \label{4.30}
\end{equation}

For $\mu =0$, equation (\ref{4.30}) has the above-mentioned solution%
\begin{equation*}
\lambda _{1,2}=\pm i\omega ,\quad \lambda _{3}=-a_{1}\quad \left( \func{Re}%
\lambda _{1,2}>\func{Re}\lambda _{3}\right) .
\end{equation*}

Let us differentiate Eq. (\ref{4.30}) with respect to the parameter $\mu $.
For $\mu =0$, we obtain:%
\begin{equation*}
\lambda ^{\prime }\left( 0\right) =\frac{d\lambda }{d\mu }=\frac{\left(
b_{0}-a_{1}b_{1}\right) \omega +i\left( a_{1}b_{0}+b_{1}\omega ^{2}\right) }{%
2\omega \left( \omega ^{2}+a_{1}^{2}\right) },
\end{equation*}%
or%
\begin{equation}
\func{Re}\lambda ^{\prime }\left( 0\right) =\frac{b_{0}-a_{1}b_{1}}{2\left(
\omega ^{2}+a_{1}^{2}\right) },\quad \func{Im}\lambda ^{\prime }\left(
0\right) =\frac{a_{1}b_{0}+b_{1}\omega ^{2}}{2\omega \left( \omega
^{2}+a_{1}^{2}\right) }.  \label{4.31}
\end{equation}

Thus, all the conditions of Hopf's theorem are fulfilled, and we may argue
that there exists a bifurcation of the generation of a cycle from a complex
focus.

Using the change of variables%
\begin{equation*}
x_{1}=\frac{a_{1}}{b_{0}}y_{1}+\frac{\omega ^{2}}{b_{0}}y_{3},\quad x_{2}=%
\frac{y_{2}}{\omega }+y_{3},\quad x_{3}=y_{1}-a_{1}y_{3},
\end{equation*}%
we represent the system of nonlinear differential equations (\ref{4.28}) in
the form that is convenient for the construction of the normal Poincar\'{e}
form. As a result of the transformation, we obtain ($\mu =0$):%
\begin{equation*}
\dot{y}_{1}=-\omega y_{2}+F_{1}\left( y_{1},y_{2},y_{3}\right) ,
\end{equation*}%
\begin{equation}
\dot{y}_{2}=\omega y_{1}+F_{2}\left( y_{1},y_{2},y_{3}\right) ,  \label{4.32}
\end{equation}%
\begin{equation*}
\dot{y}_{3}=-a_{1}y_{3}+F_{3}\left( y_{1},y_{2},y_{3}\right) ,
\end{equation*}%
where%
\begin{equation*}
F_{i}\left( y_{1},y_{2},y_{3}\right) =A_{i}\varphi \left(
y_{1},y_{2},y_{3}\right) ,\quad i=\overline{1,3};
\end{equation*}%
\begin{equation*}
\varphi \left( y_{1},y_{2},y_{3}\right) =\frac{a_{1}}{b_{0}\omega }%
y_{1}y_{2}+\frac{a_{1}}{b_{0}}y_{1}y_{3}+\frac{\omega }{b_{0}}y_{2}y_{3}+%
\frac{\omega ^{2}}{b_{0}}y_{3}^{2};
\end{equation*}%
\begin{equation*}
A_{1}=-2\omega \func{Im}\lambda ^{\prime }\left( 0\right) ,\quad
A_{2}=-2\omega \func{Re}\lambda ^{\prime }\left( 0\right) ,\quad A_{2}=-2%
\func{Re}\lambda ^{\prime }\left( 0\right) .
\end{equation*}

Let us reduce the order of the system of differential equations by means of
the introduction of the new coordinates%
\begin{equation*}
z=y_{1}+iy_{2},\quad \bar{z}=y_{1}-iy_{2},\quad \nu =y_{3}.
\end{equation*}

As a result, equations (\ref{4.32}) are represented as follows:%
\begin{equation*}
\dot{z}=i\omega z+G\left( z,\bar{z},\nu \right) ,
\end{equation*}%
\begin{equation}
\dot{\nu}=-a_{1}\nu +H\left( z,\bar{z},\nu \right) ,  \label{4.33}
\end{equation}%
\begin{equation*}
G\left( z,\bar{z},\nu \right) =F_{1}\left( z,\bar{z},\nu \right)
+iF_{2}\left( z,\bar{z},\nu \right) ,\quad H\left( z,\bar{z},\nu \right)
=F_{3}\left( z,\bar{z},\nu \right) .
\end{equation*}

For the sake of a further analysis of the main characteristics of the limit
cycle, we use the central manifold method, which yields the relation%
\begin{equation}
\nu =W\left( z,\bar{z}\right) =w_{20}\frac{z^{2}}{2}+w_{11}z\bar{z}+w_{02}%
\frac{\bar{z}^{2}}{2}+O\left( \left\vert z\right\vert ^{3}\right) ,
\label{4.34}
\end{equation}%
where%
\begin{equation*}
w_{20}=\left( a_{1}+2i\omega \right) ^{-1}h_{20},\quad
w_{11}=a_{1}^{-1}h_{11},\quad w_{02}=\left( a_{1}-2i\omega \right)
^{-1}h_{02},
\end{equation*}%
\begin{equation*}
h_{ij}=\frac{\partial ^{i+j}}{\partial z^{i}\partial \bar{z}^{j}}H\left( z,%
\bar{z},0\right) ,\quad i+j=2.
\end{equation*}

As a result of the substitution of expressions (\ref{4.34}) into (\ref{4.33}%
), after certain necessary transformations, we obtain%
\begin{equation}
\dot{z}=i\omega z+G_{20}\frac{z^{2}}{2}+G_{11}z\bar{z}+G_{02}\frac{\bar{z}%
^{2}}{2}+G_{12}\frac{z^{2}\bar{z}}{2}+\ldots ,  \label{4.35}
\end{equation}%
where%
\begin{equation*}
G_{20}=-G_{02}=\frac{\left( A_{2}-iA_{1}\right) }{2\omega },\quad G_{11}=0,
\end{equation*}%
\begin{equation*}
G_{21}=\frac{-a_{1}\left( b_{0}-a_{1}b_{1}\right) \left( \omega \left(
2b_{0}-a_{1}b_{1}\right) +i\left( a_{1}b_{0}+2b_{1}\omega ^{2}\right)
\right) }{8\omega b_{0}^{2}\left( \omega ^{2}+a_{1}^{2}\right) \left(
4\omega ^{2}+a_{1}^{2}\right) }.
\end{equation*}

Now we possess all the necessary data for the evaluation of the first
Lyapunov quantity:%
\begin{equation}
C_{1}\left( 0\right) =\frac{G_{21}}{2}+\frac{i}{6\omega }\left\vert
G_{02}\right\vert ^{2}.  \label{4.36}
\end{equation}

From (\ref{4.35}) and (\ref{4.36}), it follows that%
\begin{equation}
\func{Re}C_{1}\left( 0\right) =\frac{-a_{1}\left( b_{0}-a_{1}b_{1}\right)
\left( 2b_{0}-a_{1}b_{1}\right) }{8b_{0}^{2}\left( \omega
^{2}+a_{1}^{2}\right) \left( 4\omega ^{2}+a_{1}^{2}\right) }<0,  \label{4.37}
\end{equation}%
since $b_{0}>a_{1}b_{1}$ and, accordingly, $2b_{0}>a_{1}b_{1}$.

Let us consider peculiarities of the limit cycle that is generated from a
complex focus when a pair of roots cross the imaginary axis. A stable focus
takes place when $\mu <0$. When $\mu $ passes through zero, the focus at the
origin loses its stability. For $\mu =0$, the focus at the origin is stable
but non-coarse: the phase curves approach zero exponentially.

For $\mu >0$, the phase curves, having moved away from the focus at a
distance proportional to $\mu ^{\frac{1}{2}}$, get wound round the stable
limit cycle. In other words, the loss of stability under the change of the
sign of $\mu $ is accompanied by the excitation of a stable limit cycle
whose radius grows as $\mu ^{\frac{1}{2}}$.

Thus, the stationary state loses stability, and a stable periodic regime is
generated in the direction $\mu >0$; its amplitude is proportional to the
square root of the deviation of the parameter from its critical value.
Corresponding excitation of self-oscillations is called soft.

Concerning the obtained limit cycle, it is not difficult to derive explicit
expressions for its main characteristics following the methods of \cite{[37]}%
:

a) the amplitude is given by%
\begin{equation}
\varepsilon =\left( -\frac{\func{Re}\lambda ^{\prime }\left( 0\right) }{%
\func{Re}C_{1}\left( 0\right) }\mu \right) ^{\frac{1}{2}}+O\left( \mu
^{2}\right) ;  \label{4.38}
\end{equation}

b) the period is given by%
\begin{equation*}
T=\frac{2\pi }{\omega }\left( 1+\tau _{2}\varepsilon ^{2}+O\left(
\varepsilon ^{4}\right) \right) ;
\end{equation*}%
\begin{equation}
\tau _{2}=\frac{1}{\omega }\left( \frac{\func{Im}\lambda ^{\prime }\left(
0\right) }{\func{Re}\lambda ^{\prime }\left( 0\right) }\func{Re}C_{1}\left(
0\right) -\func{Im}C_{1}\left( 0\right) \right) .  \label{4.39}
\end{equation}

The periodic solution itself, up to the choice of the initial phase, takes
the form%
\begin{equation*}
y_{1}=\func{Re}z,\quad y_{2}=\func{Im}z,\quad y_{3}=\func{Re}\left(
w_{20}z^{2}\right) ,
\end{equation*}%
\begin{equation}
Z=\varepsilon e^{\frac{2\pi it}{T}}+\frac{iG_{02}\varepsilon ^{2}}{6\omega }%
\left( e^{-\frac{4\pi it}{T}}-3e^{\frac{4\pi it}{T}}\right) +O\left(
\varepsilon ^{3}\right) .  \label{4.40}
\end{equation}

From (\ref{4.40}), it follows:%
\begin{equation*}
y_{1}\left( t\right) =\varepsilon \cos \left( \frac{2\pi t}{T}\right) -\frac{%
A_{1}a_{1}\varepsilon ^{2}}{3\omega ^{2}}\cos \left( \frac{4\pi t}{T}\right)
+\frac{A_{2}a_{1}\varepsilon ^{2}}{6\omega ^{2}}\sin \left( \frac{4\pi t}{T}%
\right) ,
\end{equation*}%
\begin{equation}
y_{2}\left( t\right) =\varepsilon \sin \left( \frac{2\pi t}{T}\right) -\frac{%
A_{2}a_{1}\varepsilon ^{2}}{3\omega ^{2}}\cos \left( \frac{4\pi t}{T}\right)
-\frac{A_{1}a_{1}\varepsilon ^{2}}{6\omega ^{2}}\sin \left( \frac{4\pi t}{T}%
\right) ,  \label{4.41}
\end{equation}%
\begin{equation*}
y_{3}\left( t\right) =\frac{A_{3}a_{1}}{\omega ^{2}+a_{1}^{2}}\left(
-y_{1}^{2}\left( t\right) +\frac{a_{1}}{\omega }y_{1}\left( t\right)
y_{2}\left( t\right) +y_{2}^{2}\left( t\right) \right) .
\end{equation*}

Returning to the initial variables, we obtain an approximate form of the
solution for the cumulative volume of the production of the new product:%
\begin{equation}
y_{0}\left( t\right) =r+\frac{a_{1}}{b_{0}}y_{1}\left( t\right) +\frac{%
\omega ^{2}}{b_{0}}y_{3}\left( t\right) ;  \label{4.42}
\end{equation}%
for its relative speed (rate) of growth, we have:%
\begin{equation}
\frac{\dot{y}_{0}\left( t\right) }{y_{0}\left( 0\right) }=-x_{2}\left(
t\right) =\frac{y_{2}\left( t\right) }{\omega }-y_{3}\left( t\right) .
\label{4.43}
\end{equation}

The above-described mechanism of the generation of an economic cycle has a
number of characteristic features that cannot be explained within the
framework of classical linear theory.

\textit{Firstly}, in contrast to the linear model, economic oscillation
processes in the form (\ref{4.41})-(\ref{4.43}) do not possess symmetry.
This means that, inside the cycle, its phases may be different both with
respect to their form and meaning. In particular, expansion and decline have
different duration. \textit{Secondly}, according to (\ref{4.38}), the
amplitude of oscillations depend on internal parameters of the model, and it
is not an entirely external characteristic.

\textit{Thirdly}, the period of the cycle is a function of its amplitude,
which, to a certain extent, explains the irregularity of cyclic dynamics.

Furthermore, the evolution curve of the cumulative national product is a
superposition of oscillations with different frequencies with respect to a
certain trend. Such a trend can be represented, for example, by a logistic
curve: in our case, logistics easily manifests itself when the lag $Q\left(
t\right) $ is switched off.

As an illustration of the justification of the application of the
mathematical models of economic development employed in this study, we
should remind the reader of the statement of Y. A. Shumpeter quoted in \cite%
{[17]}: "Since a long time ago, experts in political economy have a habit to
mention 'trends' and 'lags'; they are probably aware of the fact that
businessmen react not only to given quantities but also to the rate of their
changes, not only to the existing quantities but also to those that are
expected in the future. In any case, in the recent years, exact theories of
lagging adaptation, of expected actions, etc. have been developed. Technical
means have been developed or borrowed from other fields. As regards the
latter, the introduction to economic theory of functional, developed by Vito
Volterra, has been the most important event... As it seems, the new methods
point to the possibility of a colossal variety of wave-like motions in the
economic life that can be applied to the explanation of cycles without any
reference to the principle of the realization of new combinations... These
new techniques of the analysis substantially extend our abilities to explain
the forms of the manifestation of reality...".

The results obtained in this study agree with the laws of cyclic dynamics:
they demonstrate the irregularity of the evolution of the economic system, a
periodic change of the phases of the cycles of complex systems and a change
of the cycles themselves \cite{[21]}.

This, in turn, implies the relevance of the prediction of cycles and of
crisis phenomena, of timely detection and identification of negative
tendencies of development for the methodology of social-economic
forecasting. At the same time, one should always bear in mind that cyclicity
is a general property of the behavior of a large number of dynamic systems
of various nature, and, when making extrapolation to the future of the
tendencies formed in the past, it is necessary to be able to apply powerful
modern techniques of the theory of nonlinear oscillations. This approach
allows us to provide a high quality analysis of crisis states of the studied
objects and systems, it also facilitates a search for optimal, efficient
ways of leaving these states in order to create anti-crisis programs of
economic and social development of the society.

\backmatter

\chapter{Instead of the conclusion}

Nowadays, the theory of economic cycles seems to be the most disputable
section of macroeconomics. What kinds of economic or non-economic factors
generate oscillation motion? What is the mechanism of their propagation in
the economy? Do cyclic oscillations form a constituent part of economic
growth, or should they be regarded as deviations from a long-term trend? How
should the government build an anti-cyclic policy, or, generally speaking is
such a policy always necessary? What mathematical problems are encountered
by researchers in their formalized description of oscillation processes in
the economy? Exhaustive answers to these and many other important questions
have not been found so far.

From the point of view of synergetic economics, there are no economic
evolutionary systems that always exhibit stability. An evolutionary system
always experiences transformation effects of external and internal forces
that are able to realize sudden structural changes, including cycles. We
want to clarify this statement by the example of an analysis of behavioral
properties of the phenomenon of a competitive interaction between different
economic subjects \cite{[23]}.

The notion of economic competition is a complicated and complex category: it
is built on the basis of a variety of approaches. An overwhelming majority
of researchers single out price competition and structural competition. In
the first approach, the mechanism of competition is realized at the expense
of price changes, whereas in the second approach, conditions of the
production of goods undergo competition. Defining the essence of
competition, one should take into account necessary constituents of the
process that can be conditionally subdivided into three groups: behavioral,
structural, and functional.

Behavioral understanding of competition has been interpreted since the time
of A. Smith as pair (without an agreement) competition for the most
lucrative selling conditions that occurs between sellers (or buyers). At the
same time, he considered price change to be the main method of competition.
Later, the behavioral properties of competition improved in the direction of
more accurate formulation of its objective and methods.

The structural constituent of competition is formed by an analysis of the
whole market or of its segments with the aim to determine the degrees of
freedom of the seller and of the buyer. The functional filling of
competition contains innovation, that is, a competition between the old and
the new, etc.

Before turning to a direct analysis of peculiarities of the models of
economic competition, it is necessary to point out the following: economic
systems, as a rule, are far from their equilibrium, they are open to
commodity and money flows, they have complex inhomogeneous structure and a
regulation system of the endogenic medium under the influence of exogenic
factors. Therefore, mathematical formalization of the processes of economic
competition is, in itself, of considerable difficulty. This should be
contrasted with the situation in physics, chemistry and biology where
mathematics has already become a natural language of the description of the
observed processes. With regard to specific properties of economic
phenomena, one refers specifically to mathematical models in economics.
Here, by the model one should understand rather rude abstraction and
idealization that represents mathematical formalization not of the
developing system of the economic space (market) itself but only of some
qualitative and quantitative characteristics of the processes that flow
therein. A general feature of many phenomenological models of the market
economy is the presence of autocatalytic (by analogy with biophysics and
population dynamics) terms that determine the possibility of growth, of the
facts of the appearance of unstable stationary states that facilitate the
excitation of self-oscillation and quasistochastic regimes.

Complex processes in the systems of market self-regulation are stipulated by
the presence in the structural schemes of feedback contours (loops) (e.g.,
the "invisible hand" of Adam Smith); they are of both positive and negative
types, which, in turn, predetermines the formulation of the problem of the
study of structural stability of the considered objects and systems. In the
equations of local competitive interactions, the feedbacks are described by
nonlinear functionals whose character allows for the initiation of the
excitation of complex dynamic regimes accompanied by corresponding
attractors. The presence of nonlinear relations and of induced instabilities
implies the use of the synergetic paradigm as a means of the description of
competitive interactions in the economic medium. It is necessary to state
that the classical linear principle of superposition loses its validity in a
complex and nonlinear world represented by the market. In such a situation,
it is impossible to argue that the whole is equal to the sum of its
constituent parts. We should expect that the evolution of economic systems
constitutes a specific transformation of all the participants of the market
interaction by means of the establishment of a coherent relation and of
mutual adjustment of the parameters of their evolution. In this case, a
nonlinear synthesis should be understood not as the unification of rigidly
established, fixed objects, but rather as the unification of developing
structures that are characterized by different economic "ages" and by
"memory" and are positioned at different stages of the evolution.

Thus, the complexity of the economic system is related to coherence. By 
\textit{coherence} we shall understand the adjustment of the rates of
business activity of the participants of the market by means of diffusive
(mixing) dissipative processes that manifest themselves macroscopically as
seeming economic chaos.

The construction of a complex market competitive organization requires
coherent unification of the constituent substructures, the adjustment of
time constants of their evolution. As a result of such a substantially
nonlinear synthesis, different structures get into a unified temporal world,
that is, they acquire one and the same sharpening impetus and begin to
function at one and the same economic rate.

The development of the concept of the construction of the system of perfect
market competition leads to the understanding of the fact that arbitrary
structures under arbitrary conditions or mutual relations, positioned at
arbitrary stages of development, cannot be included in a unified, complex
social-economic formation. It seems that there exist a restricted choice of
formations and of methods of the construction of a complex evolutionary
whole. The selectivity and quantization of the methods of the unification of
the parts into a whole is related to an imperative demand for existence in
the whole temporal world. This is a natural, inherent basis for quantization
in the process of integration of complex dissipative economic systems. In
the case when the economic subjects, integrated into a unified competition
medium, differ in the sharpening impetus, they develop at different rates in
the vicinity of a given singularity, which, in turn, provokes undesirable
market imbalance. In the world economy, this means, for example, that the
development rate, the standard of living, information supply, etc. differ
substantially for different countries and give rise to a dangerous
difference in their potential.

To restore the efficiency of market competition, it is necessary to observe
certain topology of the "architecture" of the cross-relations. In other
words, if the overlap region is small, the economic subjects will develop
without "feeling" each other and will live in different temporal worlds. On
the contrary, if the overlap is excessively large, the structures will
quickly merge into strategic alliances in the given market and possibly, may
even form a unified dynamically growing structure with a growth limit equal
to the market volume, which leads to degeneration of competition.

Now, it is possible to proceed with the characterization of the mathematical
essence of economic competition for several participants of the market. In
this work, we have restricted ourselves to the case of the competition of
two economic subjects in a single market. To construct kinetic equations of
competitive interactions, it is necessary to make a number of assumptions
that characterize the considered phenomenon at a qualitative level. It seems
that the most important issue is an analysis of the balance of the growth
rates of the processes and factors that prevent positive development. Let us
assume that the growth rate of each competitor depends on a potential
increase in the volume of a certain type of goods and on an unrealized
possibility of growth for this type, as it has been the case for each
isolated participant of the market in the absence of competition. However,
the unused possibility of quantitative growth for this type of goods under
the condition of mixing of flows of goods is a more complex quantity. It
demonstrates the availability of a free place for this type of goods in the
presence of goods expansion by another participant of the market.

For two participants of market competition, we obtain a system of two
autonomous nonlinear differential equations that resemble models of
mathematical biophysics (of population dynamics) of the Volterra type \cite%
{[10]}, as well as their various modifications with various response
functions \cite{[33]}.

A theoretical analysis (including a bifurcation one) of such models in a
sufficiently complete manner is given in the book by A. D. Bazykin \cite{[3]}%
.

Of special interest for the practice of economic forecasting is the
formulation of criteria of closeness of the parameters of the system to
dangerous boundaries, when, on crossing these boundaries, the system in a
catastrophic way goes over to a qualitatively different state. In this case,
the character of the dynamics of the volumes of goods in the market changes
drastically: for example, a spasmodic transition from monotonous economic
growth to relaxation oscillations takes place. Such boundaries of the
regions of changes of the parameters of the considered dynamic system are
called \textit{bifurcation} boundaries.

A special position in the studies of models of the competitive economy is
occupied by processes that are characterized by cyclic behavior. The
ascertainment of hidden periodicity, a search for the so-called "economic
clocks" of various nature is always a valid issue in the studies of the
problems of economic dynamics.

Now, we shall discuss a mathematical model of economic competition that is
described by a system of two ordinary differential equations with quadratic
nonlinearity. Such systems appear in competitive dynamics when one uses a
Taylor-expansion approximation of the second order for response functions in
generalized Volterra models. The study of the issue of the excitation of
self-oscillation regimes (limit cycles) is a very interesting and difficult
problem of qualitative theory of differential equations. Up to now, there is
still no solution to Hilbert's sixteenth problem, posed in 1900: Find the
maximum number of limit cycles and determine their mutual arrangement in a
system of two differential equations with quadratic nonlinearity. Among the
main results for this system, we may note the following \cite{[15]}:

1) a complete classification of their singular points (a node, a focus, a
saddle, and a center) is given;

2) a complete qualitative analysis of the systems with a center-type
singular point is carried out. A topological classification of phase
portraits is given, and a corresponding partition of the parameter space of
such systems is performed;

3) it is proved that limit cycles of quadratic systems are convex;

4) limit cycles cannot surround a node-type singular point;

5) a system that has an algebraic limit cycle in the form of an ellipse has
no other limit cycles;

6) a quadratic system that has a non-coarse focus and a phase straight line
or two singular points with zero divergence has no limit cycles;

7) a system with four singular points, two of which are focuses, with one of
them being non-coarse, may have limit cycles only around one of the focuses;

8) the maximum number of limit cycles generated by a focus or a center is
equal to three;

9) a quadratic system may have at least four limit cycles arranged as 3:1,
i.e., three limit cycles around one focus and one limit cycle around the
other focus;

10) the total number of limit cycles in a quadratic system is finite.

Three bifurcations of limit cycles are known:

1) the bifurcation of the generation (annihilation) of a limit cycle from a
complex focus;

2) the bifurcation of a separatrix cycle from a homoclinic or heteroclinic
closed trajectory;

3) the bifurcation of a multiple limit cycle.

\textit{The first bifurcation} is studied completely only for the case of
quadratic systems: the number of limit cycles generated from the singular
point is equal to three. \textit{For a system with cubic nonlinearity, the
cyclicity of the singular point is equal to not less than eleven!}

As regards \textit{the second bifurcation}, we may argue that, at present,
there exists a complete classification of separatrix cycles, and the
cyclicity of most of them is know.

\textit{The third bifurcation} is the most complicated one, and it is
insufficiently explored.

Unfortunately, all these bifurcations are of local character: when studying
them, one considers only a certain sufficiently small neighborhood of a
singular point, of a separatrix limit cycle or a multiple limit cycle and a
corresponding sufficiently small neighborhood of the parameters of the
system.

A final solution of Hilbert's sixteenth problem requires a complete
qualitative study of the system as a whole (both on the whole phase plane
and in the whole parameter space), that is, global theory of bifurcations is
needed. Besides, all local bifurcations of limit cycles should be joined
together.

Our detailed characterization of bifurcation properties of cyclic dynamics
of the competitive interaction is by no means casual. Exactly here, the
instability properties of the system with respect to small deviations of the
parameters manifest themselves most strikingly. Only in nonlinear systems
near bifurcation boundaries, qualitative differences in the character of the
behavior of the considered object are observed. An example of such
restructuring of topology is provided by the transition from a stable
aperiodic region to an unstable self-oscillation regime that occurs in a
catastrophic manner. On the phase plane, this is illustrated by \textit{%
separatrices}, i.e., lines that separate different attraction regions
(attractors). In other words, if a small perturbation "throws" the system
over a separatrix, it gets into the zone of influence of another attractor,
which cardinally restructures the phase portrait.

We have already studied in sufficient detail the qualitative peculiarities
of market dynamics for two participants of competition. Certainly, this is
only a particular case of complex organization of the economic system. It
seems that the perfection of the market relations should be directed towards
an increase in the quantity of the participants of the market.

As is well known, the appearance in the market of a third participant of
competition may initiate in the system a chaotic regime accompanied by the
appearance of a new type of the attractor. This "strange" attractor
radically changes the dynamics of the system of competitive relations, which
substantially narrows the horizon of economic forecasting. Therefore, the
most important idea that follows from synergetics is that stable development
and the dynamically developing process of the evolution of the market
necessitates a certain portion of chaos, the spontaneity of development and
self-organization, and a certain portion of external management on the part
of state institutions that should be adjusted to each other. Both the two
extremes, i.e., pure chaos, spontaneous market mechanisms of selection and
the "survival of the strongest", on the one hand, and total external
management, full control and a protectionist policy vis-\`{a}-vis selected
structures, on the other hand, are unacceptable.

\end{document}